%% file: 000_draft_arxiv.tex
\documentclass[aip,jcp,12pt,showkeys,reprint,onecolumn]{revtex4-1} 
\usepackage{amsmath}
\usepackage{amsthm}
\usepackage{graphicx}
\usepackage[colorlinks]{hyperref}
\usepackage{dcolumn}
\allowdisplaybreaks
\usepackage[usenames,dvipsnames,table]{xcolor}
\usepackage{fancyhdr}
\usepackage{mathptmx}
\usepackage[small,raggedright]{titlesec}
\usepackage{wrapfig}

\begin{document}

\thispagestyle{plain}

\title
{
Development of multicomponent coupled-cluster method for investigation of 
multiexcitonic interactions
}

\thispagestyle{plain}

\author{Benjamin H. Ellis}
\affiliation
{
Department of Chemistry, Syracuse University, Syracuse, New York 13244 USA
}
\author{Somil Aggarwal}
\affiliation
{
Jamesville-DeWitt High School, DeWitt, New York 13214 USA
}
\author{Arindam Chakraborty}
\email[corresponding author:]{archakra@syr.edu}
\affiliation
{
Department of Chemistry, Syracuse University, Syracuse, New York 13244 USA
}

\date{\today}


\input{sec_abstract_jctc}
\maketitle 
\section{Introduction} \label{sec:intro}
\input{sec_intro}

\section{Theory} \label{sec:theory}
\input{sec_theory}

\section{Computer-assisted implementation} \label{sec:implement}
\input{sec_implementation}

\section{Physical system and computational details} \label{sec:comp}
\input{sec_computation}

\section{Results and discussion}\label{sec:results}
\input{sec_results}
\section{Conclusions}\label{sec:conclusions}
\input{sec_conclusion}

\begin{acknowledgements}
\input{sec_acknowledgements}
\end{acknowledgements}

\appendix
\section{Appendix A}\label{sec:appdx_a}
\input{appendix_a}

\newpage
\section{Appendix B}\label{sec:appdx_b}
\input{appendix_b}

\bibliography{new_bib_ehcc_1}

\end{document}

%% file: sec_abstract_jctc.tex
\begin{center}
\begin{abstract}
Multicomponent systems are defined as chemical systems that require a quantum mechanical  description of two or more different types of particles. 
Non-Born-Oppenheimer electron-nuclear interactions in molecules, electron-hole interactions in electronically excited nanoparticles, and electron-positron interactions  are examples of physical systems that require a
multicomponent quantum mechanical formalism.  
The central challenge in the theoretical treatment of multicomponent systems is capturing the many-body correlation effects that exist not only between particles of identical types (electron-electron) but also between particles of different types (electron-nuclear and electron-hole).
In this work, the development and implementation of multicomponent coupled-cluster (mcCC) theory for treating particle-particle correlation in multicomponent systems is presented.
This method provides a balanced treatment of many-particle correlation in a general multicomponent system while maintaining a size-consistent and size-extensive formalism. 
The coupled-cluster ansatz presented here is the extension of the electronic structure CCSD formulation for multicomponent systems and is defined as
$\vert \Psi_\mathrm{mcCC} \rangle = e^{T_1^\mathrm{I}+T_2^\mathrm{I}+T_1^\mathrm{II}+T_2^\mathrm{II}+T_{11}^\mathrm{I,II}+T_{12}^\mathrm{I,II}+T_{21}^\mathrm{I,II}+T_{22}^\mathrm{I,II}}\vert 0^\mathrm{I}0^\mathrm{II}\rangle$. 
The cluster amplitudes in the mcCC wave function were obtained by projecting the mcCC Schr\"{o}dinger equation onto a direct product space of singly and doubly excited states of type I and II particles and then solving the resulting mcCC equations iteratively. 
These equations were derived using automated application of the generalized Wick's theorem and were implemented using computer-assisted source code generation approach.  
The applicability of the mcCC method was  demonstrated by computing biexciton binding energies for multiexcitonic systems and benchmarking the results against full configuration interaction calculations. 
The results demonstrated that  connected cluster operators that generate simultaneous excitation in type I and type II space  are critical for capturing electron-hole correlation in multiexcitonic systems. 
\end{abstract}
\end{center}

%% file: sec_intro.tex
Systems that are made up of more than one type of quantum mechanical particles are defined as multicomponent systems.  
Multicomponent systems are ubiquitous in chemistry. 
Atoms and molecules, which are the building blocks of complex chemical systems, contain both electrons and nuclei and are intrinsically multicomponent in nature. 
Similarly, molecules bound to other sub-atomic particles, such as positrons bound to molecular substrates~\cite{sirjoosingh2013reduced} and muon-substituted compounds~\cite{hudson2013zero} also form multicomponent systems.
Multicomponent terminology can also be extended to systems containing quasiparticles,~\cite{brus1984electron} such as multiexcitons which are useful for describing electronic excitations in many-electron systems.
\par
A multicomponent system is intrinsically a collection of interacting single component systems.  
Consequently, it inherits all the complexity associated with treatment of correlation in single component systems. 
The central challenge in theoretical investigations of multicomponent systems is accurate description of the particle-particle correlation that exists not only between identical particles but also between particles of different types.   
The multicomponent wave function is a mathematically complex quantity and it is desirable to introduce simplifications to the exact form of the wave function for practical applications. 
For molecules, the Born-Oppenheimer (BO) approximation is a well-known approximation that introduces parametric dependence of the nuclear coordinates in the electronic wave function. As a consequence of the BO approximation, the exact multicomponent Schr{\"o}dinger equation can be expressed as a set of two coupled single component equations for electrons and nuclei, respectively.  
Although the BO is a very useful approximation, it important to recognize that it is still an approximation and there is an ever-increasing collection of experimental and theoretical findings that demonstrate its limitations.~\cite{cafiero1,Kreibich2001,Ishimoto2006,Iordanov20039489, Chakraborty2008,Kylanpaa2012,Webb20024106,Udagawa2014, subotnik2015requisite,trivedi2015decoherence,white2015non, marciniak2015xuv,gherib2015mixed,hammes2015proton}  
There are also other multicomponent quantum mechanical systems where the BO  
approximation is not an useful approximation.  
For example, electron-positron~\cite{Oba20141146,Sabry2011,Tachikawa19945925,Tachikawa20112701} and multiexitonic (electron-hole) systems~\cite{Corni2003436,elward_dotsize,elward_heterojxn,Elward2012,Elward2012182,Elward20145224,Franceschetti20062191,McDonald2012,Luo2009,Narvaez2005,PhysRevB.60.1819,PhysRevLett.91.056404,Shumway20011553161,Shumway2006273, Sundholm201296,vanska20064035,Wimmer2006} are systems where introducing BO separation leads to unacceptable deviation from qualitative results. 
\par
In this work, we focus on multicomponent treatment of only two different types of particles, which are denoted as type I and II for the remainder of this article.
This approach assumes that the coordinates of the heavy nuclei are BO separated from the coordinates of the type I and II particles.
Consequently, the exact multicomponent wave function is approximately factored as 
\begin{align}
	\Psi_\mathrm{exact}(\mathbf{r}^\mathrm{I},\mathbf{r}^{\mathrm{II}},\mathbf{R}^\mathrm{BOS})
	\approx 
	\Psi(\mathbf{r}^\mathrm{I},\mathbf{r}^{\mathrm{II}};\mathbf{R}^\mathrm{BOS})
	\chi(\mathbf{R}^\mathrm{BOS}) .
\end{align}
In the above expression, BO separated coordinates are collectively represented as $\mathbf{R}^\mathrm{BOS}$. 
The exact multicomponent Hamiltonian be written as the sum of two operators
\begin{align}
	H_\mathrm{exact}(\mathbf{r}^\mathrm{I},\mathbf{r}^{\mathrm{II}},\mathbf{R}^\mathrm{BOS})
	&= H(\mathbf{r}^\mathrm{I},\mathbf{r}^\mathrm{II};\mathbf{R}^\mathrm{BOS}) + T_{\mathbf{R}^\mathrm{BOS}}
\end{align} 
where, $T_{\mathbf{R}^\mathrm{BOS}}$  is the kinetic energy operators associated with  the BO separated coordinates. 
The general form of the Hamiltonian for the multicomponent system is defined as
\begin{align}
	H(\mathbf{r}^\mathrm{I},\mathbf{r}^\mathrm{II};\mathbf{R}^\mathrm{BOS}) 
	= H^\mathrm{I}(\mathbf{r}^\mathrm{I};\mathbf{R}^\mathrm{BOS})
	+ H^\mathrm{II}(\mathbf{r}^\mathrm{II};\mathbf{R}^\mathrm{BOS}) + 
	V^\mathrm{I,II}(\mathbf{r}^\mathrm{I},\mathbf{r}^\mathrm{II})
\end{align} 
where $H^\mathrm{I}$ and $H^\mathrm{II}$ are single component Hamiltonians for particle type I and II, respectively. 
The interaction between the two types of particles is described by the potential energy term, $V^\mathrm{I,II}$.
The total multicomponent Hamiltonian also admits parametric dependence on coordinates of particles (denoted by $\mathbf{R}^\mathrm{BOS}$) which have been assumed to be BO-separated (BOS) from type I and II particles in the multicomponent wave function.
Consequently, the multicomponent Hamiltonian presented above does not contain any kinetic energy operators associated with the BOS coordinates.
The BOS coordinates in $H$ are the generators of the external potentials $v^\mathrm{I}_\mathrm{ext}(\mathbf{r}^\mathrm{I};\mathbf{R}^\mathrm{BOS})$ and  $v^\mathrm{II}_\mathrm{ext}(\mathbf{r}^{\mathrm{II}};\mathbf{R}^\mathrm{BOS})$ experienced by the type I and II particles.
For example, in case of multicomponent molecular systems, type I and II particles can be represented by electrons and protons and the remaining nuclei in the molecules will be treated by the BOS coordinates.
In case of quasiparticle systems, the (quasi) electrons and holes are treated as type I and II particles while all the nuclei in the system are treated using BOS coordinates.
In both cases, the BOS coordinates will be responsible for the generation of the external potential  experienced by the type I and II particles. 
This form of the multicomponent Hamiltonian has been used to describe various multicomponent interactions such as  
electron-proton interaction,~\cite{Ko2011,Sirjoosingh2012,Swalina2012,Sirjoosingh2013,Tubman2014,Brorsen2015,Sirjoosingh2015,Ishimoto2006,Ishimoto20092677}
electron-positron interaction,\cite{Oba20141146,Tachikawa19945925,Tachikawa20112701}
electron-muon interaction,\cite{Moncada201316} and electron-hole quasiparticle interaction.
\cite{elward_dotsize,elward_heterojxn,Elward2012,Elward2012182,Elward20145224,vanska20064035,Sundholm201296,Wimmer2006}
It is important to note that this approach is different from other multicomponent approaches such as the exact factorization method~\cite{Abedi2010,Abedi2012,Agostini20133625,Abedi2013,Albareda2014,Suzuki2014,Abedi2014,Chiang2014,Cederbaum2015129}
and multicomponent ($N$-particle) density functional theory~\cite{Kreibich2001} that involve non-BO treatment of all particle types in a chemical system.
\par
The overarching goal of this work is to present an approximate solution of the following multicomponent Schr{\"o}dinger equation
\begin{align}
	H 	\Psi(\mathbf{r}^\mathrm{I},\mathbf{r}^{\mathrm{II}};\mathbf{R}^\mathrm{BOS})
	= E(\mathbf{R}^\mathrm{BOS}) \Psi(\mathbf{r}^\mathrm{I},\mathbf{r}^{\mathrm{II}};\mathbf{R}^\mathrm{BOS})
\end{align}
using a multicomponent coupled-cluster (mcCC) ansatz for the many-particle wave function.
Coupled-cluster (CC) theory has been used successfully for studying electron-electron correlation in many-electron systems. \cite{Feng20151636,Klopper200911679,Krogh2001578,Laidig1986887,Mahapatra2010,Miliordos20147568,peterson_dunning,Samdal20153375,Dutta20031610,czako2014646, Bartlett19891697,Hirata20013919,Li20108591,Liu2015,Sous2014616,Stanton19937029}
In the context of multicomponent systems, the CC ansatz provides a balanced framework for a size-consistent and size-extensive treatment of  many-particle correlation.
Application of CC to excitons has been demonstrated earlier by  Sundholm et al.~\cite{Sundholm201296} and V{\"a}nsk{\"a} et al.~\cite{vanska20064035} for a two-band effective mass approximation Hamiltonian model. 
In this work we  present a  systematic derivation of the mcCC equation which is based on a multicomponent Hartree-Fock vacuum state. 
This is a general formulation that does not assume any \textit{a priori} knowledge of particle types in the derivation. 
CC is a well known method with large amounts of information available in literature that describes the theory and implementation in electronic structure theory.
~\cite{szabo1996modern,shavitt2009many,Bartlett2007291,eltit,PurvisIII19821910,Scuseria19887382,Shiozaki2009,VanVoorhis20008873}
The focus of this article is not to repeat the well-known derivations, but to highlight the key differences between the single component (electronic structure) and multicomponent coupled-cluster equations.   
The developed mcCC method was used to investigate electron-hole interactions in multiexcitonic systems. 
Multiexcitonic systems are of particular interest due to the tremendous potential they have in photovoltaic and light harvesting applications. 
Theoretical investigations have been performed  using an array of methods including configuration interaction,\cite{Shumway20011553161,Corni2003436,vanska20064035,Narvaez2005,Fischer201110006} quantum Monte Carlo,\cite{Shumway20011553161,Shumway2006273} path integral Monte Carlo,\cite{McDonald2012,Wimmer2006} Green's functions,\cite{Rabani2010227} pseudopotentials,\cite{PhysRevLett.91.056404,Franceschetti2008} and coupled-cluster theory.  \cite{Henderson2014,Sundholm201296,vanska20064035}
In this work, we used the mcCC method for calculation of total biexciton energies and biexciton binding energies in multiexcitonic systems and the results were benchmarked against full configuration interaction method.  
The derivation of the mcCC equations, computer implementation of the t-amplitude equations, and details of the multiexcitonic systems are presented in the following sections.

%% file: sec_theory.tex
\input{subsec_theory_mchf_normal}

\input{subsec_mccc_eq}

%% file: subsec_theory_mchf_normal.tex
\subsection{Construction of the vacuum states}
\label{subsection:ref=ehHF}
The multicomponent Hamiltonian in second quantized notation is defined as
\begin{align}
\label{eq:sq_H}
  H &= \sum_{pq} \langle p \vert h^\mathrm{I} \vert q \rangle p^\dagger q \\ \notag
    &+ \sum_{pqrs} \langle pq \vert v^\mathrm{I,I} \vert rs \rangle p^\dagger q^\dagger s r \\ \notag
    &+ \sum_{p'q'} \langle p' \vert h^\mathrm{II} \vert q' \rangle p'^\dagger q' \\ \notag
    &+ \sum_{p'q'r's'} \langle p'q' \vert v^\mathrm{II,II} \vert r's' \rangle p'^\dagger q'^\dagger s' r' \\ \notag
    &+ \sum_{pqp'q'} \langle pp' \vert v^\mathrm{I,II} \vert qq' \rangle p^\dagger p'^\dagger q q' 
\end{align}
where, the unprimed and primed operators represent type I and II particles, respectively.
The form of the 1-particle and 2-particle operators are given
(in atomic units) as
\begin{align}
    \label{eq:hcore_alp}
	h^\alpha (\mathbf{r}^\alpha,\mathbf{R}^\mathrm{BOS})
	&=
	\frac{-\hbar}{2 m_\alpha} \nabla^2_\alpha + v_\mathrm{ext}^\alpha(\mathbf{r}^\alpha,\mathbf{R}^\mathrm{BOS})
	\quad \quad \alpha = \mathrm{I,II}, \\
    \label{eq:v_alpalp}
	v^{\alpha,\alpha}(\mathbf{r}^\alpha) &= q^{\alpha} q^{\alpha} \epsilon^{-1} r_{\alpha \alpha}^{-1}, \\
    \label{eq:v_one_two}
	v^\mathrm{I,II}(\mathbf{r}^\mathrm{I},\mathbf{r}^\mathrm{II}) &= q^{\mathrm{I}} q^{\mathrm{II}} \epsilon^{-1} r_\mathrm{I,II}^{-1}.
\end{align}
For electron-nuclear and electron-positron systems, the dielectric function is a constant ($\epsilon = 1$).
For multiexcitonic systems, the electron-hole interaction is screened, and the screening 
can be described either by a constant dielectric value\cite{elward_heterojxn,elward_dotsize,blanton_stark} or by a 
position dependent dielectric function.~\cite{PhysRevB.60.1819,PhysRevLett.91.056404}
\par
We define the mcCC wave function by the following exponential ansatz,
\begin{align}
	\Psi^\mathrm{I,II}_\mathrm{mcCC} (\mathbf{r}^\mathrm{I},\mathbf{r}^\mathrm{II};\mathbf{R}^\mathrm{BOS})  
	= e^{T^\mathrm{I,II}} \Phi^\mathrm{I,II}_0 (\mathbf{r}^\mathrm{I},\mathbf{r}^\mathrm{II};\mathbf{R}^\mathrm{BOS}),
\end{align}
where the BOS coordinates are shown explicitly using the standard semicolon convention.
Traditionally in single-reference electronic structure coupled-cluster theory, the vacuum state $\Phi_0$  is obtained from a Hartree-Fock (HF) calculation. 
For multicomponent systems, there are two different ways to construct the single component vacuum state.
The first method involves Hartree-Fock solution of the single component Hamiltonian as shown below with
the states labeled as $\vert \tilde{0}^{\mathrm{I}} \rangle$ and $\vert \tilde{0}^{\mathrm{II}}\rangle$ for type I and II particles,
\begin{align}
	\langle \tilde{0}^{\mathrm{I}} \vert H^\mathrm{I} \vert \tilde{0}^{\mathrm{I}} \rangle 
	&=
	\min_{\Phi^\mathrm{I}_\mathrm{SD} } 
	\langle \Phi^\mathrm{I}_\mathrm{SD} \vert H^\mathrm{I} \vert \Phi^\mathrm{I}_\mathrm{SD} \rangle, \\
	\langle \tilde{0}^{\mathrm{II}} \vert H^\mathrm{II} \vert \tilde{0}^{\mathrm{II}} \rangle 
	&=
	\min_{\Phi^\mathrm{II}_\mathrm{SD} } 
	\langle \Phi^\mathrm{II}_\mathrm{SD} \vert H^\mathrm{II} \vert \Phi^\mathrm{II}_\mathrm{SD} \rangle.
\end{align}  
The minimization in the above equations is performed over a set of single Slater determinants ($\Phi_\mathrm{SD}$). 
The total vacuum energy the above  determinants is given as
\begin{align}
	\langle \tilde{0}^{\mathrm{I}} \tilde{0}^{\mathrm{II}} \vert H \vert \tilde{0}^{\mathrm{I}} \tilde{0}^{\mathrm{II}} \rangle
	&= 
	\langle \tilde{0}^{\mathrm{I}} \vert H^\mathrm{I} \vert \tilde{0}^{\mathrm{I}} \rangle
	+\langle \tilde{0}^{\mathrm{II}} \vert H^\mathrm{II} \vert \tilde{0}^{\mathrm{II}} \rangle
	+\langle \tilde{0}^{\mathrm{I}} \tilde{0}^{\mathrm{II}} \vert V^\mathrm{I,II} \vert \tilde{0}^{\mathrm{I}} \tilde{0}^{\mathrm{II}} \rangle .
\end{align}
The second method for construction of the single component vacuum states
is by solution of the multicomponent HF equation. We label these determinants by 
as $\vert 0^{\mathrm{I}} \rangle$ and $\vert 0^{\mathrm{II}} \rangle$ for type I and II particles, respectively and they 
are determined using the following energy minimization procedure
\begin{align}
    \label{eq:mcHF_min}
	\langle 0^{\mathrm{I}} 0^{\mathrm{II}} \vert H \vert 0^{\mathrm{I}} 0^{\mathrm{II}} \rangle 
	&=
	\min_{\Phi^\mathrm{I}_\mathrm{SD},\Phi^\mathrm{II}_\mathrm{SD} } 
	\langle \Phi^\mathrm{I}_\mathrm{SD} \Phi^\mathrm{II}_\mathrm{SD} \vert H \vert 
	\Phi^\mathrm{I}_\mathrm{SD} \Phi^\mathrm{II}_\mathrm{SD} \rangle 
	.
\end{align}
The vacuum state $\vert 0^{\mathrm{I}}0^{\mathrm{II}}\rangle$
constructed from the multicomponent HF should be lower in energy
than the vacuum state $\vert \tilde{0}^{\mathrm{I}} \tilde{0}^{\mathrm{II}} \rangle$ obtained using the single component HF method due to the variational principle,
\begin{align}
	\langle 0^{\mathrm{I}} 0^{\mathrm{II}} \vert H \vert 0^{\mathrm{I}} 0^{\mathrm{II}} \rangle 
	\leq \langle \tilde{0}^{\mathrm{I}} \tilde{0}^{\mathrm{II}} \vert H \vert \tilde{0}^{\mathrm{I}} \tilde{0}^{\mathrm{II}} \rangle.
\end{align}
Because of this property, we have used multicomponent HF for construction of the 
single component vacuum states in all the calculations presented in \autoref{sec:results}.
Multicomponent HF method has been used in earlier work for treating electron-proton
correlation and the details of the method is not presented here to avoid repetition. 
\cite{Webb20024106,Iordanov20039489,Ishimoto2006,Ishimoto2008}
\subsection{Effective normal-ordered Hamiltonian}
\label{subsection:ref=normalH}
The use of multicomponent HF instead of single component HF as the vacuum state has important implications on the general form of the CC equations on the vacuum energy and normal-ordering  of the operators . 
The minimization procedure for multicomponent HF (shown in Eq. \eqref{eq:mcHF_min}) results in Fock operators for type I and II particles and are given as,
\begin{align}
    \label{eq:fock_typ1}
	f^\mathrm{I} = h^\mathrm{I} 
	             + v^\mathrm{I}_\mathrm{HF}
	             + \sum_{j'=1}^{N^\mathrm{II}} \langle j' \vert v^\mathrm{I,II} \vert j' \rangle \\
    \label{eq:fock_typ2}
	f^\mathrm{II} = h^\mathrm{II} 
	             + v^\mathrm{II}_\mathrm{HF}
	             + \sum_{j=1}^{N^\mathrm{I}} \langle j \vert v^\mathrm{I,II} \vert j \rangle 
                 .
\end{align}
However, the Fock operators for the two types of particles are not independent of each other and are coupled because of the presence of the I-II coupling term. The mcHF calculation is performed by iterative solution of the coupled SCF equation 
\begin{align}
	f^\mathrm{I} \chi^\mathrm{I}_i &= \epsilon^\mathrm{I}_i \chi^\mathrm{I}_i \\
	f^\mathrm{II} \chi^\mathrm{II}_{i'} &= \epsilon^\mathrm{II}_{i'} \chi^\mathrm{II}_{i'}  .
\end{align}
The eigenfunction of the multicomponent Fock operators from Eq. \eqref{eq:fock_typ1} and Eq. \eqref{eq:fock_typ2}
are used as the single-particle states for representing the creation and annihilation operators. 
The total vacuum energy obtained from the SCF step is additive and is made up of three components
\begin{align}
	\langle 0^\mathrm{I} 0^\mathrm{II} \vert H \vert 0^\mathrm{I} 0^\mathrm{II} \rangle =
	\langle 0^\mathrm{I} 0^\mathrm{II} \vert H^\mathrm{I} \vert 0^\mathrm{I} 0^\mathrm{II} \rangle  +
	\langle 0^\mathrm{I} 0^\mathrm{II} \vert H^\mathrm{II} \vert 0^\mathrm{I} 0^\mathrm{II} \rangle +
	\langle 0^\mathrm{I} 0^\mathrm{II} \vert V^\mathrm{I,II} \vert 0^\mathrm{I} 0^\mathrm{II} \rangle.
\end{align}
This relationship will be used in the definition of the normal-ordered Hamiltonian.
Using Eq. \eqref{eq:sq_H}, the single component normal ordered Hamiltonian is defined as
\begin{align}
	H^\mathrm{I}
	&=
	H^\mathrm{I}_\mathrm{N} + \langle 0^{\mathrm{I}} \vert H^\mathrm{I} \vert 0^{\mathrm{I}} \rangle
\end{align} 
where
\begin{align}
    \label{eq:h_typ1}
	H^\mathrm{I}_\mathrm{N}
	&=
	\sum_{pq}
	\langle p \vert h^\mathrm{I} \vert q \rangle \{p^\dagger q\}
	+ \sum_{pqi} \langle p i \vert v^\mathrm{I,I} \vert q i \rangle_\mathrm{A}
	\{ p^\dagger q \}
	+ \frac{1}{4}\sum_{pqrs} 
	\langle pq \vert v^\mathrm{I,I} \vert rs \rangle_\mathrm{A}
	\{ p^\dagger q^\dagger s r \} 
\end{align}
The subscript "A" in the above expression implies that both symmetric and antisymmetric combination of the integral
\begin{align}
	\langle pq \vert v^\mathrm{I,I} \vert rs \rangle_\mathrm{A}
	&=
	\langle pq \vert v^\mathrm{I,I} \vert rs \rangle
	-\langle pq \vert v^\mathrm{I,I} \vert sr \rangle .
\end{align} 
The type I-II coupling operator in Eq. \eqref{eq:v_alpalp} can be expressed as a sum of three normal-ordered operators as shown below
\begin{align}
    \label{eq:v12n}
	V^\mathrm{I,II} 
	&=
	V^\mathrm{I,II}_\mathrm{N} + W^\mathrm{I}_\mathrm{N} + W^\mathrm{II}_\mathrm{N} + \langle 0 0' \vert V^\mathrm{I,II} \vert 00' \rangle 
\end{align}
where
\begin{align}
    \label{eq:v12n_ws}
	V^\mathrm{I,II}_\mathrm{N}
	&=
	\sum_{pq} \sum_{p'q'} \langle pp'\vert v^{\mathrm{I,II}} \vert qq' \rangle
 \{ p^\dagger q \} \{p'^\dagger q'\} \\
    \label{eq:v12n_w1}
  W^\mathrm{I}_\mathrm{N}
  &=
  \sum_{pq} \sum_{i'} \langle pi'\vert v^{\mathrm{I,II}} \vert qi' \rangle
 \{ p^\dagger q \}  \\
    \label{eq:v12n_w2}
  W^\mathrm{II}_\mathrm{N}
  &=
  \sum_{i} \sum_{p'q'} \langle ip'\vert v^{\mathrm{I,II}} \vert iq' \rangle
 \{p'^\dagger q'\}
\end{align}
The total multicomponent Hamiltonian can then be written as sum of following normal-ordered operators,
\begin{align}
	H = H^\mathrm{I}_\mathrm{N} + H^\mathrm{II}_\mathrm{N} +
	    V^\mathrm{I,II}_\mathrm{N} + W^\mathrm{I}_\mathrm{N} + W^\mathrm{II}_\mathrm{N} + \langle 0^\mathrm{I} 0^\mathrm{II} \vert H \vert 0^\mathrm{I} 0^\mathrm{II} \rangle
\end{align}
It is useful to define the following effective single component Hamiltonians $ \tilde{H}^\mathrm{I}_\mathrm{N}$ and $\tilde{H}^\mathrm{II}_\mathrm{N}$ by incorporating $W_\mathrm{N}$ terms in expressions
\begin{align}
	\tilde{H}^\mathrm{I}_\mathrm{N} 
	&= H^\mathrm{I}_\mathrm{N} + W^\mathrm{I}_\mathrm{N} \\
	\tilde{H}^\mathrm{II}_\mathrm{N} 
	&= H^\mathrm{II}_\mathrm{N} + W^\mathrm{II}_\mathrm{N}.
\end{align}
Using the above relationship, the total normal-ordered Hamiltonian is defined as
\begin{align}
	\label{eq:no_mc_hamil}
	\tilde{H}_\mathrm{N} 
	&= 
	H - \langle 0^{\mathrm{I}} 0^{\mathrm{II}} \vert H \vert 0^{\mathrm{I}} 0^{\mathrm{II}} \rangle \\
	&= 
	\tilde{H}^\mathrm{I}_\mathrm{N} +  \tilde{H}^\mathrm{II}_\mathrm{N}
	+ V^\mathrm{I,II}_\mathrm{N} .
\end{align}

%% file: subsec_mccc_eq.tex
\subsection{The mcCC Equations}
Using the normal-ordered Hamiltonian, the CC equation can be written in terms of the total correlation energy $\Delta E_{\mathrm{mcCC}}$ 
\begin{align} 
        \label{eq:mc_cc_schrod} 
        \tilde{H}_{\mathrm{N}} 
    e^{T} 
    \vert 
    0^{\mathrm{I}} 
    0^{\mathrm{II}} 
    \rangle 
    & 
    = 
    \Delta 
    E_{\mathrm{mcCC}} 
    e^{T} 
    \vert 
    0^{\mathrm{I}} 
    0^{\mathrm{II}} 
    \rangle 
\end{align} 
where $\Delta E_{\mathrm{mcCC}} = E_{\mathrm{mcCC}}-\langle 0^{\mathrm{I}} 0^{\mathrm{II}} \vert H \vert 0^{\mathrm{I}} 0^{\mathrm{II}} \rangle$ and $T$ is the cluster operator. 
In this work, we restrict the form of the cluster operator to include only single and double excitations as shown below 
\begin{align} 
        \label{eq:mc_cluster_op_sum} 
    T 
    &= 
        \sum_{ij}^{2} \left[ 
        T_{ij}^{\mathrm{I,II}} + (T_{i}^{\mathrm{I}} + T_{i}^{\mathrm{II}})\delta_{ij} 
        \right] \\ 
        \label{eq:mc_cluster_op} 
    &= 
    T^{\mathrm{I}}_{1} 
    +  
    T^{\mathrm{I}}_{2} 
    +  
    T^{\mathrm{II}}_{1} 
    +  
    T^{\mathrm{II}}_{2} 
    +  
    T^{\mathrm{I,II}}_{11} 
    +  
    T^{\mathrm{I,II}}_{12} 
    +  
    T^{\mathrm{I,II}}_{21} 
    +  
    T^{\mathrm{I,II}}_{22} 
    . 
\end{align} 
The $T^{\mathrm{I}}_{1}$ and $T^{\mathrm{I}}_{2}$ operators are identical to the single component CCSD operators (and operate only on the type I space). 
The same holds true for the corresponding type II operators.   
The indices $i, j$ are summed over the occupied space denoted by $N$ while the $a, b$ are summed over the virtual space, $M$. 
The type II operators have  primed indices to distinguish them from their type I counterparts. 
\begin{align} 
    T^{\mathrm{I}}_{1} 
    = 
    \sum_{i}^{N} 
    \sum_{a}^{M} 
    t_{i}^{a} 
    \{ 
    a^{\dagger} 
    i 
    \} 
\end{align} 
\begin{align} 
    T^{\mathrm{I}}_{2} 
    = 
    \frac{1}{4} 
    \sum_{ij}^{N} 
    \sum_{ab}^{M} 
    t_{ij}^{ab} 
    \{ 
    a^{\dagger} 
    b^{\dagger} 
    j 
    i 
    \} 
\end{align} 
\begin{align} 
    T^{\mathrm{II}}_{1} 
    = 
    \sum_{i'}^{N'} 
    \sum_{a'}^{M'} 
    t_{i'}^{a'} 
    \{ 
    a'^{\dagger} 
    i' 
    \} 
\end{align} 
\begin{align} 
    T^{\mathrm{II}}_{2} 
    = 
    \frac{1}{4} 
    \sum_{i'j'}^{N'} 
    \sum_{a'b'}^{M'} 
    t_{i'j'}^{a'b'} 
    \{ 
    a'^{\dagger} 
    b'^{\dagger} 
    j' 
    i' 
    \} 
\end{align} 
The type I,II operators produce connected excitations in both the type I and II spaces.   
These operators are expanded below, 
\begin{align} 
    T^{\mathrm{I,II}}_{11} 
    = 
    \sum_{i}^{N} 
    \sum_{a}^{M} 
    \sum_{i'}^{N'} 
    \sum_{a'}^{M'} 
    t_{ii'}^{aa'} 
    \{ 
    a^{\dagger} 
    i 
    \}\{ 
    a'^{\dagger} 
    i' 
    \} 
\end{align} 
\begin{align} 
    T^{\mathrm{I,II}}_{12} 
    = 
    \frac{1}{4} 
    \sum_{i}^{N} 
    \sum_{a}^{M} 
    \sum_{i'j'}^{N'} 
    \sum_{a'b'}^{M'} 
    t_{ii'j'}^{aa'b'} 
    \{ 
    a^{\dagger} 
    i 
    \} 
    \{ 
    a'^{\dagger} 
    b'^{\dagger} 
    j' 
    i' 
    \} 
\end{align} 
\begin{align} 
    T^{\mathrm{I,II}}_{21} 
    = 
    \frac{1}{4} 
    \sum_{ij}^{N} 
    \sum_{ab}^{M} 
    \sum_{i'}^{N'} 
    \sum_{a'}^{M'} 
    t_{iji'}^{aa'b'} 
    \{ 
    a^{\dagger} 
    b^{\dagger} 
    j 
    i 
    \} 
    \{ 
    a'^{\dagger} 
    i' 
    \} 
\end{align} 
\begin{align} 
    T^{\mathrm{I,II}}_{22} 
    = 
    \frac{1}{16} 
    \sum_{ij}^{N} 
    \sum_{ab}^{M} 
    \sum_{i'j'}^{N'} 
    \sum_{a'b'}^{M'} 
    t_{iji'j'}^{aba'b'} 
    \{ 
    a^{\dagger} 
    b^{\dagger} 
    j 
    i 
    \} 
    \{ 
    a'^{\dagger} 
    b'^{\dagger} 
    j' 
    i' 
    \} 
\end{align} 
The cluster operator, as presented above, defines a multicomponent CCSD space. 
The energy and the $t$ amplitude expressions can be obtained by performing similarity-transformation on the normal-ordered Hamiltonian from Eq. \eqref{eq:no_mc_hamil}, 
\begin{align} 
    e^{-T} 
        \tilde{H}_{\mathrm{N}} 
    e^{T} 
    \vert 
    0^{\mathrm{I}} 
    0^{\mathrm{II}} 
    \rangle 
    & 
    = 
    \Delta E_{\mathrm{mcCC}} 
    \vert 
    0^{\mathrm{I}} 
    0^{\mathrm{II}} 
    \rangle 
\end{align} 
which can be evaluated using the standard Baker-Campbell-Hausdorff (BCH) expansion\cite{eltit,shavitt2009many}  shown below, 
 \begin{align} 
        e^{-T} \tilde{H}_{\mathrm{N}} e^{T} 
        &= 
        \tilde{H}_{\mathrm{N}} +  
        [\tilde{H}_{\mathrm{N}},T] + 
        \frac{1}{2} [[\tilde{H}_{\mathrm{N}},T],T] + 
        \frac{1}{3!} [[[\tilde{H}_{\mathrm{N}},T],T],T] +  
        \frac{1}{4!} [[[[\tilde{H}_{\mathrm{N}},T],T],T],T] 
\end{align} 
However, because of multicomponent nature of the system, the number of terms in the BCH expansion is much larger 
than the single component expression. 
Substituting the expression with the Hamiltonian in Eq. \eqref{eq:no_mc_hamil} and the  $t$ amplitudes from Eq. \eqref{eq:mc_cluster_op}, we arrive at the following set of equations, 
\begin{align} 
        [\tilde{H}_{\mathrm{N}},T]  
        &=  
        [\sum_{K_1}^{3} \tilde{H}_{\mathrm{N}}^{(K_1)},\sum_{L_1}^{8} T^{(L_1)}] 
        = \sum_{K_1}^{3}  \sum_{L_1}^{8} [\tilde{H}_{\mathrm{N}}^{(K_1)},T^{(L_1)}] \\ 
        [[\tilde{H}_{\mathrm{N}},T],T]  
        &= 
        \sum_{K_1}^{3}  \sum_{L_1,L_2}^{8} [[\tilde{H}_{\mathrm{N}}^{(K_1)},T^{(L_1)}],T^{(L_2)}]  \\ 
        [[[\tilde{H}_{\mathrm{N}},T],T],T] 
        &= 
        \sum_{K_1}^{3}  \sum_{L_1,L_2,L_3}^{8} [[[\tilde{H}_{\mathrm{N}}^{(K_1)},T^{(L_1)}],T^{(L_2)}],T^{(L_3)}] \\ 
        [[[[\tilde{H}_{\mathrm{N}},T],T],T],T] 
        &= 
        \sum_{K_1}^{2}  \sum_{L_1,L_2,L_3,L_4}^{8} [[[[\tilde{H}_{\mathrm{N}}^{(K_1)},T^{(L_1)}],T^{(L_2)}],T^{(L_3)}],T^{(L_4)}]                 
\end{align} 
where $\tilde{H}_{\mathrm{N}}^{(K)}$ and $T^{(L)}$ are compact notations for the Hamiltonian and cluster operator and are defined as 
\begin{align} 
 \tilde{H}_{\mathrm{N}} &= 
 \sum_{K}^3 \tilde{H}_{\mathrm{N}}^{(K)} = \tilde{H}_\mathrm{N}^{\mathrm{I}} + \tilde{H}_\mathrm{N}^{\mathrm{II}} + V_\mathrm{N}^{\mathrm{I,II}} \\ 
 T &= \sum_{L}^8 T^{(L)} = T_1^{\mathrm{I}} + T_2^{\mathrm{I}} + T_1^{\mathrm{II}} + T_2^{\mathrm{II}}  
                     + T_{11}^{\mathrm{I,II}} + T_{12}^{\mathrm{I,II}} + T_{11}^{\mathrm{I,II}} 
                     + T_{22}^{\mathrm{I,II}}  
                     . 
\end{align} 
For the present ansatz of the mCC wave function, the number of commutators in the above expression is 9940 terms.  
However, not all of these terms contribute in all the equations mentioned above, and the list of contributing terms in Eqs. \eqref{eq:eng_eq}-\eqref{eq:typ12_t22_eq} are presented in \autoref{sec:appdx_a}. 
\par 
The solution of the BCH expansion for the total multicomponent correlation energy results in the following expression, 
\begin{align} 
        \Delta E_\mathrm{mcCC} 
        &= 
        \langle 0^{\mathrm{I}}0^{\mathrm{II}} \vert 
            \tilde{H}_{\mathrm{N}}^{\mathrm{I}}   
                [  T_{1}^{\mathrm{I}}  
                   + \frac{1}{2!}T_{1}^{\mathrm{I}^2}  
                   + T_{2}^{\mathrm{I}}  
            ]          
        \vert 0^{\mathrm{I}}0^{\mathrm{II}} \rangle \\ \notag 
        &+ 
        \langle 0^{\mathrm{I}}0^{\mathrm{II}} \vert 
            \tilde{H}_{\mathrm{N}}^{\mathrm{II}} 
              [  T_{1}^{\mathrm{II}} +  
                      \frac{1}{2!}T_{1}^{\mathrm{II}^2}  
                      + T_{2}^{\mathrm{II}}  
              ] 
        \vert 0^{\mathrm{I}}0^{\mathrm{II}} \rangle \\ \notag         
        &+ 
        \langle 0^{\mathrm{I}}0^{\mathrm{II}} \vert 
            V_{\mathrm{N}}^{\mathrm{I,II}}  
           [ T_{11}^{\mathrm{I,II}}  
             + T_{1}^{\mathrm{I}} + 
             T_{1}^{\mathrm{I}}T_{1}^{\mathrm{II}}  
             + T_{1}^{\mathrm{II}}  
           ]  
        \vert 0^{\mathrm{I}}0^{\mathrm{II}} \rangle . 
\end{align} 
As expected, the total correlation energy can be written as a sum of correlation energies from the two components and the interaction between the two components. Analogous to the electronic CCSD expression, the correlation energies for type I and type II particles depend only on $t_1$ and $t_2$  amplitudes. 
In contrast, the contribution from I-II interaction depends only on $t_1$ and $t_{11}$ amplitudes.  
It is important to note that although the expression for the type I correlation energy is identical to electronic CCSD, the $t$ amplitudes in the expressions are not independent of type~II $t$ amplitudes and depend on type II amplitudes via Eq. \eqref{eq:typ1_t1_eq} and Eq. \eqref{eq:typ1_t2_eq}. 
The determination of the total correlation energy depends on the converged $t$ amplitude equations which require the simultaneous solution of a set of non-linear, coupled equations. 
Obtaining the $t$ amplitude equations require projecting the similarity-transformed Hamiltonian onto a series of excited states denoted by $\langle K^{\mathrm{I}} K^{\mathrm{II}} \vert$.  
The following set of equations is the result of choosing $\langle K^{\mathrm{I}} K^{\mathrm{II}} \vert$ to be all combinations of vacuum, singly, and doubly excited states (represented as $0, S,$ and $D$, respectively) in both type I and type II space. 
\begin{align} 
        \label{eq:eng_eq} 
    \langle 
    0^{\mathrm{I}} 
    0^{\mathrm{II}} 
    \vert 
    e^{-T} 
        \tilde{H}_{\mathrm{N}} 
    e^{T} 
    \vert 
    0^{\mathrm{I}} 
    0^{\mathrm{II}} 
    \rangle_{\mathrm{c}} 
    & 
    = 
    \Delta 
    E_{\mathrm{mcCC}} & \textrm{correlation energy equation} 
\\ 
        \label{eq:typ1_t1_eq} 
    \langle 
    S^{\mathrm{I}} 
    0^{\mathrm{II}} 
    \vert 
    e^{-T} 
        \tilde{H}_{\mathrm{N}} 
    e^{T} 
    \vert 
    0^{\mathrm{I}} 
    0^{\mathrm{II}} 
    \rangle_{\mathrm{c}} 
    & 
    = 
    0  & t_1^{\mathrm{I}} \quad \textrm{amplitude equation} 
\\ 
        \label{eq:typ1_t2_eq} 
    \langle 
    D^{\mathrm{I}} 
    0^{\mathrm{II}} 
    \vert 
    e^{-T} 
        \tilde{H}_{\mathrm{N}} 
    e^{T} 
    \vert 
    0^{\mathrm{I}} 
    0^{\mathrm{II}} 
    \rangle_{\mathrm{c}} 
    & 
    = 
    0 & t_2^{\mathrm{I}} \quad \textrm{amplitude equation} 
\\ 
        \label{eq:typ2_t1_eq} 
    \langle 
    0^{\mathrm{I}} 
    S^{\mathrm{II}} 
    \vert 
    e^{-T} 
        \tilde{H}_{\mathrm{N}} 
    e^{T} 
    \vert 
    0^{\mathrm{I}} 
    0^{\mathrm{II}} 
    \rangle_{\mathrm{c}} 
    & 
    = 
    0 & t_1^{\mathrm{II}} \quad \textrm{amplitude equation} 
\\ 
        \label{eq:typ2_t2_eq} 
    \langle 
    0^{\mathrm{I}} 
    D^{\mathrm{II}} 
    \vert 
    e^{-T} 
        \tilde{H}_{\mathrm{N}} 
    e^{T} 
    \vert 
    0^{\mathrm{I}} 
    0^{\mathrm{II}} 
    \rangle_{\mathrm{c}} 
    & 
    = 
    0 & t_2^{\mathrm{II}} \quad \textrm{amplitude equation} 
\\ 
        \label{eq:typ12_t11_eq} 
    \langle 
    S^{\mathrm{I}} 
    S^{\mathrm{II}} 
    \vert 
    e^{-T} 
        \tilde{H}_{\mathrm{N}} 
    e^{T} 
    \vert 
    0^{\mathrm{I}} 
    0^{\mathrm{II}} 
    \rangle_{\mathrm{c}} 
    & 
    = 
    0 & t_{11}^{\mathrm{I,II}} \quad \textrm{amplitude equation} 
\\ 
        \label{eq:typ12_t12_eq} 
    \langle 
    S^{\mathrm{I}} 
    D^{\mathrm{II}} 
    \vert 
    e^{-T} 
        \tilde{H}_{\mathrm{N}} 
    e^{T} 
    \vert 
    0^{\mathrm{I}} 
    0^{\mathrm{II}} 
    \rangle_{\mathrm{c}} 
    & 
    = 
    0 & t_{12}^{\mathrm{I,II}} \quad \textrm{amplitude equation} 
\\ 
        \label{eq:typ12_t21_eq} 
    \langle 
    D^{\mathrm{I}} 
    S^{\mathrm{II}} 
    \vert 
    e^{-T} 
        \tilde{H}_{\mathrm{N}} 
    e^{T} 
    \vert 
    0^{\mathrm{I}} 
    0^{\mathrm{II}} 
    \rangle_{\mathrm{c}} 
    & 
    = 
    0 & t_{21}^{\mathrm{I,II}} \quad \textrm{amplitude equation} 
\\ 
        \label{eq:typ12_t22_eq} 
    \langle 
    D^{\mathrm{I}} 
    D^{\mathrm{II}} 
    \vert 
    e^{-T} 
        \tilde{H}_{\mathrm{N}} 
    e^{T} 
    \vert 
    0^{\mathrm{I}} 
    0^{\mathrm{II}} 
    \rangle_{\mathrm{c}} 
    & 
    = 
    0 & t_{22}^{\mathrm{I,II}}  \quad \textrm{amplitude equation} 
\end{align}

%% file: sec_implementation.tex
\begin{figure}[h!]
  \centering
    \includegraphics[width=0.9\textwidth]{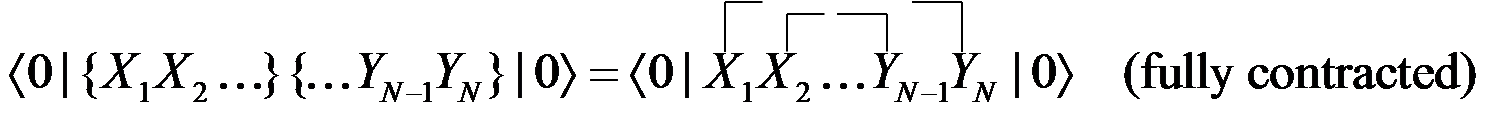}
 \caption{Wick's theorem}
 \label{fig:wick}
\end{figure}

The implementation of the $t$ amplitude equations was performed sequentially by first performing the contraction  over one particle type followed  by contraction over the other particle type.
For a general $t$ amplitude equation, this procedure is shown by the following equation
\begin{align}
	\langle 
		K^{\mathrm{I}}
		K^{\mathrm{II}} 
		\vert 
			e^{-T} H_{\mathrm{N}} e^{T} 
		\vert 
		0^{\mathrm{I}}		
		0^{\mathrm{II}} 
   \rangle_c
   &= 
   	\langle 
		K^{\mathrm{II}} \vert
		\langle 
		K^{\mathrm{I}}
		\vert 
			e^{-T} \tilde{H}_{\mathrm{N}}^{I} e^{T} 
		\vert 
		0^{\mathrm{I}}
		\rangle_c \vert 
		0^{\mathrm{II}}
   \rangle_c \\ \notag
   &+
   \langle 
		K^{\mathrm{I}} \vert
		\langle 
		K^{\mathrm{II}}
		\vert 
			e^{-T} \tilde{H}_{\mathrm{N}}^{II} e^{T} 
		\vert 
		0^{\mathrm{II}}
		\rangle_c \vert 
		0^{\mathrm{I}}
   \rangle_c \\ \notag
   &+
   \langle 
		K^{\mathrm{I}} \vert
		\langle 
		K^{\mathrm{II}}
		\vert 
			e^{-T} V_{\mathrm{N}} e^{T} 
		\vert 
		0^{\mathrm{II}}
		\rangle_c \vert 
		0^{\mathrm{I}}
   \rangle_c .
\end{align}
Contraction over indicies of one particle type resulted in a set of single component operators.
These operators (labeled as $A^{\mathrm{I}}, A^{\mathrm{II}}$, and $B$) are analogous to the the effective operators in electronic structure CC 
theory (pg. 328 of Ref. \citenum{shavitt2009many}) and are defined in Eqs. \eqref{eq:eff_op_a}-\eqref{eq:eff_op_c}
\begin{align}
	\label{eq:eff_op_a}
	\langle K^{\mathrm{II}} \vert e^{-T} \tilde{H}_{\mathrm{N}}^{\mathrm{II}} e^{T} \vert 0^{\mathrm{II}} \rangle_c
	&=
	A_0^{\mathrm{I}} \\ 
	&+ \sum_{ia} A_{ia}^{\mathrm{I}} \{ a^\dagger i \} \notag \\
	&+ \sum_{ijab} A_{ijab}^{\mathrm{I}} \{ a^\dagger b^\dagger j i \} \notag \\
	&+ \dots  \textrm{(n-body terms)} \notag  .
\end{align}
In general, the number of operators in the above expression is infinite because of the presence of the exponential operators. 
However, because of the operator expression was projected onto a space that contains only upto doubly excited spaces
$\vert D \rangle$, all three-body and higher operators had zero contribution to the $t$ amplitude equations.
Analogous to the type I particles, the operators for the type II particles are defined as 
\begin{align}
	\label{eq:eff_op_b}
	\langle K^{\mathrm{I}} \vert e^{-T} \tilde{H}_{\mathrm{N}}^{\mathrm{I}} e^{T} \vert 0^{\mathrm{I}} \rangle_c
	&=
	A_0^{\mathrm{II}} \\ 
	&+ \sum_{i'a'} A_{i'a'}^{\mathrm{II}} \{ {a'}^\dagger i' \} \notag \\
	&+ \sum_{i'j'a'b'} A_{i'j'a'b'}^{\mathrm{II}} \{ {a'}^\dagger {b'}^\dagger j' i' \} \notag \\
	&+ \dots  \textrm{(n-body terms)} \notag 
\end{align}
The contraction over the $V^{\mathrm{I,II}}_\mathrm{N}$ operators is defined as
\begin{align}
	\label{eq:eff_op_c}
	\langle K^{\mathrm{II}} \vert e^{-T} V^{\mathrm{I,II}}_\mathrm{N} e^{T} \vert 0^{\mathrm{II}} \rangle_c
	&= \sum_{pq} B_{pq}^{\mathrm{I}} \{ p^\dagger q \} \notag \\
	&+ \sum_{pqia}  B_{pqia}^{\mathrm{I}} \{ p^\dagger  a^\dagger i q \}  \notag \\
	&+ \dots  \textrm{(n-body terms)} \notag
\end{align}
The main advantage of defining the single component effective operators $A^{\mathrm{I}},A^{\mathrm{II}}$, and $B$  is that the form of the operators are similar (but not identical) to the electronic structure CCSD equations. 
As a consequence, the evaluation and implementation of these operators benefited from the development in computer assisted source code generation of electronic structure CC methods.\cite{Lyakh2005,Janssen19911,Piecuch200679,Piecuch200271}   
In this work we have developed a program to perform algebraic manipulation of the strings of second-quantized (SQ) operators.
This program was largely inspired by the tensor contraction engine (TCE) \cite{Janssen19911,Li19948812,Harris1999593,Nooijen20004549,Nooijen2000494,Head-Gordon1998616,Auer2006211,Lu2012338,Hartono200912715} which has been used extensively for implementation of electronic structure CC methods. 
A subset of techniques used in the TCE method \cite{Hirata20039887} was used for source-code generation and implementation of the $t$ amplitude equations.   
The program uses the well-known results of the Wick's expansion theorem that only fully contracted terms have non-zero contribution for a vacuum expectation value  of the a string of SQ operators (\autoref{fig:wick}).\cite{shavitt2009many} 
In the first step, for a given string of SQ operators, the  program generated a list of all valid fully-contracted terms.
In the second step, reduction operations were performed to reduce the number of terms. 
The final step involved generation of the  mcCC source code.
The reduction operation is the key step for increasing the efficiency of the implementation.  
The SQ program performed index permutation  operations and consolidated equivalent expressions to minimize the number of terms in the fully contracted list.
\par
\begin{figure}[h!]
  \centering
	\includegraphics[width=0.5\textwidth]{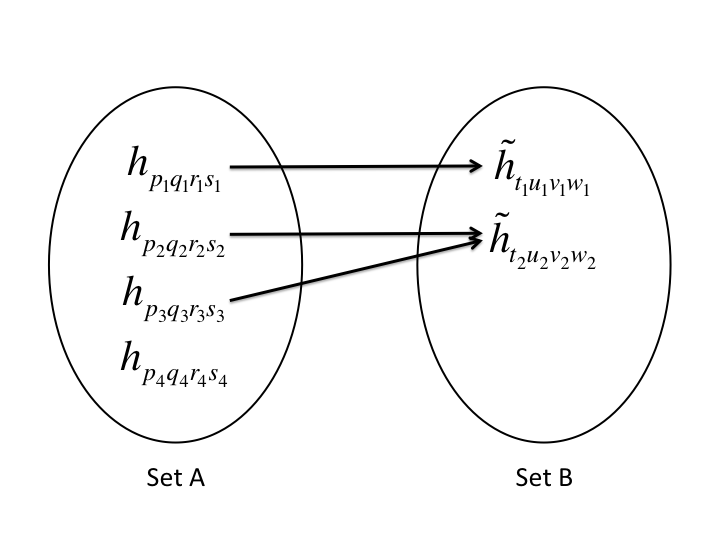}
    \caption{Many-to-one mapping of the molecular integrals. Set A contains the list of the input integrals. Set B is a subset of A that contains the list on non-zero unique integrals. 
The terms $h_{p_2q_2r_2s_2}$ and $h_{p_3q_3r_3s_3}$ are assumed to have identical numerical values and are mapped to a single term in set B. The term $h_{p_4q_4r_4s_4}$ is assumed to be zero and is not  included in set B. }
\label{fig:bbe}
\end{figure}

In addition to performing index manipulations, the SQ program  performed additional reduction operations by using the numerical value of the molecular integrals ($h_{pq}$ and $h_{pqrs}$).
This was performed by eliminating all numerically zero terms from the set of molecular integrals and mapping the non-unique terms to a set of unique terms. 
This mapping of terms is illustrated in \autoref{fig:bbe},  where set A is the set of original integrals and set B is a subset of set A containing only unique, non-zero terms. 
In the reduction step, terms with $h_{pq}=0$ and $h_{pqrs}=0$
were eliminated at a very early stage of the calculation.
The reduction process also performed additional consolidation for expressions with identical numerical values (for example $\tilde{h}_{t_2u_2v_2w_2}$ in \autoref{fig:bbe}).
\par
There are both advantages and disadvantages of using information from the molecular integrals for the reduction step. 
The main limitation is that the source code generation becomes system dependent and therefore the CC source code needs to be generated for each unique system. However, the advantage of system dependent source code is that the generated code is highly optimized for the specific system under investigation and uses the intrinsic symmetries associated with given system. 
The mapping of the integrals to the set of unique non-zero terms also helps in reducing the memory footprint.
\par
The present version of the SQ program is still in the first iteration of its development cycle and is neither as general nor as efficient as the TCE. 
In future work we envision interfacing the SQ program with the open-source version of the TCE for  extending the implementation of mcCC theory.

%% file: sec_computation.tex
The mcCC method was used to investigate electron-hole interaction in biexcitonic systems. Biexcitons are bound two-electron, two-hole quasiparticles that are formed from two weakly interacting excitons.
Using the mcCC method, we calculated total correlation energy and biexciton binding energy for a set of eight model systems.
Total correlation energy of the biexciton systems was calculated by subtracting the vacuum energy from total energy,
\begin{align}
    \Delta E_{\mathrm{mcCC}} = E_{\mathrm{mcCC}} - E_{\mathrm{ref}}
\end{align}
and biexciton binding energy $E_\mathrm{BBE}$ was defined as the difference between the energies of the free excitons ($E_\mathrm{X}$) and the total biexciton energy $E_{\mathrm{X}_2}$.
\begin{align}
	E_\mathrm{BBE}
	&=
	E_{\mathrm{X}_2} - 2E_{\mathrm{X}} .
\end{align}
Each of the eight biexcitonic systems consisted of electron-hole pairs with unit charges and masses ($m^{\mathrm{I}}=m^{\mathrm{II}}=1 \mathrm{a.u.}$).
The dielectric function in Eq. \eqref{eq:v_one_two} was assumed to be constant and was set to $\epsilon = 1$. 
The external potential was approximated using the 3D parabolic potential defined below
\begin{align}
	v^{\alpha}_{\mathrm{ext}} = \frac{1}{2} k^{\alpha} \vert \mathbf{r}^\alpha \vert^2 \quad \alpha = \mathrm{I}, \mathrm{II}.
\end{align}
A parabolic potential was chosen because they have been used extensively~\cite{RefWorks:44,RefWorks:43,RefWorks:45,RefWorks:46,RefWorks:49,RefWorks:48,RefWorks:47,Karimi20114423,Nammas20114671,Rezaei20114596,Elward2012,Elward2012182,elward_dotsize} to represent the confinement potential experienced by excitons in nanoparticles. 
The values of the force constant parameter, $k$, were selected such that the confining potential spanned from weakly confining region ($k=1.0 \times 10^{-4}$ a.u.) to strongly confining region ($k=5.0$ a.u.).
The single particle basis was constructed from 10 s even-tempered Gaussian functions whose $\alpha$ values were determined variationally with respect to $\alpha_{v}$ by minimizing the multicomponent Hartree-Fock energy using the following scheme,
\begin{align}
	\alpha_{i} = \alpha_{v} \beta^{i-1} \quad i = 1, ..., 10
\end{align}
and the exponents for Gaussian functions for each values of $k$ are given in \autoref{sec:appdx_b}.
\par
We also performed multicomponent full configuration interaction (mcFCI) calculations on the eight biexcitonic systems in order to verify validity of the mcCC method. 
The mcFCI energy was obtained from the solution of the following equations 
\begin{align}
	\Psi_\mathrm{mcFCI} 
	&= 
	\sum_{i}^{N_\mathrm{FCI}^\mathrm{I}} 
	\sum_{i'}^{N_\mathrm{FCI}^\mathrm{II}} 
	c_{ii'} \Phi^\mathrm{I}_{i} \Phi^\mathrm{II}_{i'} \\
	E_\mathrm{mcFCI} &= \min_{\mathbf{c}} \langle \Psi_{\mathrm{mcFCI}} \vert
	H
	\vert \Psi_{\mathrm{mcFCI}} \rangle
\end{align}
and the mcFCI correlation energy ($\Delta E_\mathrm{mcFCI}=E_\mathrm{mcFCI}-E_\mathrm{mcHF}$)  was compared with the mcCC results.

%% file: sec_results.tex
\begin{figure}[ht] 
\centerline{\includegraphics[width=1.0\textwidth]{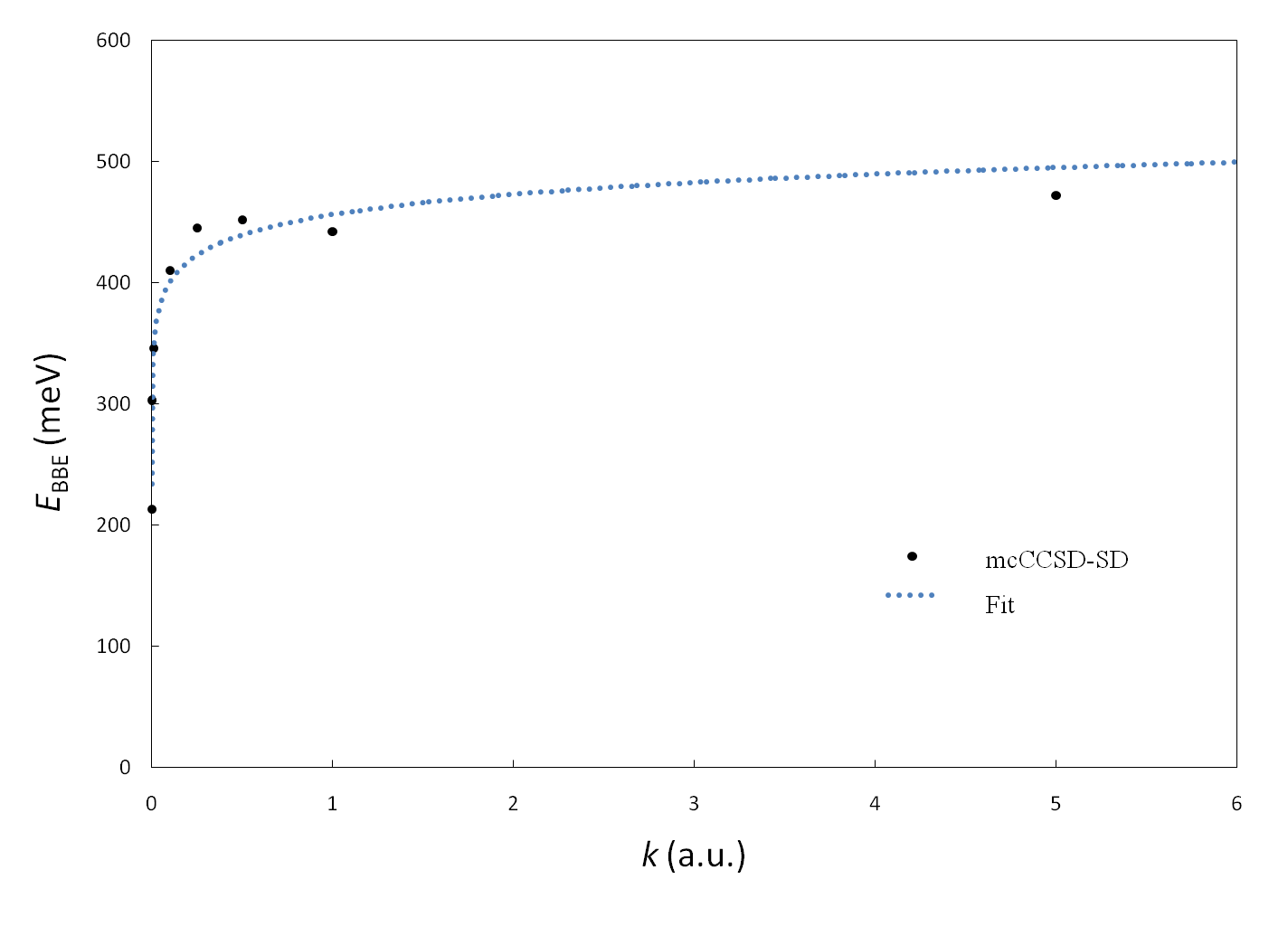}} 
\caption{Dependence of biexciton binding energy calculated using mcCCSD-SD on the strength of the confinement potential.}  
\label{fig:bbe} 
\end{figure}

The mcCC method was applied using two different ansatz for the $T$ operators as shown in the following equation 
\begin{align} 
        \vert \Psi_\mathrm{mcCCSD-SD} \rangle 
        &=e^{ 
        T_1^{\mathrm{I}} + T_2^{\mathrm{I}} + T_1^{\mathrm{II}} + T_2^{\mathrm{II}}  
                     + T_{11}^{\mathrm{I,II}} + T_{12}^{\mathrm{I,II}} + T_{11}^{\mathrm{I,II}} 
                     + T_{22}^{\mathrm{I,II}} } 
        \vert 0^\mathrm{I} 0^\mathrm{II} \rangle  \\  
        \vert \Psi_\mathrm{mcCCSD}  \rangle 
        &=e^{ 
        T_1^{\mathrm{I}} + T_2^{\mathrm{I}} + T_1^{\mathrm{II}} + T_2^{\mathrm{II}} } 
        \vert 0^\mathrm{I} 0^\mathrm{II} \rangle    .              
\end{align} 
The main difference between the two forms of the CC wave function is that $\Psi_\mathrm{mcCCSD}$ does not include the connected excitation operators ($T^\mathrm{I,II}$) that are present in $\Psi_\mathrm{mcCCSD-SD}$.  
The suffix "-SD" in the above expression is used to denote that both connected single and double excitation operators were included in the calculation. 
The correlation energy calculated using mcCCSD, mcCCSD-SD and mcFCI for the biexcitonic systems are presented in \autoref{tab:TBE}.  
Comparison between the mcCCSD-SD and mcFCI results show very good agreement between the two methods for all values of $k$. 
The maximum difference between mcCCSD-SD and mcFCI correlation energy was found to be in the order of $10^{-5} \mathrm{Hartrees}$. 
In contrast, the mcCCSD calculations were found to underestimate the correlation energy by at least a factor of $2$ for all values of the confinement potential. 
These results highlight the importance of including connected excitation operator in the mcCC wave function.  
The results also show that connected operators are important in both weak and strong confinement regions.  
\begin{table}[ht] 
  \begin{center} 
   \caption{\textbf{Total biexciton energy calculated from mcHF and correlation energy from mcCCSD, mcCCSD-SD, and mcFCI methods reported in Hartrees as function of $k$.}} 
   \label{tab:TBE} 
    \begin{tabular}{c c c c c} 
    \hline 
          $k$      &  
          $E_{\mathrm{mcHF}}$     &  
          $\Delta E_{\mathrm{mcCCSD}}$ &  
          $\Delta E_{\mathrm{mcCCSD-SD}}$ &   
          $\Delta E_{\mathrm{mcFCI}}$               \\   
      \hline 
      0.0001   & -0.21231    & -0.00667 &    -0.01633  &    -0.01634  \\ 
      0.0010   & -0.17993    & -0.01018 &    -0.02224  &    -0.02224  \\ 
      0.0100   &  0.02441    & -0.00950 &    -0.02286  &    -0.02286  \\ 
      0.1000   &  0.93611    & -0.00960 &    -0.02524  &    -0.02524  \\  
      0.2500   &  1.80941    & -0.00975 &    -0.02682  &    -0.02682  \\  
      0.5000   &  2.83955    & -0.00967 &    -0.02687  &    -0.02688  \\ 
      1.0000   &  4.34384    & -0.00936 &    -0.02595  &    -0.02595  \\  
      5.0000   & 10.97117    & -0.00948 &    -0.02715  &    -0.02716  \\ 
     \hline  
    \end{tabular} 
  \end{center}   
\end{table}  
The biexciton binding energies calculated using the methods described above are presented in \autoref{tab:BBE}.  
The mcCCSD-SD binding energies were found to be in very good agreement with the mcFCI results. 
The mcCCSD results deviated from the mcFCI results and recovered only 39\%-85\% of the mcFCI binding energy. 
\begin{table}[ht] 
  \begin{center} 
          \caption{\textbf{Biexciton binding energy calculated using mcHF, mcCCSD, mcCCSD-SD, and mcFCI methods reported in meV as function of $k$.}} 
   \label{tab:BBE} 
    \begin{tabular}{c c c c c} 
    \hline 
      $k \mathrm{(a.u.)}$      & mcHF     & mcCCSD & mcCCSD-SD &  mcFCI              \\   
      \hline 
      0.0001 &   0  & 182 &   213 &   213 \\ 
      0.0010 &   0  & 277 &   303 &   303 \\ 
      0.0100 &   1  & 260 &   346 &   346 \\ 
      0.1000 &   0  & 261 &   410 &   410 \\ 
      0.2500 &   0  & 265 &   445 &   445 \\ 
      0.5000 &   0  & 263 &   452 &   452 \\ 
      1.0000 &   0  & 254 &   442 &   442 \\ 
      5.0000 &   0  & 258 &   472 &   472 \\ 
     \hline  
    \end{tabular} 
  \end{center}   
\end{table}  
Interestingly, the deviation from the mcCCSD energy from the mcCCSD-SD values was found to be dependent on the strength of the confinement potentials.  
Based on these trends we hypothesize that the contributions from the connected excitation operators to $E_\mathrm{BBE}$ are more important in strongly confined as compared to weakly confined systems.  
The results in \autoref{tab:BBE}  also highlight the importance of particle-particle correlation in calculations of biexciton binding energy. 
Specifically, comparison of the mcHF and mcCCSD-SD results shows that the Hartree-Fock approximation is qualitatively inadequate capturing biexcitonic interactions. 
Comparing with mcCCSD results, we find that inclusion of e-e and h-h correlation can significantly improve the Hartree-Fock results.  
However, the quality of the improvement is a function of confinement potential.  
The dependence of the biexciton binding energy on the strength of the potential is shown in \autoref{fig:bbe}. 
As expected, we find the biexciton to be bound strongly in high confinement region. 
However, the binding energy was found to decrease in the region of weak confinement. 
This trend is consistent with the quantum confinement effect observed in optical properties of semiconductor nanoparticles.   

%% file: sec_conclusion.tex
In this work, we present the theoretical development and implementation details
of the multicomponent coupled-cluster theory. This method 
was developed to investigate particle-particle correlation in many-particle 
multicomponent systems such as electron-proton, electron-positron, and electron-hole systems.
Specifically, the construction of the 
multicomponent vacuum state, form of the cluster operator, and the 
expressions for the cluster amplitudes were discussed.
The coupled-cluster equations were implemented using a 
computer-assisted source code generation strategy employing an integral-driven approach.
The developed method was used to investigate  electron-hole quasiparticle interactions in 
biexcitonic systems by calculating total correlation energy and biexciton binding energies. 
These quantities were found to be in very good agreement with full configuration interaction results.
Based on these calculations we conclude that  (1)  the mean-field approximation severely underestimates the biexciton binding energies 
and is not a suitable approximation for treating biexcitonic systems and (2) the inclusion of connected excitation operators in the 
multicomponent coupled-cluster wave function is crucial for accurate calculation of electron-hole correlation energy.
The method presented here provides a foundation for extending  
multicomponent coupled-cluster theory and future work will explore
perturbative approaches to connected excitation and linearized version
of this method.

%% file: sec_acknowledgements.tex
We are grateful to Syracuse University and National Science Foundation (CHE-1349892) for the financial support. We also thank Jeremy Scher for his assistance  with the illustrations. AC will also like to thank Filipp Furche for useful discussions about this work. 

%% file: appendix_a.tex
\subsection{Energy Expression ($\langle 0^{\mathrm{I}} 0^{\mathrm{II}} \vert e^{-T} H_{\mathrm{N}} e^{T} \vert 0^{\mathrm{I}} 0^{\mathrm{II}} \rangle = \Delta E_{\mathrm{mcCC}}$)}
\input{00_H_00}
\subsection{Type I Single Amplitude ($\langle S^{\mathrm{I}} 0^{\mathrm{II}} \vert e^{-T} H_{\mathrm{N}} e^{T} \vert 0^{\mathrm{I}} 0^{\mathrm{II}} \rangle = 0$)}
\input{S0_H_00}
\subsection{Type I Double Amplitude ($\langle D^{\mathrm{I}} 0^{\mathrm{II}} \vert e^{-T} H_{\mathrm{N}} e^{T} \vert 0^{\mathrm{I}} 0^{\mathrm{II}} \rangle = 0$)}
\input{D0_H_00}
\subsection{Type II Single Amplitude ($\langle 0^{\mathrm{I}} S^{\mathrm{II}} \vert e^{-T} H_{\mathrm{N}} e^{T} \vert 0^{\mathrm{I}} 0^{\mathrm{II}} \rangle = 0$)}
\input{0S_H_00}
\subsection{Type II Double Amplitude ($\langle 0^{\mathrm{I}} D^{\mathrm{II}} \vert e^{-T} H_{\mathrm{N}} e^{T} \vert 0^{\mathrm{I}} 0^{\mathrm{II}} \rangle = 0$)}
\input{0D_H_00}
\subsection{Type I, II Single, Single Amplitude ($\langle S^{\mathrm{I}} S^{\mathrm{II}} \vert e^{-T} H_{\mathrm{N}} e^{T} \vert 0^{\mathrm{I}} 0^{\mathrm{II}} \rangle = 0$)}
\input{SS_H_00}
\subsection{Type I, II Single, Double Amplitude ($\langle S^{\mathrm{I}} D^{\mathrm{II}} \vert e^{-T} H_{\mathrm{N}} e^{T} \vert 0^{\mathrm{I}} 0^{\mathrm{II}} \rangle = 0$)}
\input{SD_H_00}
\subsection{Type I, II Double, Single Amplitude ($\langle D^{\mathrm{I}} S^{\mathrm{II}} \vert e^{-T} H_{\mathrm{N}} e^{T} \vert 0^{\mathrm{I}} 0^{\mathrm{II}} \rangle = 0$)}
\input{DS_H_00}
\subsection{Type I, II Double, Double Amplitude ($\langle D^{\mathrm{I}} D^{\mathrm{II}} \vert e^{-T} H_{\mathrm{N}} e^{T} \vert 0^{\mathrm{I}} 0^{\mathrm{II}} \rangle = 0$)}
\input{DD_H_00}

%% file: 00_H_00.tex
\noindent
$\langle 0^{\mathrm{I}}0^{\mathrm{II}} \vert   V_{\mathrm{N}}^{\mathrm{I,II}}  [  1 + T_{11}^{\mathrm{I,II}} + T_{1}^{\mathrm{I}} + T_{1}^{\mathrm{I}}T_{1}^{\mathrm{II}} + T_{1}^{\mathrm{II}}  ]  +    \tilde{H}_{\mathrm{N}}^{\mathrm{II}}  [  1 + T_{1}^{\mathrm{II}} + \frac{1}{2!}T_{1}^{\mathrm{II}^2} + T_{2}^{\mathrm{II}}  ]  +    \tilde{H}_{\mathrm{N}}^{\mathrm{I}}  [  1 + T_{1}^{\mathrm{I}} + \frac{1}{2!}T_{1}^{\mathrm{I}^2} + T_{2}^{\mathrm{I}}  ] \vert 0^{\mathrm{I}}0^{\mathrm{II}} \rangle = \Delta E_{\mathrm{mcCC}}$

%% file: S0_H_00.tex
\noindent
$\langle S^{\mathrm{I}}0^{\mathrm{II}} \vert   V_{\mathrm{N}}^{\mathrm{I,II}}  [  1 \allowbreak + \allowbreak T_{11}^{\mathrm{I,II}} \allowbreak + \allowbreak T_{1}^{\mathrm{I}} \allowbreak + \allowbreak T_{1}^{\mathrm{I}}T_{11}^{\mathrm{I,II}} \allowbreak + \allowbreak T_{1}^{\mathrm{I}}T_{1}^{\mathrm{II}} \allowbreak + \allowbreak \frac{1}{2!}T_{1}^{\mathrm{I}^2} \allowbreak + \allowbreak \frac{1}{2!}T_{1}^{\mathrm{I}^2}T_{1}^{\mathrm{II}} \allowbreak + \allowbreak T_{1}^{\mathrm{II}} \allowbreak + \allowbreak T_{21}^{\mathrm{I,II}} \allowbreak + \allowbreak T_{2}^{\mathrm{I}} \allowbreak + \allowbreak T_{2}^{\mathrm{I}}T_{1}^{\mathrm{II}}  ]  \allowbreak + \allowbreak  [  (-T_{1}^{\mathrm{I}})  ]  V_{\mathrm{N}}^{\mathrm{I,II}}  [  1 \allowbreak + \allowbreak T_{11}^{\mathrm{I,II}} \allowbreak + \allowbreak T_{1}^{\mathrm{I}} \allowbreak + \allowbreak T_{1}^{\mathrm{I}}T_{1}^{\mathrm{II}} \allowbreak + \allowbreak T_{1}^{\mathrm{II}}  ]  \allowbreak + \allowbreak    \tilde{H}_{\mathrm{N}}^{\mathrm{II}}  [  T_{11}^{\mathrm{I,II}} \allowbreak + \allowbreak T_{12}^{\mathrm{I,II}} \allowbreak + \allowbreak T_{1}^{\mathrm{I}} \allowbreak + \allowbreak T_{1}^{\mathrm{I}}T_{1}^{\mathrm{II}} \allowbreak + \allowbreak T_{1}^{\mathrm{I}}\frac{1}{2!}T_{1}^{\mathrm{II}^2} \allowbreak + \allowbreak T_{1}^{\mathrm{I}}T_{2}^{\mathrm{II}} \allowbreak + \allowbreak T_{1}^{\mathrm{II}}T_{11}^{\mathrm{I,II}}  ]  \allowbreak + \allowbreak  [  (-T_{1}^{\mathrm{I}})  ]  \tilde{H}_{\mathrm{N}}^{\mathrm{II}}  [  1 \allowbreak + \allowbreak T_{1}^{\mathrm{II}} \allowbreak + \allowbreak \frac{1}{2!}T_{1}^{\mathrm{II}^2} \allowbreak + \allowbreak T_{2}^{\mathrm{II}}  ]  \allowbreak + \allowbreak    \tilde{H}_{\mathrm{N}}^{\mathrm{I}}  [  1 \allowbreak + \allowbreak T_{1}^{\mathrm{I}} \allowbreak + \allowbreak T_{1}^{\mathrm{I}}T_{2}^{\mathrm{I}} \allowbreak + \allowbreak \frac{1}{2!}T_{1}^{\mathrm{I}^2} \allowbreak + \allowbreak \frac{1}{3!}T_{1}^{\mathrm{I}^3} \allowbreak + \allowbreak T_{2}^{\mathrm{I}}  ]  \allowbreak + \allowbreak  [  (-T_{1}^{\mathrm{I}})  ]  \tilde{H}_{\mathrm{N}}^{\mathrm{I}}  [  1 \allowbreak + \allowbreak T_{1}^{\mathrm{I}} \allowbreak + \allowbreak \frac{1}{2!}T_{1}^{\mathrm{I}^2} \allowbreak + \allowbreak T_{2}^{\mathrm{I}}  ] \vert 0^{\mathrm{I}}0^{\mathrm{II}} \rangle = 0$

%% file: D0_H_00.tex
\noindent
$\langle D^{\mathrm{I}}0^{\mathrm{II}} \vert   V_{\mathrm{N}}^{\mathrm{I,II}}  [  T_{11}^{\mathrm{I,II}} \allowbreak + \allowbreak T_{1}^{\mathrm{I}} \allowbreak + \allowbreak T_{1}^{\mathrm{I}}T_{11}^{\mathrm{I,II}} \allowbreak + \allowbreak T_{1}^{\mathrm{I}}T_{1}^{\mathrm{II}} \allowbreak + \allowbreak T_{1}^{\mathrm{I}}T_{21}^{\mathrm{I,II}} \allowbreak + \allowbreak T_{1}^{\mathrm{I}}T_{2}^{\mathrm{I}} \allowbreak + \allowbreak T_{1}^{\mathrm{I}}T_{2}^{\mathrm{I}}T_{1}^{\mathrm{II}} \allowbreak + \allowbreak \frac{1}{2!}T_{1}^{\mathrm{I}^2} \allowbreak + \allowbreak \frac{1}{2!}T_{1}^{\mathrm{I}^2}T_{11}^{\mathrm{I,II}} \allowbreak + \allowbreak \frac{1}{2!}T_{1}^{\mathrm{I}^2}T_{1}^{\mathrm{II}} \allowbreak + \allowbreak \frac{1}{3!}T_{1}^{\mathrm{I}^3} \allowbreak + \allowbreak \frac{1}{3!}T_{1}^{\mathrm{I}^3}T_{1}^{\mathrm{II}} \allowbreak + \allowbreak T_{21}^{\mathrm{I,II}} \allowbreak + \allowbreak T_{2}^{\mathrm{I}} \allowbreak + \allowbreak T_{2}^{\mathrm{I}}T_{11}^{\mathrm{I,II}} \allowbreak + \allowbreak T_{2}^{\mathrm{I}}T_{1}^{\mathrm{II}}  ]  \allowbreak + \allowbreak  [  (-T_{2}^{\mathrm{I}})  ]  V_{\mathrm{N}}^{\mathrm{I,II}}  [  1 \allowbreak + \allowbreak T_{11}^{\mathrm{I,II}} \allowbreak + \allowbreak T_{1}^{\mathrm{I}} \allowbreak + \allowbreak T_{1}^{\mathrm{I}}T_{1}^{\mathrm{II}} \allowbreak + \allowbreak T_{1}^{\mathrm{II}}  ]  \allowbreak + \allowbreak  [  (-T_{1}^{\mathrm{I}})  ]  V_{\mathrm{N}}^{\mathrm{I,II}}  [  1 \allowbreak + \allowbreak T_{11}^{\mathrm{I,II}} \allowbreak + \allowbreak T_{1}^{\mathrm{I}} \allowbreak + \allowbreak T_{1}^{\mathrm{I}}T_{11}^{\mathrm{I,II}} \allowbreak + \allowbreak T_{1}^{\mathrm{I}}T_{1}^{\mathrm{II}} \allowbreak + \allowbreak \frac{1}{2!}T_{1}^{\mathrm{I}^2} \allowbreak + \allowbreak \frac{1}{2!}T_{1}^{\mathrm{I}^2}T_{1}^{\mathrm{II}} \allowbreak + \allowbreak T_{1}^{\mathrm{II}} \allowbreak + \allowbreak T_{21}^{\mathrm{I,II}} \allowbreak + \allowbreak T_{2}^{\mathrm{I}} \allowbreak + \allowbreak T_{2}^{\mathrm{I}}T_{1}^{\mathrm{II}}  ]  \allowbreak + \allowbreak  [  \frac{1}{2!}T_{1}^{\mathrm{I}^2}  ]  V_{\mathrm{N}}^{\mathrm{I,II}}  [  1 \allowbreak + \allowbreak T_{11}^{\mathrm{I,II}} \allowbreak + \allowbreak T_{1}^{\mathrm{I}} \allowbreak + \allowbreak T_{1}^{\mathrm{I}}T_{1}^{\mathrm{II}} \allowbreak + \allowbreak T_{1}^{\mathrm{II}}  ]  \allowbreak + \allowbreak    \tilde{H}_{\mathrm{N}}^{\mathrm{II}}  [  T_{1}^{\mathrm{I}}T_{11}^{\mathrm{I,II}} \allowbreak + \allowbreak T_{1}^{\mathrm{I}}T_{12}^{\mathrm{I,II}} \allowbreak + \allowbreak T_{1}^{\mathrm{I}}T_{1}^{\mathrm{II}}T_{11}^{\mathrm{I,II}} \allowbreak + \allowbreak \frac{1}{2!}T_{1}^{\mathrm{I}^2} \allowbreak + \allowbreak \frac{1}{2!}T_{1}^{\mathrm{I}^2}T_{1}^{\mathrm{II}} \allowbreak + \allowbreak \frac{1}{2!}T_{1}^{\mathrm{I}^2}\frac{1}{2!}T_{1}^{\mathrm{II}^2} \allowbreak + \allowbreak \frac{1}{2!}T_{1}^{\mathrm{I}^2}T_{2}^{\mathrm{II}} \allowbreak + \allowbreak T_{1}^{\mathrm{II}}T_{21}^{\mathrm{I,II}} \allowbreak + \allowbreak T_{21}^{\mathrm{I,II}} \allowbreak + \allowbreak T_{22}^{\mathrm{I,II}} \allowbreak + \allowbreak T_{2}^{\mathrm{I}} \allowbreak + \allowbreak T_{2}^{\mathrm{I}}T_{1}^{\mathrm{II}} \allowbreak + \allowbreak T_{2}^{\mathrm{I}}\frac{1}{2!}T_{1}^{\mathrm{II}^2} \allowbreak + \allowbreak T_{2}^{\mathrm{I}}T_{2}^{\mathrm{II}}  ]  \allowbreak + \allowbreak  [  (-T_{2}^{\mathrm{I}})  ]  \tilde{H}_{\mathrm{N}}^{\mathrm{II}}  [  1 \allowbreak + \allowbreak T_{1}^{\mathrm{II}} \allowbreak + \allowbreak \frac{1}{2!}T_{1}^{\mathrm{II}^2} \allowbreak + \allowbreak T_{2}^{\mathrm{II}}  ]  \allowbreak + \allowbreak  [  (-T_{1}^{\mathrm{I}})  ]  \tilde{H}_{\mathrm{N}}^{\mathrm{II}}  [  T_{11}^{\mathrm{I,II}} \allowbreak + \allowbreak T_{12}^{\mathrm{I,II}} \allowbreak + \allowbreak T_{1}^{\mathrm{I}} \allowbreak + \allowbreak T_{1}^{\mathrm{I}}T_{1}^{\mathrm{II}} \allowbreak + \allowbreak T_{1}^{\mathrm{I}}\frac{1}{2!}T_{1}^{\mathrm{II}^2} \allowbreak + \allowbreak T_{1}^{\mathrm{I}}T_{2}^{\mathrm{II}} \allowbreak + \allowbreak T_{1}^{\mathrm{II}}T_{11}^{\mathrm{I,II}}  ]  \allowbreak + \allowbreak  [  \frac{1}{2!}T_{1}^{\mathrm{I}^2}  ]  \tilde{H}_{\mathrm{N}}^{\mathrm{II}}  [  1 \allowbreak + \allowbreak T_{1}^{\mathrm{II}} \allowbreak + \allowbreak \frac{1}{2!}T_{1}^{\mathrm{II}^2} \allowbreak + \allowbreak T_{2}^{\mathrm{II}}  ]  \allowbreak + \allowbreak    \allowbreak \tilde{H}_{\mathrm{N}}^{\mathrm{I}} \allowbreak  [  1 \allowbreak + \allowbreak T_{1}^{\mathrm{I}} \allowbreak + \allowbreak T_{1}^{\mathrm{I}}T_{2}^{\mathrm{I}} \allowbreak + \allowbreak \frac{1}{2!}T_{1}^{\mathrm{I}^2} \allowbreak + \allowbreak \frac{1}{2!}T_{1}^{\mathrm{I}^2}T_{2}^{\mathrm{I}} \allowbreak + \allowbreak \frac{1}{3!}T_{1}^{\mathrm{I}^3} \allowbreak + \allowbreak \frac{1}{4!}T_{1}^{\mathrm{I}^4} \allowbreak + \allowbreak T_{2}^{\mathrm{I}} \allowbreak + \allowbreak \frac{1}{2!}T_{2}^{\mathrm{I}^2}  ]  \allowbreak + \allowbreak  [  (-T_{2}^{\mathrm{I}})  ]  \allowbreak \tilde{H}_{\mathrm{N}}^{\mathrm{I}} \allowbreak  [  1 \allowbreak + \allowbreak T_{1}^{\mathrm{I}} \allowbreak + \allowbreak \frac{1}{2!}T_{1}^{\mathrm{I}^2} \allowbreak + \allowbreak T_{2}^{\mathrm{I}}  ]  \allowbreak + \allowbreak  [  (-T_{1}^{\mathrm{I}})  ]  \allowbreak \tilde{H}_{\mathrm{N}}^{\mathrm{I}} \allowbreak  [  1 \allowbreak + \allowbreak T_{1}^{\mathrm{I}} \allowbreak + \allowbreak T_{1}^{\mathrm{I}}T_{2}^{\mathrm{I}} \allowbreak + \allowbreak \frac{1}{2!}T_{1}^{\mathrm{I}^2} \allowbreak + \allowbreak \frac{1}{3!}T_{1}^{\mathrm{I}^3} \allowbreak + \allowbreak T_{2}^{\mathrm{I}}  ]  \allowbreak + \allowbreak  [  \frac{1}{2!}T_{1}^{\mathrm{I}^2}  ]  \allowbreak \tilde{H}_{\mathrm{N}}^{\mathrm{I}} \allowbreak  [  1 \allowbreak + \allowbreak T_{1}^{\mathrm{I}} \allowbreak + \allowbreak \frac{1}{2!}T_{1}^{\mathrm{I}^2} \allowbreak + \allowbreak T_{2}^{\mathrm{I}}  ] \vert 0^{\mathrm{I}}0^{\mathrm{II}} \rangle = 0$

%% file: 0S_H_00.tex
\noindent
$\langle 0^{\mathrm{I}}S^{\mathrm{II}} \vert   V_{\mathrm{N}}^{\mathrm{I,II}}  [  1 \allowbreak + \allowbreak T_{11}^{\mathrm{I,II}} \allowbreak + \allowbreak T_{12}^{\mathrm{I,II}} \allowbreak + \allowbreak T_{1}^{\mathrm{I}} \allowbreak + \allowbreak T_{1}^{\mathrm{I}}T_{1}^{\mathrm{II}} \allowbreak + \allowbreak T_{1}^{\mathrm{I}}\frac{1}{2!}T_{1}^{\mathrm{II}^2} \allowbreak + \allowbreak T_{1}^{\mathrm{I}}T_{2}^{\mathrm{II}} \allowbreak + \allowbreak T_{1}^{\mathrm{II}} \allowbreak + \allowbreak T_{1}^{\mathrm{II}}T_{11}^{\mathrm{I,II}} \allowbreak + \allowbreak \frac{1}{2!}T_{1}^{\mathrm{II}^2} \allowbreak + \allowbreak T_{2}^{\mathrm{II}}  ]  \allowbreak + \allowbreak  [  (-T_{1}^{\mathrm{II}})  ]  V_{\mathrm{N}}^{\mathrm{I,II}}  [  1 \allowbreak + \allowbreak T_{11}^{\mathrm{I,II}} \allowbreak + \allowbreak T_{1}^{\mathrm{I}} \allowbreak + \allowbreak T_{1}^{\mathrm{I}}T_{1}^{\mathrm{II}} \allowbreak + \allowbreak T_{1}^{\mathrm{II}}  ]  \allowbreak + \allowbreak    \tilde{H}_{\mathrm{N}}^{\mathrm{II}}  [  1 \allowbreak + \allowbreak T_{1}^{\mathrm{II}} \allowbreak + \allowbreak T_{1}^{\mathrm{II}}T_{2}^{\mathrm{II}} \allowbreak + \allowbreak \frac{1}{2!}T_{1}^{\mathrm{II}^2} \allowbreak + \allowbreak \frac{1}{3!}T_{1}^{\mathrm{II}^3} \allowbreak + \allowbreak T_{2}^{\mathrm{II}}  ]  \allowbreak + \allowbreak  [  (-T_{1}^{\mathrm{II}})  ]  \tilde{H}_{\mathrm{N}}^{\mathrm{II}}  [  1 \allowbreak + \allowbreak T_{1}^{\mathrm{II}} \allowbreak + \allowbreak \frac{1}{2!}T_{1}^{\mathrm{II}^2} \allowbreak + \allowbreak T_{2}^{\mathrm{II}}  ]  \allowbreak + \allowbreak    \tilde{H}_{\mathrm{N}}^{\mathrm{I}}  [  T_{11}^{\mathrm{I,II}} \allowbreak + \allowbreak T_{1}^{\mathrm{I}}T_{11}^{\mathrm{I,II}} \allowbreak + \allowbreak T_{1}^{\mathrm{I}}T_{1}^{\mathrm{II}} \allowbreak + \allowbreak \frac{1}{2!}T_{1}^{\mathrm{I}^2}T_{1}^{\mathrm{II}} \allowbreak + \allowbreak T_{1}^{\mathrm{II}} \allowbreak + \allowbreak T_{21}^{\mathrm{I,II}} \allowbreak + \allowbreak T_{2}^{\mathrm{I}}T_{1}^{\mathrm{II}}  ]  \allowbreak + \allowbreak  [  (-T_{1}^{\mathrm{II}})  ]  \tilde{H}_{\mathrm{N}}^{\mathrm{I}}  [  1 \allowbreak + \allowbreak T_{1}^{\mathrm{I}} \allowbreak + \allowbreak \frac{1}{2!}T_{1}^{\mathrm{I}^2} \allowbreak + \allowbreak T_{2}^{\mathrm{I}}  ] \vert 0^{\mathrm{I}}0^{\mathrm{II}} \rangle = 0$

%% file: 0D_H_00.tex
\noindent
$\langle 0^{\mathrm{I}}D^{\mathrm{II}} \vert   V_{\mathrm{N}}^{\mathrm{I,II}}  [  T_{11}^{\mathrm{I,II}} \allowbreak + \allowbreak T_{12}^{\mathrm{I,II}} \allowbreak + \allowbreak T_{1}^{\mathrm{I}}T_{1}^{\mathrm{II}} \allowbreak + \allowbreak T_{1}^{\mathrm{I}}T_{1}^{\mathrm{II}}T_{2}^{\mathrm{II}} \allowbreak + \allowbreak T_{1}^{\mathrm{I}}\frac{1}{2!}T_{1}^{\mathrm{II}^2} \allowbreak + \allowbreak T_{1}^{\mathrm{I}}\frac{1}{3!}T_{1}^{\mathrm{II}^3} \allowbreak + \allowbreak T_{1}^{\mathrm{I}}T_{2}^{\mathrm{II}} \allowbreak + \allowbreak T_{1}^{\mathrm{II}} \allowbreak + \allowbreak T_{1}^{\mathrm{II}}T_{11}^{\mathrm{I,II}} \allowbreak + \allowbreak T_{1}^{\mathrm{II}}T_{12}^{\mathrm{I,II}} \allowbreak + \allowbreak T_{1}^{\mathrm{II}}T_{2}^{\mathrm{II}} \allowbreak + \allowbreak \frac{1}{2!}T_{1}^{\mathrm{II}^2} \allowbreak + \allowbreak \frac{1}{2!}T_{1}^{\mathrm{II}^2}T_{11}^{\mathrm{I,II}} \allowbreak + \allowbreak \frac{1}{3!}T_{1}^{\mathrm{II}^3} \allowbreak + \allowbreak T_{2}^{\mathrm{II}} \allowbreak + \allowbreak T_{2}^{\mathrm{II}}T_{11}^{\mathrm{I,II}}  ]  \allowbreak + \allowbreak  [  (-T_{2}^{\mathrm{II}})  ]  V_{\mathrm{N}}^{\mathrm{I,II}}  [  1 \allowbreak + \allowbreak T_{11}^{\mathrm{I,II}} \allowbreak + \allowbreak T_{1}^{\mathrm{I}} \allowbreak + \allowbreak T_{1}^{\mathrm{I}}T_{1}^{\mathrm{II}} \allowbreak + \allowbreak T_{1}^{\mathrm{II}}  ]  \allowbreak + \allowbreak  [  (-T_{1}^{\mathrm{II}})  ]  V_{\mathrm{N}}^{\mathrm{I,II}}  [  1 \allowbreak + \allowbreak T_{11}^{\mathrm{I,II}} \allowbreak + \allowbreak T_{12}^{\mathrm{I,II}} \allowbreak + \allowbreak T_{1}^{\mathrm{I}} \allowbreak + \allowbreak T_{1}^{\mathrm{I}}T_{1}^{\mathrm{II}} \allowbreak + \allowbreak T_{1}^{\mathrm{I}}\frac{1}{2!}T_{1}^{\mathrm{II}^2} \allowbreak + \allowbreak T_{1}^{\mathrm{I}}T_{2}^{\mathrm{II}} \allowbreak + \allowbreak T_{1}^{\mathrm{II}} \allowbreak + \allowbreak T_{1}^{\mathrm{II}}T_{11}^{\mathrm{I,II}} \allowbreak + \allowbreak \frac{1}{2!}T_{1}^{\mathrm{II}^2} \allowbreak + \allowbreak T_{2}^{\mathrm{II}}  ]  \allowbreak + \allowbreak  [  \frac{1}{2!}T_{1}^{\mathrm{II}^2}  ]  V_{\mathrm{N}}^{\mathrm{I,II}}  [  1 \allowbreak + \allowbreak T_{11}^{\mathrm{I,II}} \allowbreak + \allowbreak T_{1}^{\mathrm{I}} \allowbreak + \allowbreak T_{1}^{\mathrm{I}}T_{1}^{\mathrm{II}} \allowbreak + \allowbreak T_{1}^{\mathrm{II}}  ]  \allowbreak + \allowbreak    \tilde{H}_{\mathrm{N}}^{\mathrm{II}}  [  1 \allowbreak + \allowbreak T_{1}^{\mathrm{II}} \allowbreak + \allowbreak T_{1}^{\mathrm{II}}T_{2}^{\mathrm{II}} \allowbreak + \allowbreak \frac{1}{2!}T_{1}^{\mathrm{II}^2} \allowbreak + \allowbreak \frac{1}{2!}T_{1}^{\mathrm{II}^2}T_{2}^{\mathrm{II}} \allowbreak + \allowbreak \frac{1}{3!}T_{1}^{\mathrm{II}^3} \allowbreak + \allowbreak \frac{1}{4!}T_{1}^{\mathrm{II}^4} \allowbreak + \allowbreak T_{2}^{\mathrm{II}} \allowbreak + \allowbreak \frac{1}{2!}T_{2}^{\mathrm{II}^2}  ]  \allowbreak + \allowbreak  [  (-T_{2}^{\mathrm{II}})  ]  \tilde{H}_{\mathrm{N}}^{\mathrm{II}}  [  1 \allowbreak + \allowbreak T_{1}^{\mathrm{II}} \allowbreak + \allowbreak \frac{1}{2!}T_{1}^{\mathrm{II}^2} \allowbreak + \allowbreak T_{2}^{\mathrm{II}}  ]  \allowbreak + \allowbreak  [  (-T_{1}^{\mathrm{II}})  ]  \tilde{H}_{\mathrm{N}}^{\mathrm{II}}  [  1 \allowbreak + \allowbreak T_{1}^{\mathrm{II}} \allowbreak + \allowbreak T_{1}^{\mathrm{II}}T_{2}^{\mathrm{II}} \allowbreak + \allowbreak \frac{1}{2!}T_{1}^{\mathrm{II}^2} \allowbreak + \allowbreak \frac{1}{3!}T_{1}^{\mathrm{II}^3} \allowbreak + \allowbreak T_{2}^{\mathrm{II}}  ]  \allowbreak + \allowbreak  [  \frac{1}{2!}T_{1}^{\mathrm{II}^2}  ]  \tilde{H}_{\mathrm{N}}^{\mathrm{II}}  [  1 \allowbreak + \allowbreak T_{1}^{\mathrm{II}} \allowbreak + \allowbreak \frac{1}{2!}T_{1}^{\mathrm{II}^2} \allowbreak + \allowbreak T_{2}^{\mathrm{II}}  ]  \allowbreak + \allowbreak    \allowbreak  \tilde{H}_{\mathrm{N}}^{\mathrm{I}} \allowbreak    [  T_{12}^{\mathrm{I,II}} \allowbreak + \allowbreak T_{1}^{\mathrm{I}}T_{12}^{\mathrm{I,II}} \allowbreak + \allowbreak T_{1}^{\mathrm{I}}T_{1}^{\mathrm{II}}T_{11}^{\mathrm{I,II}} \allowbreak + \allowbreak T_{1}^{\mathrm{I}}\frac{1}{2!}T_{1}^{\mathrm{II}^2} \allowbreak + \allowbreak T_{1}^{\mathrm{I}}T_{2}^{\mathrm{II}} \allowbreak + \allowbreak \frac{1}{2!}T_{1}^{\mathrm{I}^2}\frac{1}{2!}T_{1}^{\mathrm{II}^2} \allowbreak + \allowbreak \frac{1}{2!}T_{1}^{\mathrm{I}^2}T_{2}^{\mathrm{II}} \allowbreak + \allowbreak T_{1}^{\mathrm{II}}T_{11}^{\mathrm{I,II}} \allowbreak + \allowbreak T_{1}^{\mathrm{II}}T_{21}^{\mathrm{I,II}} \allowbreak + \allowbreak \frac{1}{2!}T_{1}^{\mathrm{II}^2} \allowbreak + \allowbreak T_{22}^{\mathrm{I,II}} \allowbreak + \allowbreak T_{2}^{\mathrm{I}}\frac{1}{2!}T_{1}^{\mathrm{II}^2} \allowbreak + \allowbreak T_{2}^{\mathrm{I}}T_{2}^{\mathrm{II}} \allowbreak + \allowbreak T_{2}^{\mathrm{II}}  ]  \allowbreak + \allowbreak  [  (-T_{2}^{\mathrm{II}})  ]  \allowbreak  \tilde{H}_{\mathrm{N}}^{\mathrm{I}} \allowbreak    [  1 \allowbreak + \allowbreak T_{1}^{\mathrm{I}} \allowbreak + \allowbreak \frac{1}{2!}T_{1}^{\mathrm{I}^2} \allowbreak + \allowbreak T_{2}^{\mathrm{I}}  ]  \allowbreak + \allowbreak  [  (-T_{1}^{\mathrm{II}})  ]  \allowbreak  \tilde{H}_{\mathrm{N}}^{\mathrm{I}} \allowbreak    [  T_{11}^{\mathrm{I,II}} \allowbreak + \allowbreak T_{1}^{\mathrm{I}}T_{11}^{\mathrm{I,II}} \allowbreak + \allowbreak T_{1}^{\mathrm{I}}T_{1}^{\mathrm{II}} \allowbreak + \allowbreak \frac{1}{2!}T_{1}^{\mathrm{I}^2}T_{1}^{\mathrm{II}} \allowbreak + \allowbreak T_{1}^{\mathrm{II}} \allowbreak + \allowbreak T_{21}^{\mathrm{I,II}} \allowbreak + \allowbreak T_{2}^{\mathrm{I}}T_{1}^{\mathrm{II}}  ]  \allowbreak + \allowbreak  [  \frac{1}{2!}T_{1}^{\mathrm{II}^2}  ]  \allowbreak  \tilde{H}_{\mathrm{N}}^{\mathrm{I}} \allowbreak   [  1 \allowbreak + \allowbreak T_{1}^{\mathrm{I}} \allowbreak + \allowbreak \frac{1}{2!}T_{1}^{\mathrm{I}^2} \allowbreak + \allowbreak T_{2}^{\mathrm{I}}  ] \vert 0^{\mathrm{I}}0^{\mathrm{II}} \rangle = 0$

%% file: SS_H_00.tex
\noindent
$\langle S^{\mathrm{I}}S^{\mathrm{II}} \vert   V_{\mathrm{N}}^{\mathrm{I,II}}  [  1 \allowbreak + \allowbreak T_{11}^{\mathrm{I,II}} \allowbreak + \allowbreak T_{12}^{\mathrm{I,II}} \allowbreak + \allowbreak T_{1}^{\mathrm{I}} \allowbreak + \allowbreak T_{1}^{\mathrm{I}}T_{11}^{\mathrm{I,II}} \allowbreak + \allowbreak T_{1}^{\mathrm{I}}T_{12}^{\mathrm{I,II}} \allowbreak + \allowbreak T_{1}^{\mathrm{I}}T_{1}^{\mathrm{II}} \allowbreak + \allowbreak T_{1}^{\mathrm{I}}T_{1}^{\mathrm{II}}T_{11}^{\mathrm{I,II}} \allowbreak + \allowbreak T_{1}^{\mathrm{I}}\frac{1}{2!}T_{1}^{\mathrm{II}^2} \allowbreak + \allowbreak T_{1}^{\mathrm{I}}T_{2}^{\mathrm{II}} \allowbreak + \allowbreak \frac{1}{2!}T_{1}^{\mathrm{I}^2} \allowbreak + \allowbreak \frac{1}{2!}T_{1}^{\mathrm{I}^2}T_{1}^{\mathrm{II}} \allowbreak + \allowbreak \frac{1}{2!}T_{1}^{\mathrm{I}^2}\frac{1}{2!}T_{1}^{\mathrm{II}^2} \allowbreak + \allowbreak \frac{1}{2!}T_{1}^{\mathrm{I}^2}T_{2}^{\mathrm{II}} \allowbreak + \allowbreak T_{1}^{\mathrm{II}} \allowbreak + \allowbreak T_{1}^{\mathrm{II}}T_{11}^{\mathrm{I,II}} \allowbreak + \allowbreak T_{1}^{\mathrm{II}}T_{21}^{\mathrm{I,II}} \allowbreak + \allowbreak \frac{1}{2!}T_{1}^{\mathrm{II}^2} \allowbreak + \allowbreak T_{21}^{\mathrm{I,II}} \allowbreak + \allowbreak T_{22}^{\mathrm{I,II}} \allowbreak + \allowbreak T_{2}^{\mathrm{I}} \allowbreak + \allowbreak T_{2}^{\mathrm{I}}T_{1}^{\mathrm{II}} \allowbreak + \allowbreak T_{2}^{\mathrm{I}}\frac{1}{2!}T_{1}^{\mathrm{II}^2} \allowbreak + \allowbreak T_{2}^{\mathrm{I}}T_{2}^{\mathrm{II}} \allowbreak + \allowbreak T_{2}^{\mathrm{II}}  ]  \allowbreak + \allowbreak  [  (-T_{11}^{\mathrm{I,II}})  ]  V_{\mathrm{N}}^{\mathrm{I,II}}  [  1 \allowbreak + \allowbreak T_{11}^{\mathrm{I,II}} \allowbreak + \allowbreak T_{1}^{\mathrm{I}} \allowbreak + \allowbreak T_{1}^{\mathrm{I}}T_{1}^{\mathrm{II}} \allowbreak + \allowbreak T_{1}^{\mathrm{II}}  ]  \allowbreak + \allowbreak  [  (-T_{1}^{\mathrm{II}})  ]  V_{\mathrm{N}}^{\mathrm{I,II}}  [  1 \allowbreak + \allowbreak T_{11}^{\mathrm{I,II}} \allowbreak + \allowbreak T_{1}^{\mathrm{I}} \allowbreak + \allowbreak T_{1}^{\mathrm{I}}T_{11}^{\mathrm{I,II}} \allowbreak + \allowbreak T_{1}^{\mathrm{I}}T_{1}^{\mathrm{II}} \allowbreak + \allowbreak \frac{1}{2!}T_{1}^{\mathrm{I}^2} \allowbreak + \allowbreak \frac{1}{2!}T_{1}^{\mathrm{I}^2}T_{1}^{\mathrm{II}} \allowbreak + \allowbreak T_{1}^{\mathrm{II}} \allowbreak + \allowbreak T_{21}^{\mathrm{I,II}} \allowbreak + \allowbreak T_{2}^{\mathrm{I}} \allowbreak + \allowbreak T_{2}^{\mathrm{I}}T_{1}^{\mathrm{II}}  ]  \allowbreak + \allowbreak  [  (-T_{1}^{\mathrm{I}})  ]  V_{\mathrm{N}}^{\mathrm{I,II}}  [  1 \allowbreak + \allowbreak T_{11}^{\mathrm{I,II}} \allowbreak + \allowbreak T_{12}^{\mathrm{I,II}} \allowbreak + \allowbreak T_{1}^{\mathrm{I}} \allowbreak + \allowbreak T_{1}^{\mathrm{I}}T_{1}^{\mathrm{II}} \allowbreak + \allowbreak T_{1}^{\mathrm{I}}\frac{1}{2!}T_{1}^{\mathrm{II}^2} \allowbreak + \allowbreak T_{1}^{\mathrm{I}}T_{2}^{\mathrm{II}} \allowbreak + \allowbreak T_{1}^{\mathrm{II}} \allowbreak + \allowbreak T_{1}^{\mathrm{II}}T_{11}^{\mathrm{I,II}} \allowbreak + \allowbreak \frac{1}{2!}T_{1}^{\mathrm{II}^2} \allowbreak + \allowbreak T_{2}^{\mathrm{II}}  ]  \allowbreak + \allowbreak  [  (-T_{1}^{\mathrm{I}})(-T_{1}^{\mathrm{II}})  ]  V_{\mathrm{N}}^{\mathrm{I,II}}  [  1 \allowbreak + \allowbreak T_{11}^{\mathrm{I,II}} \allowbreak + \allowbreak T_{1}^{\mathrm{I}} \allowbreak + \allowbreak T_{1}^{\mathrm{I}}T_{1}^{\mathrm{II}} \allowbreak + \allowbreak T_{1}^{\mathrm{II}}  ]  \allowbreak + \allowbreak    \tilde{H}_{\mathrm{N}}^{\mathrm{II}}  [  T_{11}^{\mathrm{I,II}} \allowbreak + \allowbreak T_{12}^{\mathrm{I,II}} \allowbreak + \allowbreak T_{1}^{\mathrm{I}} \allowbreak + \allowbreak T_{1}^{\mathrm{I}}T_{1}^{\mathrm{II}} \allowbreak + \allowbreak T_{1}^{\mathrm{I}}T_{1}^{\mathrm{II}}T_{2}^{\mathrm{II}} \allowbreak + \allowbreak T_{1}^{\mathrm{I}}\frac{1}{2!}T_{1}^{\mathrm{II}^2} \allowbreak + \allowbreak T_{1}^{\mathrm{I}}\frac{1}{3!}T_{1}^{\mathrm{II}^3} \allowbreak + \allowbreak T_{1}^{\mathrm{I}}T_{2}^{\mathrm{II}} \allowbreak + \allowbreak T_{1}^{\mathrm{II}}T_{11}^{\mathrm{I,II}} \allowbreak + \allowbreak T_{1}^{\mathrm{II}}T_{12}^{\mathrm{I,II}} \allowbreak + \allowbreak \frac{1}{2!}T_{1}^{\mathrm{II}^2}T_{11}^{\mathrm{I,II}} \allowbreak + \allowbreak T_{2}^{\mathrm{II}}T_{11}^{\mathrm{I,II}}  ]  \allowbreak + \allowbreak  [  (-T_{11}^{\mathrm{I,II}})  ]  \tilde{H}_{\mathrm{N}}^{\mathrm{II}}  [  1 \allowbreak + \allowbreak T_{1}^{\mathrm{II}} \allowbreak + \allowbreak \frac{1}{2!}T_{1}^{\mathrm{II}^2} \allowbreak + \allowbreak T_{2}^{\mathrm{II}}  ]  \allowbreak + \allowbreak  [  (-T_{1}^{\mathrm{II}})  ]  \tilde{H}_{\mathrm{N}}^{\mathrm{II}}  [  T_{11}^{\mathrm{I,II}} \allowbreak + \allowbreak T_{12}^{\mathrm{I,II}} \allowbreak + \allowbreak T_{1}^{\mathrm{I}} \allowbreak + \allowbreak T_{1}^{\mathrm{I}}T_{1}^{\mathrm{II}} \allowbreak + \allowbreak T_{1}^{\mathrm{I}}\frac{1}{2!}T_{1}^{\mathrm{II}^2} \allowbreak + \allowbreak T_{1}^{\mathrm{I}}T_{2}^{\mathrm{II}} \allowbreak + \allowbreak T_{1}^{\mathrm{II}}T_{11}^{\mathrm{I,II}}  ]  \allowbreak + \allowbreak  [  (-T_{1}^{\mathrm{I}})  ]  \tilde{H}_{\mathrm{N}}^{\mathrm{II}}  [  1 \allowbreak + \allowbreak T_{1}^{\mathrm{II}} \allowbreak + \allowbreak T_{1}^{\mathrm{II}}T_{2}^{\mathrm{II}} \allowbreak + \allowbreak \frac{1}{2!}T_{1}^{\mathrm{II}^2} \allowbreak + \allowbreak \frac{1}{3!}T_{1}^{\mathrm{II}^3} \allowbreak + \allowbreak T_{2}^{\mathrm{II}}  ]  \allowbreak + \allowbreak  [  (-T_{1}^{\mathrm{I}})(-T_{1}^{\mathrm{II}})  ]  \tilde{H}_{\mathrm{N}}^{\mathrm{II}}  [  1 \allowbreak + \allowbreak T_{1}^{\mathrm{II}} \allowbreak + \allowbreak \frac{1}{2!}T_{1}^{\mathrm{II}^2} \allowbreak + \allowbreak T_{2}^{\mathrm{II}}  ]  \allowbreak + \allowbreak    \tilde{H}_{\mathrm{N}}^{\mathrm{I}}  [  T_{11}^{\mathrm{I,II}} \allowbreak + \allowbreak T_{1}^{\mathrm{I}}T_{11}^{\mathrm{I,II}} \allowbreak + \allowbreak T_{1}^{\mathrm{I}}T_{1}^{\mathrm{II}} \allowbreak + \allowbreak T_{1}^{\mathrm{I}}T_{21}^{\mathrm{I,II}} \allowbreak + \allowbreak T_{1}^{\mathrm{I}}T_{2}^{\mathrm{I}}T_{1}^{\mathrm{II}} \allowbreak + \allowbreak \frac{1}{2!}T_{1}^{\mathrm{I}^2}T_{11}^{\mathrm{I,II}} \allowbreak + \allowbreak \frac{1}{2!}T_{1}^{\mathrm{I}^2}T_{1}^{\mathrm{II}} \allowbreak + \allowbreak \frac{1}{3!}T_{1}^{\mathrm{I}^3}T_{1}^{\mathrm{II}} \allowbreak + \allowbreak T_{1}^{\mathrm{II}} \allowbreak + \allowbreak T_{21}^{\mathrm{I,II}} \allowbreak + \allowbreak T_{2}^{\mathrm{I}}T_{11}^{\mathrm{I,II}} \allowbreak + \allowbreak T_{2}^{\mathrm{I}}T_{1}^{\mathrm{II}}  ]  \allowbreak + \allowbreak  [  (-T_{11}^{\mathrm{I,II}})  ]  \tilde{H}_{\mathrm{N}}^{\mathrm{I}}  [  1 \allowbreak + \allowbreak T_{1}^{\mathrm{I}} \allowbreak + \allowbreak \frac{1}{2!}T_{1}^{\mathrm{I}^2} \allowbreak + \allowbreak T_{2}^{\mathrm{I}}  ]  \allowbreak + \allowbreak  [  (-T_{1}^{\mathrm{II}})  ]  \tilde{H}_{\mathrm{N}}^{\mathrm{I}}  [  1 \allowbreak + \allowbreak T_{1}^{\mathrm{I}} \allowbreak + \allowbreak T_{1}^{\mathrm{I}}T_{2}^{\mathrm{I}} \allowbreak + \allowbreak \frac{1}{2!}T_{1}^{\mathrm{I}^2} \allowbreak + \allowbreak \frac{1}{3!}T_{1}^{\mathrm{I}^3} \allowbreak + \allowbreak T_{2}^{\mathrm{I}}  ]  \allowbreak + \allowbreak  [  (-T_{1}^{\mathrm{I}})  ]  \tilde{H}_{\mathrm{N}}^{\mathrm{I}}  [  T_{11}^{\mathrm{I,II}} \allowbreak + \allowbreak T_{1}^{\mathrm{I}}T_{11}^{\mathrm{I,II}} \allowbreak + \allowbreak T_{1}^{\mathrm{I}}T_{1}^{\mathrm{II}} \allowbreak + \allowbreak \frac{1}{2!}T_{1}^{\mathrm{I}^2}T_{1}^{\mathrm{II}} \allowbreak + \allowbreak T_{1}^{\mathrm{II}} \allowbreak + \allowbreak T_{21}^{\mathrm{I,II}} \allowbreak + \allowbreak T_{2}^{\mathrm{I}}T_{1}^{\mathrm{II}}  ]  \allowbreak + \allowbreak  [  (-T_{1}^{\mathrm{I}})(-T_{1}^{\mathrm{II}})  ]  \tilde{H}_{\mathrm{N}}^{\mathrm{I}}  [  1 \allowbreak + \allowbreak T_{1}^{\mathrm{I}} \allowbreak + \allowbreak \frac{1}{2!}T_{1}^{\mathrm{I}^2} \allowbreak + \allowbreak T_{2}^{\mathrm{I}}  ] \vert 0^{\mathrm{I}}0^{\mathrm{II}} \rangle = 0$

%% file: SD_H_00.tex
\noindent
$\langle S^{\mathrm{I}}D^{\mathrm{II}} \vert   V_{\mathrm{N}}^{\mathrm{I,II}}  [  T_{11}^{\mathrm{I,II}} \allowbreak + \allowbreak T_{11}^{\mathrm{I,II}}T_{12}^{\mathrm{I,II}} \allowbreak + \allowbreak T_{12}^{\mathrm{I,II}} \allowbreak + \allowbreak T_{1}^{\mathrm{I}}T_{11}^{\mathrm{I,II}} \allowbreak + \allowbreak T_{1}^{\mathrm{I}}T_{12}^{\mathrm{I,II}} \allowbreak + \allowbreak T_{1}^{\mathrm{I}}T_{1}^{\mathrm{II}} \allowbreak + \allowbreak T_{1}^{\mathrm{I}}T_{1}^{\mathrm{II}}T_{11}^{\mathrm{I,II}} \allowbreak + \allowbreak T_{1}^{\mathrm{I}}T_{1}^{\mathrm{II}}T_{12}^{\mathrm{I,II}} \allowbreak + \allowbreak T_{1}^{\mathrm{I}}T_{1}^{\mathrm{II}}T_{2}^{\mathrm{II}} \allowbreak + \allowbreak T_{1}^{\mathrm{I}}\frac{1}{2!}T_{1}^{\mathrm{II}^2} \allowbreak + \allowbreak T_{1}^{\mathrm{I}}\frac{1}{2!}T_{1}^{\mathrm{II}^2}T_{11}^{\mathrm{I,II}} \allowbreak + \allowbreak T_{1}^{\mathrm{I}}\frac{1}{3!}T_{1}^{\mathrm{II}^3} \allowbreak + \allowbreak T_{1}^{\mathrm{I}}T_{2}^{\mathrm{II}} \allowbreak + \allowbreak T_{1}^{\mathrm{I}}T_{2}^{\mathrm{II}}T_{11}^{\mathrm{I,II}} \allowbreak + \allowbreak \frac{1}{2!}T_{1}^{\mathrm{I}^2}T_{1}^{\mathrm{II}} \allowbreak + \allowbreak \allowbreak  \frac{1}{2!}T_{1}^{\mathrm{I}^2}T_{1}^{\mathrm{II}}T_{2}^{\mathrm{II}} \allowbreak + \allowbreak \frac{1}{2!}T_{1}^{\mathrm{I}^2}\frac{1}{2!}T_{1}^{\mathrm{II}^2} \allowbreak + \allowbreak \frac{1}{2!}T_{1}^{\mathrm{I}^2}\frac{1}{3!}T_{1}^{\mathrm{II}^3} \allowbreak + \allowbreak \frac{1}{2!}T_{1}^{\mathrm{I}^2}T_{2}^{\mathrm{II}} \allowbreak + \allowbreak T_{1}^{\mathrm{II}} \allowbreak + \allowbreak T_{1}^{\mathrm{II}}T_{11}^{\mathrm{I,II}} \allowbreak + \allowbreak T_{1}^{\mathrm{II}}T_{12}^{\mathrm{I,II}} \allowbreak + \allowbreak T_{1}^{\mathrm{II}}T_{21}^{\mathrm{I,II}} \allowbreak + \allowbreak T_{1}^{\mathrm{II}}T_{22}^{\mathrm{I,II}} \allowbreak + \allowbreak T_{1}^{\mathrm{II}}T_{2}^{\mathrm{II}} \allowbreak + \allowbreak \frac{1}{2!}T_{1}^{\mathrm{II}^2} \allowbreak + \allowbreak \frac{1}{2!}T_{1}^{\mathrm{II}^2}T_{11}^{\mathrm{I,II}} \allowbreak + \allowbreak \frac{1}{2!}T_{1}^{\mathrm{II}^2}T_{21}^{\mathrm{I,II}} \allowbreak + \allowbreak \frac{1}{3!}T_{1}^{\mathrm{II}^3} \allowbreak + \allowbreak T_{21}^{\mathrm{I,II}} \allowbreak + \allowbreak T_{22}^{\mathrm{I,II}} \allowbreak + \allowbreak \allowbreak  T_{2}^{\mathrm{I}}T_{1}^{\mathrm{II}} \allowbreak + \allowbreak T_{2}^{\mathrm{I}}T_{1}^{\mathrm{II}}T_{2}^{\mathrm{II}} \allowbreak + \allowbreak T_{2}^{\mathrm{I}}\frac{1}{2!}T_{1}^{\mathrm{II}^2} \allowbreak + \allowbreak T_{2}^{\mathrm{I}}\frac{1}{3!}T_{1}^{\mathrm{II}^3} \allowbreak + \allowbreak T_{2}^{\mathrm{I}}T_{2}^{\mathrm{II}} \allowbreak + \allowbreak T_{2}^{\mathrm{II}} \allowbreak + \allowbreak T_{2}^{\mathrm{II}}T_{11}^{\mathrm{I,II}} \allowbreak + \allowbreak T_{2}^{\mathrm{II}}T_{21}^{\mathrm{I,II}}  ]  \allowbreak + \allowbreak  [  (-T_{12}^{\mathrm{I,II}})  ]  V_{\mathrm{N}}^{\mathrm{I,II}}  [  1 \allowbreak + \allowbreak T_{11}^{\mathrm{I,II}} \allowbreak + \allowbreak T_{1}^{\mathrm{I}} \allowbreak + \allowbreak T_{1}^{\mathrm{I}}T_{1}^{\mathrm{II}} \allowbreak + \allowbreak T_{1}^{\mathrm{II}}  ]  \allowbreak + \allowbreak  [  (-T_{11}^{\mathrm{I,II}})  ]  V_{\mathrm{N}}^{\mathrm{I,II}}  [  1 \allowbreak + \allowbreak T_{11}^{\mathrm{I,II}} \allowbreak + \allowbreak T_{12}^{\mathrm{I,II}} \allowbreak + \allowbreak T_{1}^{\mathrm{I}} \allowbreak + \allowbreak T_{1}^{\mathrm{I}}T_{1}^{\mathrm{II}} \allowbreak + \allowbreak T_{1}^{\mathrm{I}}\frac{1}{2!}T_{1}^{\mathrm{II}^2} \allowbreak + \allowbreak T_{1}^{\mathrm{I}}T_{2}^{\mathrm{II}} \allowbreak + \allowbreak T_{1}^{\mathrm{II}} \allowbreak + \allowbreak T_{1}^{\mathrm{II}}T_{11}^{\mathrm{I,II}} \allowbreak + \allowbreak \frac{1}{2!}T_{1}^{\mathrm{II}^2} \allowbreak + \allowbreak T_{2}^{\mathrm{II}}  ]  \allowbreak + \allowbreak \allowbreak  [  (-T_{2}^{\mathrm{II}})  ]  V_{\mathrm{N}}^{\mathrm{I,II}}  [  1 \allowbreak + \allowbreak T_{11}^{\mathrm{I,II}} \allowbreak + \allowbreak T_{1}^{\mathrm{I}} \allowbreak + \allowbreak T_{1}^{\mathrm{I}}T_{11}^{\mathrm{I,II}} \allowbreak + \allowbreak T_{1}^{\mathrm{I}}T_{1}^{\mathrm{II}} \allowbreak + \allowbreak \frac{1}{2!}T_{1}^{\mathrm{I}^2} \allowbreak + \allowbreak \frac{1}{2!}T_{1}^{\mathrm{I}^2}T_{1}^{\mathrm{II}} \allowbreak + \allowbreak T_{1}^{\mathrm{II}} \allowbreak + \allowbreak T_{21}^{\mathrm{I,II}} \allowbreak + \allowbreak T_{2}^{\mathrm{I}} \allowbreak + \allowbreak T_{2}^{\mathrm{I}}T_{1}^{\mathrm{II}}  ]  \allowbreak + \allowbreak  [  (-T_{1}^{\mathrm{II}})  ]  V_{\mathrm{N}}^{\mathrm{I,II}}  [  1 \allowbreak + \allowbreak T_{11}^{\mathrm{I,II}} \allowbreak + \allowbreak T_{12}^{\mathrm{I,II}} \allowbreak + \allowbreak T_{1}^{\mathrm{I}} \allowbreak + \allowbreak T_{1}^{\mathrm{I}}T_{11}^{\mathrm{I,II}} \allowbreak + \allowbreak T_{1}^{\mathrm{I}}T_{12}^{\mathrm{I,II}} \allowbreak + \allowbreak T_{1}^{\mathrm{I}}T_{1}^{\mathrm{II}} \allowbreak + \allowbreak T_{1}^{\mathrm{I}}T_{1}^{\mathrm{II}}T_{11}^{\mathrm{I,II}} \allowbreak + \allowbreak T_{1}^{\mathrm{I}}\frac{1}{2!}T_{1}^{\mathrm{II}^2} \allowbreak + \allowbreak T_{1}^{\mathrm{I}}T_{2}^{\mathrm{II}} \allowbreak + \allowbreak \frac{1}{2!}T_{1}^{\mathrm{I}^2} \allowbreak + \allowbreak \frac{1}{2!}T_{1}^{\mathrm{I}^2}T_{1}^{\mathrm{II}} \allowbreak + \allowbreak \frac{1}{2!}T_{1}^{\mathrm{I}^2}\frac{1}{2!}T_{1}^{\mathrm{II}^2} \allowbreak + \allowbreak \frac{1}{2!}T_{1}^{\mathrm{I}^2}T_{2}^{\mathrm{II}} \allowbreak  \allowbreak + \allowbreak T_{1}^{\mathrm{II}} \allowbreak + \allowbreak T_{1}^{\mathrm{II}}T_{11}^{\mathrm{I,II}} \allowbreak + \allowbreak T_{1}^{\mathrm{II}}T_{21}^{\mathrm{I,II}} \allowbreak + \allowbreak \frac{1}{2!}T_{1}^{\mathrm{II}^2} \allowbreak + \allowbreak T_{21}^{\mathrm{I,II}} \allowbreak + \allowbreak T_{22}^{\mathrm{I,II}} \allowbreak + \allowbreak T_{2}^{\mathrm{I}} \allowbreak + \allowbreak T_{2}^{\mathrm{I}}T_{1}^{\mathrm{II}} \allowbreak + \allowbreak T_{2}^{\mathrm{I}}\frac{1}{2!}T_{1}^{\mathrm{II}^2} \allowbreak + \allowbreak T_{2}^{\mathrm{I}}T_{2}^{\mathrm{II}} \allowbreak + \allowbreak T_{2}^{\mathrm{II}}  ]  \allowbreak + \allowbreak  [  (-T_{11}^{\mathrm{I,II}})(-T_{1}^{\mathrm{II}})  ]  V_{\mathrm{N}}^{\mathrm{I,II}}  [  1 \allowbreak + \allowbreak T_{11}^{\mathrm{I,II}} \allowbreak + \allowbreak T_{1}^{\mathrm{I}} \allowbreak + \allowbreak T_{1}^{\mathrm{I}}T_{1}^{\mathrm{II}} \allowbreak + \allowbreak T_{1}^{\mathrm{II}}  ]  \allowbreak + \allowbreak  [  \frac{1}{2!}T_{1}^{\mathrm{II}^2}  ]  V_{\mathrm{N}}^{\mathrm{I,II}}  [  1 \allowbreak + \allowbreak T_{11}^{\mathrm{I,II}} \allowbreak + \allowbreak T_{1}^{\mathrm{I}} \allowbreak + \allowbreak T_{1}^{\mathrm{I}}T_{11}^{\mathrm{I,II}} \allowbreak + \allowbreak T_{1}^{\mathrm{I}}T_{1}^{\mathrm{II}} \allowbreak + \allowbreak \frac{1}{2!}T_{1}^{\mathrm{I}^2} \allowbreak + \allowbreak \frac{1}{2!}T_{1}^{\mathrm{I}^2}T_{1}^{\mathrm{II}} \allowbreak + \allowbreak T_{1}^{\mathrm{II}} \allowbreak + \allowbreak T_{21}^{\mathrm{I,II}} \allowbreak + \allowbreak T_{2}^{\mathrm{I}} \allowbreak + \allowbreak T_{2}^{\mathrm{I}}T_{1}^{\mathrm{II}}  ]  \allowbreak + \allowbreak  [  (-T_{1}^{\mathrm{I}})  ]  V_{\mathrm{N}}^{\mathrm{I,II}}  [  T_{11}^{\mathrm{I,II}} \allowbreak + \allowbreak T_{12}^{\mathrm{I,II}} \allowbreak + \allowbreak T_{1}^{\mathrm{I}}T_{1}^{\mathrm{II}} \allowbreak + \allowbreak T_{1}^{\mathrm{I}}T_{1}^{\mathrm{II}}T_{2}^{\mathrm{II}} \allowbreak + \allowbreak T_{1}^{\mathrm{I}}\frac{1}{2!}T_{1}^{\mathrm{II}^2} \allowbreak + \allowbreak T_{1}^{\mathrm{I}}\frac{1}{3!}T_{1}^{\mathrm{II}^3} \allowbreak + \allowbreak T_{1}^{\mathrm{I}}T_{2}^{\mathrm{II}} \allowbreak + \allowbreak T_{1}^{\mathrm{II}} \allowbreak + \allowbreak T_{1}^{\mathrm{II}}T_{11}^{\mathrm{I,II}} \allowbreak + \allowbreak T_{1}^{\mathrm{II}}T_{12}^{\mathrm{I,II}} \allowbreak + \allowbreak T_{1}^{\mathrm{II}}T_{2}^{\mathrm{II}} \allowbreak + \allowbreak \frac{1}{2!}T_{1}^{\mathrm{II}^2} \allowbreak + \allowbreak \frac{1}{2!}T_{1}^{\mathrm{II}^2}T_{11}^{\mathrm{I,II}} \allowbreak + \allowbreak \frac{1}{3!}T_{1}^{\mathrm{II}^3} \allowbreak + \allowbreak T_{2}^{\mathrm{II}} \allowbreak + \allowbreak T_{2}^{\mathrm{II}}T_{11}^{\mathrm{I,II}}  ]  \allowbreak + \allowbreak  [  (-T_{1}^{\mathrm{I}})(-T_{2}^{\mathrm{II}})  ]  V_{\mathrm{N}}^{\mathrm{I,II}}  [  1 \allowbreak + \allowbreak T_{11}^{\mathrm{I,II}} \allowbreak + \allowbreak T_{1}^{\mathrm{I}} \allowbreak + \allowbreak T_{1}^{\mathrm{I}}T_{1}^{\mathrm{II}} \allowbreak + \allowbreak T_{1}^{\mathrm{II}}  ]  \allowbreak + \allowbreak  [  (-T_{1}^{\mathrm{I}})(-T_{1}^{\mathrm{II}})  ]  V_{\mathrm{N}}^{\mathrm{I,II}}  [  1 \allowbreak + \allowbreak T_{11}^{\mathrm{I,II}} \allowbreak + \allowbreak T_{12}^{\mathrm{I,II}} \allowbreak + \allowbreak T_{1}^{\mathrm{I}} \allowbreak + \allowbreak T_{1}^{\mathrm{I}}T_{1}^{\mathrm{II}} \allowbreak + \allowbreak T_{1}^{\mathrm{I}}\frac{1}{2!}T_{1}^{\mathrm{II}^2} \allowbreak + \allowbreak T_{1}^{\mathrm{I}}T_{2}^{\mathrm{II}} \allowbreak + \allowbreak T_{1}^{\mathrm{II}} \allowbreak + \allowbreak T_{1}^{\mathrm{II}}T_{11}^{\mathrm{I,II}} \allowbreak + \allowbreak \frac{1}{2!}T_{1}^{\mathrm{II}^2} \allowbreak + \allowbreak T_{2}^{\mathrm{II}}  ]  \allowbreak + \allowbreak  [  (-T_{1}^{\mathrm{I}})\frac{1}{2!}T_{1}^{\mathrm{II}^2}  ]  V_{\mathrm{N}}^{\mathrm{I,II}}  [  1 \allowbreak + \allowbreak T_{11}^{\mathrm{I,II}} \allowbreak + \allowbreak T_{1}^{\mathrm{I}} \allowbreak + \allowbreak T_{1}^{\mathrm{I}}T_{1}^{\mathrm{II}} \allowbreak + \allowbreak T_{1}^{\mathrm{II}}  ]  \allowbreak + \allowbreak    \tilde{H}_{\mathrm{N}}^{\mathrm{II}}  [  T_{11}^{\mathrm{I,II}} \allowbreak + \allowbreak T_{12}^{\mathrm{I,II}} \allowbreak + \allowbreak T_{1}^{\mathrm{I}} \allowbreak + \allowbreak T_{1}^{\mathrm{I}}T_{1}^{\mathrm{II}} \allowbreak + \allowbreak T_{1}^{\mathrm{I}}T_{1}^{\mathrm{II}}T_{2}^{\mathrm{II}} \allowbreak + \allowbreak T_{1}^{\mathrm{I}}\frac{1}{2!}T_{1}^{\mathrm{II}^2} \allowbreak + \allowbreak T_{1}^{\mathrm{I}}\frac{1}{2!}T_{1}^{\mathrm{II}^2}T_{2}^{\mathrm{II}} \allowbreak + \allowbreak T_{1}^{\mathrm{I}}\frac{1}{3!}T_{1}^{\mathrm{II}^3} \allowbreak + \allowbreak T_{1}^{\mathrm{I}}\frac{1}{4!}T_{1}^{\mathrm{II}^4} \allowbreak + \allowbreak T_{1}^{\mathrm{I}}T_{2}^{\mathrm{II}} \allowbreak + \allowbreak T_{1}^{\mathrm{I}}\frac{1}{2!}T_{2}^{\mathrm{II}^2} \allowbreak + \allowbreak T_{1}^{\mathrm{II}}T_{11}^{\mathrm{I,II}} \allowbreak + \allowbreak T_{1}^{\mathrm{II}}T_{12}^{\mathrm{I,II}} \allowbreak + \allowbreak T_{1}^{\mathrm{II}}T_{2}^{\mathrm{II}}T_{11}^{\mathrm{I,II}} \allowbreak + \allowbreak \frac{1}{2!}T_{1}^{\mathrm{II}^2}T_{11}^{\mathrm{I,II}} \allowbreak + \allowbreak \frac{1}{2!}T_{1}^{\mathrm{II}^2}T_{12}^{\mathrm{I,II}} \allowbreak + \allowbreak \frac{1}{3!}T_{1}^{\mathrm{II}^3}T_{11}^{\mathrm{I,II}} \allowbreak + \allowbreak T_{2}^{\mathrm{II}}T_{11}^{\mathrm{I,II}} \allowbreak + \allowbreak T_{2}^{\mathrm{II}}T_{12}^{\mathrm{I,II}}  ]  \allowbreak + \allowbreak  [  (-T_{12}^{\mathrm{I,II}})  ]  \tilde{H}_{\mathrm{N}}^{\mathrm{II}}  [  1 \allowbreak + \allowbreak T_{1}^{\mathrm{II}} \allowbreak + \allowbreak \frac{1}{2!}T_{1}^{\mathrm{II}^2} \allowbreak + \allowbreak T_{2}^{\mathrm{II}}  ]  \allowbreak + \allowbreak  [  (-T_{11}^{\mathrm{I,II}})  ]  \tilde{H}_{\mathrm{N}}^{\mathrm{II}}  [  1 \allowbreak + \allowbreak T_{1}^{\mathrm{II}} \allowbreak + \allowbreak T_{1}^{\mathrm{II}}T_{2}^{\mathrm{II}} \allowbreak + \allowbreak \frac{1}{2!}T_{1}^{\mathrm{II}^2} \allowbreak + \allowbreak \frac{1}{3!}T_{1}^{\mathrm{II}^3} \allowbreak + \allowbreak T_{2}^{\mathrm{II}}  ]  \allowbreak + \allowbreak  [  (-T_{2}^{\mathrm{II}})  ]  \tilde{H}_{\mathrm{N}}^{\mathrm{II}}  [  T_{11}^{\mathrm{I,II}} \allowbreak + \allowbreak T_{12}^{\mathrm{I,II}} \allowbreak + \allowbreak T_{1}^{\mathrm{I}} \allowbreak + \allowbreak T_{1}^{\mathrm{I}}T_{1}^{\mathrm{II}} \allowbreak + \allowbreak T_{1}^{\mathrm{I}}\frac{1}{2!}T_{1}^{\mathrm{II}^2} \allowbreak + \allowbreak T_{1}^{\mathrm{I}}T_{2}^{\mathrm{II}} \allowbreak + \allowbreak T_{1}^{\mathrm{II}}T_{11}^{\mathrm{I,II}}  ]  \allowbreak + \allowbreak  [  (-T_{1}^{\mathrm{II}})  ]  \tilde{H}_{\mathrm{N}}^{\mathrm{II}}  [  T_{11}^{\mathrm{I,II}} \allowbreak + \allowbreak T_{12}^{\mathrm{I,II}} \allowbreak + \allowbreak T_{1}^{\mathrm{I}} \allowbreak + \allowbreak T_{1}^{\mathrm{I}}T_{1}^{\mathrm{II}} \allowbreak + \allowbreak T_{1}^{\mathrm{I}}T_{1}^{\mathrm{II}}T_{2}^{\mathrm{II}} \allowbreak + \allowbreak T_{1}^{\mathrm{I}}\frac{1}{2!}T_{1}^{\mathrm{II}^2} \allowbreak + \allowbreak T_{1}^{\mathrm{I}}\frac{1}{3!}T_{1}^{\mathrm{II}^3} \allowbreak + \allowbreak T_{1}^{\mathrm{I}}T_{2}^{\mathrm{II}} \allowbreak + \allowbreak T_{1}^{\mathrm{II}}T_{11}^{\mathrm{I,II}} \allowbreak + \allowbreak T_{1}^{\mathrm{II}}T_{12}^{\mathrm{I,II}} \allowbreak + \allowbreak \frac{1}{2!}T_{1}^{\mathrm{II}^2}T_{11}^{\mathrm{I,II}} \allowbreak + \allowbreak T_{2}^{\mathrm{II}}T_{11}^{\mathrm{I,II}}  ]  \allowbreak + \allowbreak  [  (-T_{11}^{\mathrm{I,II}})(-T_{1}^{\mathrm{II}})  ]  \tilde{H}_{\mathrm{N}}^{\mathrm{II}}  [  1 \allowbreak + \allowbreak T_{1}^{\mathrm{II}} \allowbreak + \allowbreak \frac{1}{2!}T_{1}^{\mathrm{II}^2} \allowbreak + \allowbreak T_{2}^{\mathrm{II}}  ]  \allowbreak + \allowbreak  [  \frac{1}{2!}T_{1}^{\mathrm{II}^2}  ]  \tilde{H}_{\mathrm{N}}^{\mathrm{II}}  [  T_{11}^{\mathrm{I,II}} \allowbreak + \allowbreak T_{12}^{\mathrm{I,II}} \allowbreak + \allowbreak T_{1}^{\mathrm{I}} \allowbreak + \allowbreak T_{1}^{\mathrm{I}}T_{1}^{\mathrm{II}} \allowbreak + \allowbreak T_{1}^{\mathrm{I}}\frac{1}{2!}T_{1}^{\mathrm{II}^2} \allowbreak + \allowbreak T_{1}^{\mathrm{I}}T_{2}^{\mathrm{II}} \allowbreak + \allowbreak T_{1}^{\mathrm{II}}T_{11}^{\mathrm{I,II}}  ]  \allowbreak + \allowbreak  [  (-T_{1}^{\mathrm{I}})  ]  \tilde{H}_{\mathrm{N}}^{\mathrm{II}}  [  1 \allowbreak + \allowbreak T_{1}^{\mathrm{II}} \allowbreak + \allowbreak T_{1}^{\mathrm{II}}T_{2}^{\mathrm{II}} \allowbreak + \allowbreak \frac{1}{2!}T_{1}^{\mathrm{II}^2} \allowbreak + \allowbreak \frac{1}{2!}T_{1}^{\mathrm{II}^2}T_{2}^{\mathrm{II}} \allowbreak + \allowbreak \frac{1}{3!}T_{1}^{\mathrm{II}^3} \allowbreak + \allowbreak \frac{1}{4!}T_{1}^{\mathrm{II}^4} \allowbreak + \allowbreak T_{2}^{\mathrm{II}} \allowbreak + \allowbreak \frac{1}{2!}T_{2}^{\mathrm{II}^2}  ]  \allowbreak + \allowbreak  [  (-T_{1}^{\mathrm{I}})(-T_{2}^{\mathrm{II}})  ]  \tilde{H}_{\mathrm{N}}^{\mathrm{II}}  [  1 \allowbreak + \allowbreak T_{1}^{\mathrm{II}} \allowbreak + \allowbreak \frac{1}{2!}T_{1}^{\mathrm{II}^2} \allowbreak + \allowbreak T_{2}^{\mathrm{II}}  ]  \allowbreak + \allowbreak  [  (-T_{1}^{\mathrm{I}})(-T_{1}^{\mathrm{II}})  ]  \tilde{H}_{\mathrm{N}}^{\mathrm{II}}  [  1 \allowbreak + \allowbreak T_{1}^{\mathrm{II}} \allowbreak + \allowbreak T_{1}^{\mathrm{II}}T_{2}^{\mathrm{II}} \allowbreak + \allowbreak \frac{1}{2!}T_{1}^{\mathrm{II}^2} \allowbreak + \allowbreak \frac{1}{3!}T_{1}^{\mathrm{II}^3} \allowbreak + \allowbreak T_{2}^{\mathrm{II}}  ]  \allowbreak + \allowbreak  [  (-T_{1}^{\mathrm{I}})\frac{1}{2!}T_{1}^{\mathrm{II}^2}  ]  \tilde{H}_{\mathrm{N}}^{\mathrm{II}}  [  1 \allowbreak + \allowbreak T_{1}^{\mathrm{II}} \allowbreak + \allowbreak \frac{1}{2!}T_{1}^{\mathrm{II}^2} \allowbreak + \allowbreak T_{2}^{\mathrm{II}}  ]  \allowbreak + \allowbreak    \allowbreak \tilde{H}_{\mathrm{N}}^{\mathrm{I}} \allowbreak  [  T_{11}^{\mathrm{I,II}}T_{21}^{\mathrm{I,II}} \allowbreak + \allowbreak T_{12}^{\mathrm{I,II}} \allowbreak + \allowbreak T_{1}^{\mathrm{I}}T_{12}^{\mathrm{I,II}} \allowbreak + \allowbreak T_{1}^{\mathrm{I}}T_{1}^{\mathrm{II}}T_{11}^{\mathrm{I,II}} \allowbreak + \allowbreak T_{1}^{\mathrm{I}}T_{1}^{\mathrm{II}}T_{21}^{\mathrm{I,II}} \allowbreak + \allowbreak T_{1}^{\mathrm{I}}\frac{1}{2!}T_{1}^{\mathrm{II}^2} \allowbreak + \allowbreak T_{1}^{\mathrm{I}}T_{22}^{\mathrm{I,II}} \allowbreak + \allowbreak T_{1}^{\mathrm{I}}T_{2}^{\mathrm{I}}\frac{1}{2!}T_{1}^{\mathrm{II}^2} \allowbreak + \allowbreak T_{1}^{\mathrm{I}}T_{2}^{\mathrm{I}}T_{2}^{\mathrm{II}} \allowbreak + \allowbreak T_{1}^{\mathrm{I}}T_{2}^{\mathrm{II}} \allowbreak + \allowbreak \frac{1}{2!}T_{1}^{\mathrm{I}^2}T_{12}^{\mathrm{I,II}} \allowbreak + \allowbreak \frac{1}{2!}T_{1}^{\mathrm{I}^2}T_{1}^{\mathrm{II}}T_{11}^{\mathrm{I,II}} \allowbreak + \allowbreak \frac{1}{2!}T_{1}^{\mathrm{I}^2}\frac{1}{2!}T_{1}^{\mathrm{II}^2} \allowbreak + \allowbreak \frac{1}{2!}T_{1}^{\mathrm{I}^2}T_{2}^{\mathrm{II}} \allowbreak + \allowbreak \frac{1}{3!}T_{1}^{\mathrm{I}^3}\frac{1}{2!}T_{1}^{\mathrm{II}^2} \allowbreak + \allowbreak \frac{1}{3!}T_{1}^{\mathrm{I}^3}T_{2}^{\mathrm{II}} \allowbreak + \allowbreak T_{1}^{\mathrm{II}}T_{11}^{\mathrm{I,II}} \allowbreak + \allowbreak T_{1}^{\mathrm{II}}T_{21}^{\mathrm{I,II}} \allowbreak + \allowbreak \frac{1}{2!}T_{1}^{\mathrm{II}^2} \allowbreak + \allowbreak T_{22}^{\mathrm{I,II}} \allowbreak + \allowbreak T_{2}^{\mathrm{I}}T_{12}^{\mathrm{I,II}} \allowbreak + \allowbreak T_{2}^{\mathrm{I}}T_{1}^{\mathrm{II}}T_{11}^{\mathrm{I,II}} \allowbreak + \allowbreak T_{2}^{\mathrm{I}}\frac{1}{2!}T_{1}^{\mathrm{II}^2} \allowbreak + \allowbreak T_{2}^{\mathrm{I}}T_{2}^{\mathrm{II}} \allowbreak + \allowbreak T_{2}^{\mathrm{II}}  ]  \allowbreak + \allowbreak  [  (-T_{12}^{\mathrm{I,II}})  ]  \allowbreak \tilde{H}_{\mathrm{N}}^{\mathrm{I}} \allowbreak \allowbreak  [  1 \allowbreak + \allowbreak T_{1}^{\mathrm{I}} \allowbreak + \allowbreak \frac{1}{2!}T_{1}^{\mathrm{I}^2} \allowbreak + \allowbreak T_{2}^{\mathrm{I}}  ]  \allowbreak + \allowbreak  [  (-T_{11}^{\mathrm{I,II}})  ]  \allowbreak \tilde{H}_{\mathrm{N}}^{\mathrm{I}} \allowbreak  [  T_{11}^{\mathrm{I,II}} \allowbreak + \allowbreak T_{1}^{\mathrm{I}}T_{11}^{\mathrm{I,II}} \allowbreak + \allowbreak T_{1}^{\mathrm{I}}T_{1}^{\mathrm{II}} \allowbreak + \allowbreak \frac{1}{2!}T_{1}^{\mathrm{I}^2}T_{1}^{\mathrm{II}} \allowbreak + \allowbreak T_{1}^{\mathrm{II}} \allowbreak + \allowbreak T_{21}^{\mathrm{I,II}} \allowbreak + \allowbreak T_{2}^{\mathrm{I}}T_{1}^{\mathrm{II}}  ]  \allowbreak + \allowbreak  [  (-T_{2}^{\mathrm{II}})  ]  \allowbreak \tilde{H}_{\mathrm{N}}^{\mathrm{I}} \allowbreak  [  1 \allowbreak + \allowbreak T_{1}^{\mathrm{I}} \allowbreak + \allowbreak T_{1}^{\mathrm{I}}T_{2}^{\mathrm{I}} \allowbreak + \allowbreak \frac{1}{2!}T_{1}^{\mathrm{I}^2} \allowbreak + \allowbreak \frac{1}{3!}T_{1}^{\mathrm{I}^3} \allowbreak + \allowbreak T_{2}^{\mathrm{I}}  ]  \allowbreak + \allowbreak  [  (-T_{1}^{\mathrm{II}})  ]  \allowbreak \tilde{H}_{\mathrm{N}}^{\mathrm{I}} \allowbreak  [  T_{11}^{\mathrm{I,II}} \allowbreak + \allowbreak T_{1}^{\mathrm{I}}T_{11}^{\mathrm{I,II}} \allowbreak + \allowbreak T_{1}^{\mathrm{I}}T_{1}^{\mathrm{II}} \allowbreak + \allowbreak T_{1}^{\mathrm{I}}T_{21}^{\mathrm{I,II}} \allowbreak + \allowbreak T_{1}^{\mathrm{I}}T_{2}^{\mathrm{I}}T_{1}^{\mathrm{II}} \allowbreak + \allowbreak \frac{1}{2!}T_{1}^{\mathrm{I}^2}T_{11}^{\mathrm{I,II}} \allowbreak + \allowbreak \frac{1}{2!}T_{1}^{\mathrm{I}^2}T_{1}^{\mathrm{II}} \allowbreak + \allowbreak \frac{1}{3!}T_{1}^{\mathrm{I}^3}T_{1}^{\mathrm{II}} \allowbreak + \allowbreak T_{1}^{\mathrm{II}} \allowbreak + \allowbreak T_{21}^{\mathrm{I,II}} \allowbreak + \allowbreak T_{2}^{\mathrm{I}}T_{11}^{\mathrm{I,II}} \allowbreak + \allowbreak T_{2}^{\mathrm{I}}T_{1}^{\mathrm{II}}  ]  \allowbreak + \allowbreak  [  (-T_{11}^{\mathrm{I,II}})(-T_{1}^{\mathrm{II}})  ]  \allowbreak \tilde{H}_{\mathrm{N}}^{\mathrm{I}} \allowbreak  [  1 \allowbreak + \allowbreak T_{1}^{\mathrm{I}} \allowbreak + \allowbreak \frac{1}{2!}T_{1}^{\mathrm{I}^2} \allowbreak + \allowbreak T_{2}^{\mathrm{I}}  ]  \allowbreak + \allowbreak  [  \frac{1}{2!}T_{1}^{\mathrm{II}^2}  ]  \allowbreak \tilde{H}_{\mathrm{N}}^{\mathrm{I}} \allowbreak  [  1 \allowbreak + \allowbreak T_{1}^{\mathrm{I}} \allowbreak + \allowbreak T_{1}^{\mathrm{I}}T_{2}^{\mathrm{I}} \allowbreak + \allowbreak \frac{1}{2!}T_{1}^{\mathrm{I}^2} \allowbreak + \allowbreak \frac{1}{3!}T_{1}^{\mathrm{I}^3} \allowbreak + \allowbreak T_{2}^{\mathrm{I}}  ]  \allowbreak + \allowbreak  [  (-T_{1}^{\mathrm{I}})  ] \allowbreak  \allowbreak \tilde{H}_{\mathrm{N}}^{\mathrm{I}} \allowbreak  [  T_{12}^{\mathrm{I,II}} \allowbreak + \allowbreak T_{1}^{\mathrm{I}}T_{12}^{\mathrm{I,II}} \allowbreak + \allowbreak T_{1}^{\mathrm{I}}T_{1}^{\mathrm{II}}T_{11}^{\mathrm{I,II}} \allowbreak + \allowbreak T_{1}^{\mathrm{I}}\frac{1}{2!}T_{1}^{\mathrm{II}^2} \allowbreak + \allowbreak T_{1}^{\mathrm{I}}T_{2}^{\mathrm{II}} \allowbreak + \allowbreak \frac{1}{2!}T_{1}^{\mathrm{I}^2}\frac{1}{2!}T_{1}^{\mathrm{II}^2} \allowbreak + \allowbreak \frac{1}{2!}T_{1}^{\mathrm{I}^2}T_{2}^{\mathrm{II}} \allowbreak + \allowbreak T_{1}^{\mathrm{II}}T_{11}^{\mathrm{I,II}} \allowbreak + \allowbreak T_{1}^{\mathrm{II}}T_{21}^{\mathrm{I,II}} \allowbreak + \allowbreak \allowbreak  \frac{1}{2!}T_{1}^{\mathrm{II}^2} \allowbreak + \allowbreak T_{22}^{\mathrm{I,II}} \allowbreak + \allowbreak T_{2}^{\mathrm{I}}\frac{1}{2!}T_{1}^{\mathrm{II}^2} \allowbreak + \allowbreak T_{2}^{\mathrm{I}}T_{2}^{\mathrm{II}} \allowbreak + \allowbreak T_{2}^{\mathrm{II}}  ]  \allowbreak + \allowbreak  [  (-T_{1}^{\mathrm{I}})(-T_{2}^{\mathrm{II}})  ]  \allowbreak \tilde{H}_{\mathrm{N}}^{\mathrm{I}} \allowbreak \allowbreak [  1 \allowbreak + \allowbreak T_{1}^{\mathrm{I}} \allowbreak + \allowbreak \allowbreak  \frac{1}{2!}T_{1}^{\mathrm{I}^2} \allowbreak + \allowbreak T_{2}^{\mathrm{I}}  ]  \allowbreak + \allowbreak  [  (-T_{1}^{\mathrm{I}})(-T_{1}^{\mathrm{II}})  ]  \allowbreak \tilde{H}_{\mathrm{N}}^{\mathrm{I}} \allowbreak  [  T_{11}^{\mathrm{I,II}}  \allowbreak + \allowbreak T_{1}^{\mathrm{I}}T_{11}^{\mathrm{I,II}} \allowbreak + \allowbreak T_{1}^{\mathrm{I}}T_{1}^{\mathrm{II}} \allowbreak + \allowbreak \frac{1}{2!}T_{1}^{\mathrm{I}^2}T_{1}^{\mathrm{II}} \allowbreak + \allowbreak T_{1}^{\mathrm{II}} \allowbreak + \allowbreak T_{21}^{\mathrm{I,II}} \allowbreak + \allowbreak T_{2}^{\mathrm{I}}T_{1}^{\mathrm{II}}  ]  \allowbreak + \allowbreak  [  (-T_{1}^{\mathrm{I}})\frac{1}{2!}T_{1}^{\mathrm{II}^2}  ]  \allowbreak \tilde{H}_{\mathrm{N}}^{\mathrm{I}} \allowbreak  [  1 \allowbreak + \allowbreak T_{1}^{\mathrm{I}} \allowbreak + \allowbreak 
\frac{1}{2!}T_{1}^{\mathrm{I}^2} \allowbreak + \allowbreak T_{2}^{\mathrm{I}}  ] \vert 0^{\mathrm{I}}0^{\mathrm{II}} \rangle = 0$

%% file: DS_H_00.tex
\noindent
$\langle D^{\mathrm{I}}S^{\mathrm{II}} \vert   \allowbreak  V_{\mathrm{N}}^{\mathrm{I,II}} \allowbreak  [  T_{11}^{\mathrm{I,II}} \allowbreak + \allowbreak T_{11}^{\mathrm{I,II}}T_{21}^{\mathrm{I,II}} \allowbreak + \allowbreak T_{12}^{\mathrm{I,II}} \allowbreak + \allowbreak T_{1}^{\mathrm{I}} \allowbreak + \allowbreak T_{1}^{\mathrm{I}}T_{11}^{\mathrm{I,II}} \allowbreak + \allowbreak T_{1}^{\mathrm{I}}T_{12}^{\mathrm{I,II}} \allowbreak + \allowbreak T_{1}^{\mathrm{I}}T_{1}^{\mathrm{II}} \allowbreak + \allowbreak T_{1}^{\mathrm{I}}T_{1}^{\mathrm{II}}T_{11}^{\mathrm{I,II}} \allowbreak + \allowbreak T_{1}^{\mathrm{I}}T_{1}^{\mathrm{II}}T_{21}^{\mathrm{I,II}} \allowbreak + \allowbreak T_{1}^{\mathrm{I}}\frac{1}{2!}T_{1}^{\mathrm{II}^2} \allowbreak + \allowbreak T_{1}^{\mathrm{I}}T_{21}^{\mathrm{I,II}} \allowbreak + \allowbreak T_{1}^{\mathrm{I}}T_{22}^{\mathrm{I,II}} \allowbreak + \allowbreak T_{1}^{\mathrm{I}}T_{2}^{\mathrm{I}} \allowbreak + \allowbreak T_{1}^{\mathrm{I}}T_{2}^{\mathrm{I}}T_{1}^{\mathrm{II}} \allowbreak + \allowbreak T_{1}^{\mathrm{I}}T_{2}^{\mathrm{I}}\frac{1}{2!}T_{1}^{\mathrm{II}^2} \allowbreak + \allowbreak T_{1}^{\mathrm{I}}T_{2}^{\mathrm{I}}T_{2}^{\mathrm{II}} \allowbreak + \allowbreak T_{1}^{\mathrm{I}}T_{2}^{\mathrm{II}} \allowbreak + \allowbreak \frac{1}{2!}T_{1}^{\mathrm{I}^2} \allowbreak + \allowbreak \frac{1}{2!}T_{1}^{\mathrm{I}^2}T_{11}^{\mathrm{I,II}} \allowbreak + \allowbreak \frac{1}{2!}T_{1}^{\mathrm{I}^2}T_{12}^{\mathrm{I,II}} \allowbreak + \allowbreak \frac{1}{2!}T_{1}^{\mathrm{I}^2}T_{1}^{\mathrm{II}} \allowbreak + \allowbreak \frac{1}{2!}T_{1}^{\mathrm{I}^2}T_{1}^{\mathrm{II}}T_{11}^{\mathrm{I,II}} \allowbreak + \allowbreak \frac{1}{2!}T_{1}^{\mathrm{I}^2}\frac{1}{2!}T_{1}^{\mathrm{II}^2} \allowbreak + \allowbreak \frac{1}{2!}T_{1}^{\mathrm{I}^2}T_{2}^{\mathrm{II}} \allowbreak + \allowbreak \frac{1}{3!}T_{1}^{\mathrm{I}^3} \allowbreak + \allowbreak \frac{1}{3!}T_{1}^{\mathrm{I}^3}T_{1}^{\mathrm{II}} \allowbreak + \allowbreak \frac{1}{3!}T_{1}^{\mathrm{I}^3}\frac{1}{2!}T_{1}^{\mathrm{II}^2} \allowbreak + \allowbreak \frac{1}{3!}T_{1}^{\mathrm{I}^3}T_{2}^{\mathrm{II}} \allowbreak + \allowbreak T_{1}^{\mathrm{II}}T_{11}^{\mathrm{I,II}} \allowbreak + \allowbreak T_{1}^{\mathrm{II}}T_{21}^{\mathrm{I,II}} \allowbreak + \allowbreak T_{21}^{\mathrm{I,II}} \allowbreak + \allowbreak T_{22}^{\mathrm{I,II}} \allowbreak + \allowbreak T_{2}^{\mathrm{I}} \allowbreak + \allowbreak T_{2}^{\mathrm{I}}T_{11}^{\mathrm{I,II}} \allowbreak + \allowbreak T_{2}^{\mathrm{I}}T_{12}^{\mathrm{I,II}} \allowbreak + \allowbreak T_{2}^{\mathrm{I}}T_{1}^{\mathrm{II}} \allowbreak + \allowbreak T_{2}^{\mathrm{I}}T_{1}^{\mathrm{II}}T_{11}^{\mathrm{I,II}} \allowbreak + \allowbreak T_{2}^{\mathrm{I}}\frac{1}{2!}T_{1}^{\mathrm{II}^2} \allowbreak + \allowbreak T_{2}^{\mathrm{I}}T_{2}^{\mathrm{II}}  ]  \allowbreak + \allowbreak  [  (-T_{21}^{\mathrm{I,II}})  ]  \allowbreak  V_{\mathrm{N}}^{\mathrm{I,II}} \allowbreak  [  1 \allowbreak + \allowbreak T_{11}^{\mathrm{I,II}} \allowbreak + \allowbreak T_{1}^{\mathrm{I}} \allowbreak + \allowbreak T_{1}^{\mathrm{I}}T_{1}^{\mathrm{II}} \allowbreak + \allowbreak T_{1}^{\mathrm{II}}  ]  \allowbreak + \allowbreak  [  (-T_{11}^{\mathrm{I,II}})  ]  \allowbreak  V_{\mathrm{N}}^{\mathrm{I,II}} \allowbreak  [  1 \allowbreak + \allowbreak T_{11}^{\mathrm{I,II}} \allowbreak + \allowbreak T_{1}^{\mathrm{I}} \allowbreak + \allowbreak T_{1}^{\mathrm{I}}T_{11}^{\mathrm{I,II}} \allowbreak + \allowbreak T_{1}^{\mathrm{I}}T_{1}^{\mathrm{II}} \allowbreak + \allowbreak \frac{1}{2!}T_{1}^{\mathrm{I}^2} \allowbreak + \allowbreak \frac{1}{2!}T_{1}^{\mathrm{I}^2}T_{1}^{\mathrm{II}} \allowbreak + \allowbreak T_{1}^{\mathrm{II}} \allowbreak + \allowbreak T_{21}^{\mathrm{I,II}} \allowbreak + \allowbreak T_{2}^{\mathrm{I}} \allowbreak + \allowbreak T_{2}^{\mathrm{I}}T_{1}^{\mathrm{II}}  ]  \allowbreak + \allowbreak  [  (-T_{1}^{\mathrm{II}})  ]  \allowbreak  V_{\mathrm{N}}^{\mathrm{I,II}} \allowbreak  [  T_{11}^{\mathrm{I,II}} \allowbreak + \allowbreak T_{1}^{\mathrm{I}} \allowbreak + \allowbreak T_{1}^{\mathrm{I}}T_{11}^{\mathrm{I,II}} \allowbreak + \allowbreak T_{1}^{\mathrm{I}}T_{1}^{\mathrm{II}} \allowbreak + \allowbreak T_{1}^{\mathrm{I}}T_{21}^{\mathrm{I,II}} \allowbreak + \allowbreak T_{1}^{\mathrm{I}}T_{2}^{\mathrm{I}} \allowbreak + \allowbreak T_{1}^{\mathrm{I}}T_{2}^{\mathrm{I}}T_{1}^{\mathrm{II}} \allowbreak + \allowbreak \frac{1}{2!}T_{1}^{\mathrm{I}^2} \allowbreak + \allowbreak \frac{1}{2!}T_{1}^{\mathrm{I}^2}T_{11}^{\mathrm{I,II}} \allowbreak + \allowbreak \frac{1}{2!}T_{1}^{\mathrm{I}^2}T_{1}^{\mathrm{II}} \allowbreak + \allowbreak \frac{1}{3!}T_{1}^{\mathrm{I}^3} \allowbreak + \allowbreak \frac{1}{3!}T_{1}^{\mathrm{I}^3}T_{1}^{\mathrm{II}} \allowbreak + \allowbreak T_{21}^{\mathrm{I,II}} \allowbreak + \allowbreak T_{2}^{\mathrm{I}} \allowbreak + \allowbreak T_{2}^{\mathrm{I}}T_{11}^{\mathrm{I,II}} \allowbreak + \allowbreak T_{2}^{\mathrm{I}}T_{1}^{\mathrm{II}}  ]  \allowbreak + \allowbreak  [  (-T_{2}^{\mathrm{I}})  ]  \allowbreak  V_{\mathrm{N}}^{\mathrm{I,II}} \allowbreak  [  1 \allowbreak + \allowbreak T_{11}^{\mathrm{I,II}} \allowbreak + \allowbreak T_{12}^{\mathrm{I,II}} \allowbreak + \allowbreak T_{1}^{\mathrm{I}} \allowbreak + \allowbreak T_{1}^{\mathrm{I}}T_{1}^{\mathrm{II}} \allowbreak + \allowbreak T_{1}^{\mathrm{I}}\frac{1}{2!}T_{1}^{\mathrm{II}^2} \allowbreak + \allowbreak T_{1}^{\mathrm{I}}T_{2}^{\mathrm{II}} \allowbreak + \allowbreak T_{1}^{\mathrm{II}} \allowbreak + \allowbreak T_{1}^{\mathrm{II}}T_{11}^{\mathrm{I,II}} \allowbreak + \allowbreak \frac{1}{2!}T_{1}^{\mathrm{II}^2} \allowbreak + \allowbreak T_{2}^{\mathrm{II}}  ]  \allowbreak + \allowbreak  [  (-T_{1}^{\mathrm{II}})(-T_{2}^{\mathrm{I}})  ]  \allowbreak  V_{\mathrm{N}}^{\mathrm{I,II}} \allowbreak  [  1 \allowbreak + \allowbreak T_{11}^{\mathrm{I,II}} \allowbreak + \allowbreak T_{1}^{\mathrm{I}} \allowbreak + \allowbreak T_{1}^{\mathrm{I}}T_{1}^{\mathrm{II}} \allowbreak + \allowbreak T_{1}^{\mathrm{II}}  ]  \allowbreak + \allowbreak  [  (-T_{1}^{\mathrm{I}})  ]  \allowbreak  V_{\mathrm{N}}^{\mathrm{I,II}} \allowbreak  [  1 \allowbreak + \allowbreak T_{11}^{\mathrm{I,II}} \allowbreak + \allowbreak T_{12}^{\mathrm{I,II}} \allowbreak + \allowbreak T_{1}^{\mathrm{I}} \allowbreak + \allowbreak T_{1}^{\mathrm{I}}T_{11}^{\mathrm{I,II}} \allowbreak + \allowbreak T_{1}^{\mathrm{I}}T_{12}^{\mathrm{I,II}} \allowbreak + \allowbreak T_{1}^{\mathrm{I}}T_{1}^{\mathrm{II}} \allowbreak + \allowbreak T_{1}^{\mathrm{I}}T_{1}^{\mathrm{II}}T_{11}^{\mathrm{I,II}} \allowbreak + \allowbreak T_{1}^{\mathrm{I}}\frac{1}{2!}T_{1}^{\mathrm{II}^2} \allowbreak + \allowbreak T_{1}^{\mathrm{I}}T_{2}^{\mathrm{II}} \allowbreak + \allowbreak \frac{1}{2!}T_{1}^{\mathrm{I}^2} \allowbreak + \allowbreak \frac{1}{2!}T_{1}^{\mathrm{I}^2}T_{1}^{\mathrm{II}} \allowbreak + \allowbreak \frac{1}{2!}T_{1}^{\mathrm{I}^2}\frac{1}{2!}T_{1}^{\mathrm{II}^2} \allowbreak + \allowbreak \frac{1}{2!}T_{1}^{\mathrm{I}^2}T_{2}^{\mathrm{II}} \allowbreak + \allowbreak T_{1}^{\mathrm{II}} \allowbreak + \allowbreak T_{1}^{\mathrm{II}}T_{11}^{\mathrm{I,II}} \allowbreak + \allowbreak T_{1}^{\mathrm{II}}T_{21}^{\mathrm{I,II}} \allowbreak + \allowbreak \frac{1}{2!}T_{1}^{\mathrm{II}^2} \allowbreak + \allowbreak T_{21}^{\mathrm{I,II}} \allowbreak + \allowbreak T_{22}^{\mathrm{I,II}} \allowbreak + \allowbreak T_{2}^{\mathrm{I}} \allowbreak + \allowbreak T_{2}^{\mathrm{I}}T_{1}^{\mathrm{II}} \allowbreak + \allowbreak T_{2}^{\mathrm{I}}\frac{1}{2!}T_{1}^{\mathrm{II}^2} \allowbreak + \allowbreak T_{2}^{\mathrm{I}}T_{2}^{\mathrm{II}} \allowbreak + \allowbreak T_{2}^{\mathrm{II}}  ]  \allowbreak + \allowbreak  [  (-T_{11}^{\mathrm{I,II}})(-T_{1}^{\mathrm{I}})  ]  \allowbreak  V_{\mathrm{N}}^{\mathrm{I,II}} \allowbreak  [  1 \allowbreak + \allowbreak T_{11}^{\mathrm{I,II}} \allowbreak + \allowbreak T_{1}^{\mathrm{I}} \allowbreak + \allowbreak T_{1}^{\mathrm{I}}T_{1}^{\mathrm{II}} \allowbreak + \allowbreak T_{1}^{\mathrm{II}}  ]  \allowbreak + \allowbreak  [  (-T_{1}^{\mathrm{I}})(-T_{1}^{\mathrm{II}})  ]  \allowbreak  V_{\mathrm{N}}^{\mathrm{I,II}} \allowbreak  [  1 \allowbreak + \allowbreak T_{11}^{\mathrm{I,II}} \allowbreak + \allowbreak T_{1}^{\mathrm{I}} \allowbreak + \allowbreak T_{1}^{\mathrm{I}}T_{11}^{\mathrm{I,II}} \allowbreak + \allowbreak T_{1}^{\mathrm{I}}T_{1}^{\mathrm{II}} \allowbreak + \allowbreak \frac{1}{2!}T_{1}^{\mathrm{I}^2} \allowbreak + \allowbreak \frac{1}{2!}T_{1}^{\mathrm{I}^2}T_{1}^{\mathrm{II}} \allowbreak + \allowbreak T_{1}^{\mathrm{II}} \allowbreak + \allowbreak T_{21}^{\mathrm{I,II}} \allowbreak + \allowbreak T_{2}^{\mathrm{I}} \allowbreak + \allowbreak T_{2}^{\mathrm{I}}T_{1}^{\mathrm{II}}  ]  \allowbreak + \allowbreak  [  \frac{1}{2!}T_{1}^{\mathrm{I}^2}  ]  \allowbreak  V_{\mathrm{N}}^{\mathrm{I,II}} \allowbreak  [  1 \allowbreak + \allowbreak T_{11}^{\mathrm{I,II}} \allowbreak + \allowbreak T_{12}^{\mathrm{I,II}} \allowbreak + \allowbreak T_{1}^{\mathrm{I}} \allowbreak + \allowbreak T_{1}^{\mathrm{I}}T_{1}^{\mathrm{II}} \allowbreak + \allowbreak T_{1}^{\mathrm{I}}\frac{1}{2!}T_{1}^{\mathrm{II}^2} \allowbreak + \allowbreak T_{1}^{\mathrm{I}}T_{2}^{\mathrm{II}} \allowbreak + \allowbreak T_{1}^{\mathrm{II}} \allowbreak + \allowbreak T_{1}^{\mathrm{II}}T_{11}^{\mathrm{I,II}} \allowbreak + \allowbreak \frac{1}{2!}T_{1}^{\mathrm{II}^2} \allowbreak + \allowbreak T_{2}^{\mathrm{II}}  ]  \allowbreak + \allowbreak  [  (-T_{1}^{\mathrm{II}})\frac{1}{2!}T_{1}^{\mathrm{I}^2}  ]  \allowbreak  V_{\mathrm{N}}^{\mathrm{I,II}} \allowbreak  [  1 \allowbreak + \allowbreak T_{11}^{\mathrm{I,II}} \allowbreak + \allowbreak T_{1}^{\mathrm{I}} \allowbreak + \allowbreak T_{1}^{\mathrm{I}}T_{1}^{\mathrm{II}} \allowbreak + \allowbreak T_{1}^{\mathrm{II}}  ]  \allowbreak + \allowbreak    \allowbreak \tilde{H}_{\mathrm{N}}^{\mathrm{II}}  \allowbreak [  T_{11}^{\mathrm{I,II}}T_{12}^{\mathrm{I,II}} \allowbreak + \allowbreak T_{1}^{\mathrm{I}}T_{11}^{\mathrm{I,II}} \allowbreak + \allowbreak T_{1}^{\mathrm{I}}T_{12}^{\mathrm{I,II}} \allowbreak + \allowbreak T_{1}^{\mathrm{I}}T_{1}^{\mathrm{II}}T_{11}^{\mathrm{I,II}} \allowbreak + \allowbreak T_{1}^{\mathrm{I}}T_{1}^{\mathrm{II}}T_{12}^{\mathrm{I,II}} \allowbreak + \allowbreak T_{1}^{\mathrm{I}}\frac{1}{2!}T_{1}^{\mathrm{II}^2}T_{11}^{\mathrm{I,II}} \allowbreak + \allowbreak T_{1}^{\mathrm{I}}T_{2}^{\mathrm{II}}T_{11}^{\mathrm{I,II}} \allowbreak + \allowbreak \frac{1}{2!}T_{1}^{\mathrm{I}^2} \allowbreak + \allowbreak \frac{1}{2!}T_{1}^{\mathrm{I}^2}T_{1}^{\mathrm{II}} \allowbreak + \allowbreak \frac{1}{2!}T_{1}^{\mathrm{I}^2}T_{1}^{\mathrm{II}}T_{2}^{\mathrm{II}} \allowbreak + \allowbreak \frac{1}{2!}T_{1}^{\mathrm{I}^2}\frac{1}{2!}T_{1}^{\mathrm{II}^2} \allowbreak + \allowbreak \frac{1}{2!}T_{1}^{\mathrm{I}^2}\frac{1}{3!}T_{1}^{\mathrm{II}^3} \allowbreak + \allowbreak \frac{1}{2!}T_{1}^{\mathrm{I}^2}T_{2}^{\mathrm{II}} \allowbreak + \allowbreak T_{1}^{\mathrm{II}}T_{21}^{\mathrm{I,II}} \allowbreak + \allowbreak T_{1}^{\mathrm{II}}T_{22}^{\mathrm{I,II}} \allowbreak + \allowbreak \frac{1}{2!}T_{1}^{\mathrm{II}^2}T_{21}^{\mathrm{I,II}} \allowbreak + \allowbreak T_{21}^{\mathrm{I,II}} \allowbreak + \allowbreak T_{22}^{\mathrm{I,II}} \allowbreak + \allowbreak T_{2}^{\mathrm{I}} \allowbreak + \allowbreak T_{2}^{\mathrm{I}}T_{1}^{\mathrm{II}} \allowbreak + \allowbreak T_{2}^{\mathrm{I}}T_{1}^{\mathrm{II}}T_{2}^{\mathrm{II}} \allowbreak + \allowbreak T_{2}^{\mathrm{I}}\frac{1}{2!}T_{1}^{\mathrm{II}^2} \allowbreak + \allowbreak T_{2}^{\mathrm{I}}\frac{1}{3!}T_{1}^{\mathrm{II}^3} \allowbreak + \allowbreak T_{2}^{\mathrm{I}}T_{2}^{\mathrm{II}} \allowbreak + \allowbreak T_{2}^{\mathrm{II}}T_{21}^{\mathrm{I,II}}  ]  \allowbreak + \allowbreak  [  (-T_{21}^{\mathrm{I,II}})  ]  \allowbreak \tilde{H}_{\mathrm{N}}^{\mathrm{II}}  \allowbreak [  1 \allowbreak + \allowbreak T_{1}^{\mathrm{II}} \allowbreak + \allowbreak \frac{1}{2!}T_{1}^{\mathrm{II}^2} \allowbreak + \allowbreak T_{2}^{\mathrm{II}}  ]  \allowbreak + \allowbreak  [  (-T_{11}^{\mathrm{I,II}})  ]  \allowbreak \tilde{H}_{\mathrm{N}}^{\mathrm{II}}  \allowbreak [  T_{11}^{\mathrm{I,II}} \allowbreak + \allowbreak T_{12}^{\mathrm{I,II}} \allowbreak + \allowbreak T_{1}^{\mathrm{I}} \allowbreak + \allowbreak T_{1}^{\mathrm{I}}T_{1}^{\mathrm{II}} \allowbreak + \allowbreak T_{1}^{\mathrm{I}}\frac{1}{2!}T_{1}^{\mathrm{II}^2} \allowbreak + \allowbreak T_{1}^{\mathrm{I}}T_{2}^{\mathrm{II}} \allowbreak + \allowbreak T_{1}^{\mathrm{II}}T_{11}^{\mathrm{I,II}}  ]  \allowbreak + \allowbreak  [  (-T_{1}^{\mathrm{II}})  ]  \allowbreak \tilde{H}_{\mathrm{N}}^{\mathrm{II}}  \allowbreak [  T_{1}^{\mathrm{I}}T_{11}^{\mathrm{I,II}} \allowbreak + \allowbreak T_{1}^{\mathrm{I}}T_{12}^{\mathrm{I,II}} \allowbreak + \allowbreak T_{1}^{\mathrm{I}}T_{1}^{\mathrm{II}}T_{11}^{\mathrm{I,II}} \allowbreak + \allowbreak \frac{1}{2!}T_{1}^{\mathrm{I}^2} \allowbreak + \allowbreak \frac{1}{2!}T_{1}^{\mathrm{I}^2}T_{1}^{\mathrm{II}} \allowbreak + \allowbreak \frac{1}{2!}T_{1}^{\mathrm{I}^2}\frac{1}{2!}T_{1}^{\mathrm{II}^2} \allowbreak + \allowbreak \frac{1}{2!}T_{1}^{\mathrm{I}^2}T_{2}^{\mathrm{II}} \allowbreak + \allowbreak T_{1}^{\mathrm{II}}T_{21}^{\mathrm{I,II}} \allowbreak + \allowbreak T_{21}^{\mathrm{I,II}} \allowbreak + \allowbreak T_{22}^{\mathrm{I,II}} \allowbreak + \allowbreak T_{2}^{\mathrm{I}} \allowbreak + \allowbreak T_{2}^{\mathrm{I}}T_{1}^{\mathrm{II}} \allowbreak + \allowbreak T_{2}^{\mathrm{I}}\frac{1}{2!}T_{1}^{\mathrm{II}^2} \allowbreak + \allowbreak T_{2}^{\mathrm{I}}T_{2}^{\mathrm{II}}  ]  \allowbreak + \allowbreak  [  (-T_{2}^{\mathrm{I}})  ]  \allowbreak \tilde{H}_{\mathrm{N}}^{\mathrm{II}}  \allowbreak [  1 \allowbreak + \allowbreak T_{1}^{\mathrm{II}} \allowbreak + \allowbreak T_{1}^{\mathrm{II}}T_{2}^{\mathrm{II}} \allowbreak + \allowbreak \frac{1}{2!}T_{1}^{\mathrm{II}^2} \allowbreak + \allowbreak \frac{1}{3!}T_{1}^{\mathrm{II}^3} \allowbreak + \allowbreak T_{2}^{\mathrm{II}}  ]  \allowbreak + \allowbreak  [  (-T_{1}^{\mathrm{II}})(-T_{2}^{\mathrm{I}})  ]  \allowbreak \tilde{H}_{\mathrm{N}}^{\mathrm{II}}  \allowbreak [  1 \allowbreak + \allowbreak T_{1}^{\mathrm{II}} \allowbreak + \allowbreak \frac{1}{2!}T_{1}^{\mathrm{II}^2} \allowbreak + \allowbreak T_{2}^{\mathrm{II}}  ]  \allowbreak + \allowbreak  [  (-T_{1}^{\mathrm{I}})  ]  \allowbreak \tilde{H}_{\mathrm{N}}^{\mathrm{II}}  \allowbreak [  T_{11}^{\mathrm{I,II}} \allowbreak + \allowbreak T_{12}^{\mathrm{I,II}} \allowbreak + \allowbreak T_{1}^{\mathrm{I}} \allowbreak + \allowbreak T_{1}^{\mathrm{I}}T_{1}^{\mathrm{II}} \allowbreak + \allowbreak T_{1}^{\mathrm{I}}T_{1}^{\mathrm{II}}T_{2}^{\mathrm{II}} \allowbreak + \allowbreak T_{1}^{\mathrm{I}}\frac{1}{2!}T_{1}^{\mathrm{II}^2} \allowbreak + \allowbreak T_{1}^{\mathrm{I}}\frac{1}{3!}T_{1}^{\mathrm{II}^3} \allowbreak + \allowbreak T_{1}^{\mathrm{I}}T_{2}^{\mathrm{II}} \allowbreak + \allowbreak T_{1}^{\mathrm{II}}T_{11}^{\mathrm{I,II}} \allowbreak + \allowbreak T_{1}^{\mathrm{II}}T_{12}^{\mathrm{I,II}} \allowbreak + \allowbreak \frac{1}{2!}T_{1}^{\mathrm{II}^2}T_{11}^{\mathrm{I,II}} \allowbreak + \allowbreak T_{2}^{\mathrm{II}}T_{11}^{\mathrm{I,II}}  ]  \allowbreak + \allowbreak  [  (-T_{11}^{\mathrm{I,II}})(-T_{1}^{\mathrm{I}})  ]  \allowbreak \tilde{H}_{\mathrm{N}}^{\mathrm{II}}  \allowbreak [  1 \allowbreak + \allowbreak T_{1}^{\mathrm{II}} \allowbreak + \allowbreak \frac{1}{2!}T_{1}^{\mathrm{II}^2} \allowbreak + \allowbreak T_{2}^{\mathrm{II}}  ]  \allowbreak + \allowbreak  [  (-T_{1}^{\mathrm{I}})(-T_{1}^{\mathrm{II}})  ]  \allowbreak \tilde{H}_{\mathrm{N}}^{\mathrm{II}}  \allowbreak [  T_{11}^{\mathrm{I,II}} \allowbreak + \allowbreak T_{12}^{\mathrm{I,II}} \allowbreak + \allowbreak T_{1}^{\mathrm{I}} \allowbreak + \allowbreak T_{1}^{\mathrm{I}}T_{1}^{\mathrm{II}} \allowbreak + \allowbreak T_{1}^{\mathrm{I}}\frac{1}{2!}T_{1}^{\mathrm{II}^2} \allowbreak + \allowbreak T_{1}^{\mathrm{I}}T_{2}^{\mathrm{II}} \allowbreak + \allowbreak T_{1}^{\mathrm{II}}T_{11}^{\mathrm{I,II}}  ]  \allowbreak + \allowbreak  [  \frac{1}{2!}T_{1}^{\mathrm{I}^2}  ]  \allowbreak \tilde{H}_{\mathrm{N}}^{\mathrm{II}}  \allowbreak [  1 \allowbreak + \allowbreak T_{1}^{\mathrm{II}} \allowbreak + \allowbreak T_{1}^{\mathrm{II}}T_{2}^{\mathrm{II}} \allowbreak + \allowbreak \frac{1}{2!}T_{1}^{\mathrm{II}^2} \allowbreak + \allowbreak \frac{1}{3!}T_{1}^{\mathrm{II}^3} \allowbreak + \allowbreak T_{2}^{\mathrm{II}}  ]  \allowbreak + \allowbreak  [  (-T_{1}^{\mathrm{II}})\frac{1}{2!}T_{1}^{\mathrm{I}^2}  ]  \allowbreak \tilde{H}_{\mathrm{N}}^{\mathrm{II}}  \allowbreak [  1 \allowbreak + \allowbreak T_{1}^{\mathrm{II}} \allowbreak + \allowbreak \frac{1}{2!}T_{1}^{\mathrm{II}^2} \allowbreak + \allowbreak T_{2}^{\mathrm{II}}  ]  \allowbreak + \allowbreak    \tilde{H}_{\mathrm{N}}^{\mathrm{I}}  [  T_{11}^{\mathrm{I,II}} \allowbreak + \allowbreak T_{1}^{\mathrm{I}}T_{11}^{\mathrm{I,II}} \allowbreak + \allowbreak T_{1}^{\mathrm{I}}T_{1}^{\mathrm{II}} \allowbreak + \allowbreak T_{1}^{\mathrm{I}}T_{21}^{\mathrm{I,II}} \allowbreak + \allowbreak T_{1}^{\mathrm{I}}T_{2}^{\mathrm{I}}T_{11}^{\mathrm{I,II}} \allowbreak + \allowbreak T_{1}^{\mathrm{I}}T_{2}^{\mathrm{I}}T_{1}^{\mathrm{II}} \allowbreak + \allowbreak \frac{1}{2!}T_{1}^{\mathrm{I}^2}T_{11}^{\mathrm{I,II}} \allowbreak + \allowbreak \frac{1}{2!}T_{1}^{\mathrm{I}^2}T_{1}^{\mathrm{II}} \allowbreak + \allowbreak \frac{1}{2!}T_{1}^{\mathrm{I}^2}T_{21}^{\mathrm{I,II}} \allowbreak + \allowbreak \frac{1}{2!}T_{1}^{\mathrm{I}^2}T_{2}^{\mathrm{I}}T_{1}^{\mathrm{II}} \allowbreak + \allowbreak \frac{1}{3!}T_{1}^{\mathrm{I}^3}T_{11}^{\mathrm{I,II}} \allowbreak + \allowbreak \frac{1}{3!}T_{1}^{\mathrm{I}^3}T_{1}^{\mathrm{II}} \allowbreak + \allowbreak \frac{1}{4!}T_{1}^{\mathrm{I}^4}T_{1}^{\mathrm{II}} \allowbreak + \allowbreak T_{1}^{\mathrm{II}} \allowbreak + \allowbreak T_{21}^{\mathrm{I,II}} \allowbreak + \allowbreak T_{2}^{\mathrm{I}}T_{11}^{\mathrm{I,II}} \allowbreak + \allowbreak T_{2}^{\mathrm{I}}T_{1}^{\mathrm{II}} \allowbreak + \allowbreak T_{2}^{\mathrm{I}}T_{21}^{\mathrm{I,II}} \allowbreak + \allowbreak \frac{1}{2!}T_{2}^{\mathrm{I}^2}T_{1}^{\mathrm{II}}  ]  \allowbreak + \allowbreak  [  (-T_{21}^{\mathrm{I,II}})  ]  \tilde{H}_{\mathrm{N}}^{\mathrm{I}}  [  1 \allowbreak + \allowbreak T_{1}^{\mathrm{I}} \allowbreak + \allowbreak \frac{1}{2!}T_{1}^{\mathrm{I}^2} \allowbreak + \allowbreak T_{2}^{\mathrm{I}}  ]  \allowbreak + \allowbreak  [  (-T_{11}^{\mathrm{I,II}})  ]  \tilde{H}_{\mathrm{N}}^{\mathrm{I}}  [  1 \allowbreak + \allowbreak T_{1}^{\mathrm{I}} \allowbreak + \allowbreak T_{1}^{\mathrm{I}}T_{2}^{\mathrm{I}} \allowbreak + \allowbreak \frac{1}{2!}T_{1}^{\mathrm{I}^2} \allowbreak + \allowbreak \frac{1}{3!}T_{1}^{\mathrm{I}^3} \allowbreak + \allowbreak T_{2}^{\mathrm{I}}  ]  \allowbreak + \allowbreak  [  (-T_{1}^{\mathrm{II}})  ]  \tilde{H}_{\mathrm{N}}^{\mathrm{I}}  [  1 \allowbreak + \allowbreak T_{1}^{\mathrm{I}} \allowbreak + \allowbreak T_{1}^{\mathrm{I}}T_{2}^{\mathrm{I}} \allowbreak + \allowbreak \frac{1}{2!}T_{1}^{\mathrm{I}^2} \allowbreak + \allowbreak \frac{1}{2!}T_{1}^{\mathrm{I}^2}T_{2}^{\mathrm{I}} \allowbreak + \allowbreak \frac{1}{3!}T_{1}^{\mathrm{I}^3} \allowbreak + \allowbreak \frac{1}{4!}T_{1}^{\mathrm{I}^4} \allowbreak + \allowbreak T_{2}^{\mathrm{I}} \allowbreak + \allowbreak \frac{1}{2!}T_{2}^{\mathrm{I}^2}  ]  \allowbreak + \allowbreak  [  (-T_{2}^{\mathrm{I}})  ]  \tilde{H}_{\mathrm{N}}^{\mathrm{I}}  [  T_{11}^{\mathrm{I,II}} \allowbreak + \allowbreak T_{1}^{\mathrm{I}}T_{11}^{\mathrm{I,II}} \allowbreak + \allowbreak T_{1}^{\mathrm{I}}T_{1}^{\mathrm{II}} \allowbreak + \allowbreak \frac{1}{2!}T_{1}^{\mathrm{I}^2}T_{1}^{\mathrm{II}} \allowbreak + \allowbreak T_{1}^{\mathrm{II}} \allowbreak + \allowbreak T_{21}^{\mathrm{I,II}} \allowbreak + \allowbreak T_{2}^{\mathrm{I}}T_{1}^{\mathrm{II}}  ]  \allowbreak + \allowbreak  [  (-T_{1}^{\mathrm{II}})(-T_{2}^{\mathrm{I}})  ]  \tilde{H}_{\mathrm{N}}^{\mathrm{I}}  [  1 \allowbreak + \allowbreak T_{1}^{\mathrm{I}} \allowbreak + \allowbreak \frac{1}{2!}T_{1}^{\mathrm{I}^2} \allowbreak + \allowbreak T_{2}^{\mathrm{I}}  ]  \allowbreak + \allowbreak  [  (-T_{1}^{\mathrm{I}})  ]  \tilde{H}_{\mathrm{N}}^{\mathrm{I}}  [  T_{11}^{\mathrm{I,II}} \allowbreak + \allowbreak T_{1}^{\mathrm{I}}T_{11}^{\mathrm{I,II}} \allowbreak + \allowbreak T_{1}^{\mathrm{I}}T_{1}^{\mathrm{II}} \allowbreak + \allowbreak T_{1}^{\mathrm{I}}T_{21}^{\mathrm{I,II}} \allowbreak + \allowbreak T_{1}^{\mathrm{I}}T_{2}^{\mathrm{I}}T_{1}^{\mathrm{II}} \allowbreak + \allowbreak \frac{1}{2!}T_{1}^{\mathrm{I}^2}T_{11}^{\mathrm{I,II}} \allowbreak + \allowbreak \frac{1}{2!}T_{1}^{\mathrm{I}^2}T_{1}^{\mathrm{II}} \allowbreak + \allowbreak \frac{1}{3!}T_{1}^{\mathrm{I}^3}T_{1}^{\mathrm{II}} \allowbreak + \allowbreak T_{1}^{\mathrm{II}} \allowbreak + \allowbreak T_{21}^{\mathrm{I,II}} \allowbreak + \allowbreak T_{2}^{\mathrm{I}}T_{11}^{\mathrm{I,II}} \allowbreak + \allowbreak T_{2}^{\mathrm{I}}T_{1}^{\mathrm{II}}  ]  \allowbreak + \allowbreak  [  (-T_{11}^{\mathrm{I,II}})(-T_{1}^{\mathrm{I}})  ]  \tilde{H}_{\mathrm{N}}^{\mathrm{I}}  [  1 \allowbreak + \allowbreak T_{1}^{\mathrm{I}} \allowbreak + \allowbreak \frac{1}{2!}T_{1}^{\mathrm{I}^2} \allowbreak + \allowbreak T_{2}^{\mathrm{I}}  ]  \allowbreak + \allowbreak  [  (-T_{1}^{\mathrm{I}})(-T_{1}^{\mathrm{II}})  ]  \tilde{H}_{\mathrm{N}}^{\mathrm{I}}  [  1 \allowbreak + \allowbreak T_{1}^{\mathrm{I}} \allowbreak + \allowbreak T_{1}^{\mathrm{I}}T_{2}^{\mathrm{I}} \allowbreak + \allowbreak \frac{1}{2!}T_{1}^{\mathrm{I}^2} \allowbreak + \allowbreak \frac{1}{3!}T_{1}^{\mathrm{I}^3} \allowbreak + \allowbreak T_{2}^{\mathrm{I}}  ]  \allowbreak + \allowbreak  [  \frac{1}{2!}T_{1}^{\mathrm{I}^2}  ]  \tilde{H}_{\mathrm{N}}^{\mathrm{I}}  [  T_{11}^{\mathrm{I,II}} \allowbreak + \allowbreak T_{1}^{\mathrm{I}}T_{11}^{\mathrm{I,II}} \allowbreak + \allowbreak T_{1}^{\mathrm{I}}T_{1}^{\mathrm{II}} \allowbreak + \allowbreak \frac{1}{2!}T_{1}^{\mathrm{I}^2}T_{1}^{\mathrm{II}} \allowbreak + \allowbreak T_{1}^{\mathrm{II}} \allowbreak + \allowbreak T_{21}^{\mathrm{I,II}} \allowbreak + \allowbreak T_{2}^{\mathrm{I}}T_{1}^{\mathrm{II}}  ]  \allowbreak + \allowbreak  [  (-T_{1}^{\mathrm{II}})\frac{1}{2!}T_{1}^{\mathrm{I}^2}  ]  \tilde{H}_{\mathrm{N}}^{\mathrm{I}}  [  1 \allowbreak + \allowbreak T_{1}^{\mathrm{I}} \allowbreak + \allowbreak \frac{1}{2!}T_{1}^{\mathrm{I}^2} \allowbreak + \allowbreak T_{2}^{\mathrm{I}}  ] \vert 0^{\mathrm{I}}0^{\mathrm{II}} \rangle = 0$

%% file: DD_H_00.tex
\noindent
$\langle D^{\mathrm{I}}D^{\mathrm{II}} \vert   \allowbreak V_{\mathrm{N}}^{\mathrm{I,II}} \allowbreak   [  T_{11}^{\mathrm{I,II}} \allowbreak + \allowbreak T_{11}^{\mathrm{I,II}}T_{12}^{\mathrm{I,II}} \allowbreak + \allowbreak T_{11}^{\mathrm{I,II}}T_{21}^{\mathrm{I,II}} \allowbreak + \allowbreak T_{11}^{\mathrm{I,II}}T_{22}^{\mathrm{I,II}} \allowbreak + \allowbreak \frac{1}{2!}T_{11}^{\mathrm{I,II}^3} \allowbreak + \allowbreak T_{12}^{\mathrm{I,II}} \allowbreak + \allowbreak T_{12}^{\mathrm{I,II}}T_{21}^{\mathrm{I,II}} \allowbreak + \allowbreak T_{1}^{\mathrm{I}}T_{11}^{\mathrm{I,II}} \allowbreak + \allowbreak T_{1}^{\mathrm{I}}T_{11}^{\mathrm{I,II}}T_{12}^{\mathrm{I,II}} \allowbreak + \allowbreak T_{1}^{\mathrm{I}}T_{12}^{\mathrm{I,II}} \allowbreak + \allowbreak T_{1}^{\mathrm{I}}T_{1}^{\mathrm{II}} \allowbreak + \allowbreak T_{1}^{\mathrm{I}}T_{1}^{\mathrm{II}}T_{11}^{\mathrm{I,II}} \allowbreak + \allowbreak T_{1}^{\mathrm{I}}T_{1}^{\mathrm{II}}T_{12}^{\mathrm{I,II}} \allowbreak + \allowbreak T_{1}^{\mathrm{I}}T_{1}^{\mathrm{II}}T_{21}^{\mathrm{I,II}} \allowbreak + \allowbreak T_{1}^{\mathrm{I}}T_{1}^{\mathrm{II}}T_{22}^{\mathrm{I,II}} \allowbreak + \allowbreak T_{1}^{\mathrm{I}}T_{1}^{\mathrm{II}}T_{2}^{\mathrm{II}} \allowbreak + \allowbreak T_{1}^{\mathrm{I}}\frac{1}{2!}T_{1}^{\mathrm{II}^2} \allowbreak + \allowbreak T_{1}^{\mathrm{I}}\frac{1}{2!}T_{1}^{\mathrm{II}^2}T_{11}^{\mathrm{I,II}} \allowbreak + \allowbreak T_{1}^{\mathrm{I}}\frac{1}{2!}T_{1}^{\mathrm{II}^2}T_{21}^{\mathrm{I,II}} \allowbreak + \allowbreak T_{1}^{\mathrm{I}}\frac{1}{3!}T_{1}^{\mathrm{II}^3} \allowbreak + \allowbreak T_{1}^{\mathrm{I}}T_{21}^{\mathrm{I,II}} \allowbreak + \allowbreak T_{1}^{\mathrm{I}}T_{22}^{\mathrm{I,II}} \allowbreak + \allowbreak T_{1}^{\mathrm{I}}T_{2}^{\mathrm{I}}T_{1}^{\mathrm{II}} \allowbreak + \allowbreak T_{1}^{\mathrm{I}}T_{2}^{\mathrm{I}}T_{1}^{\mathrm{II}}T_{2}^{\mathrm{II}} \allowbreak + \allowbreak T_{1}^{\mathrm{I}}T_{2}^{\mathrm{I}}\frac{1}{2!}T_{1}^{\mathrm{II}^2} \allowbreak + \allowbreak T_{1}^{\mathrm{I}}T_{2}^{\mathrm{I}}\frac{1}{3!}T_{1}^{\mathrm{II}^3} \allowbreak + \allowbreak T_{1}^{\mathrm{I}}T_{2}^{\mathrm{I}}T_{2}^{\mathrm{II}} \allowbreak + \allowbreak T_{1}^{\mathrm{I}}T_{2}^{\mathrm{II}} \allowbreak + \allowbreak T_{1}^{\mathrm{I}}T_{2}^{\mathrm{II}}T_{11}^{\mathrm{I,II}} \allowbreak + \allowbreak T_{1}^{\mathrm{I}}T_{2}^{\mathrm{II}}T_{21}^{\mathrm{I,II}} \allowbreak + \allowbreak \frac{1}{2!}T_{1}^{\mathrm{I}^2}T_{11}^{\mathrm{I,II}} \allowbreak + \allowbreak \frac{1}{2!}T_{1}^{\mathrm{I}^2}T_{12}^{\mathrm{I,II}} \allowbreak + \allowbreak \frac{1}{2!}T_{1}^{\mathrm{I}^2}T_{1}^{\mathrm{II}} \allowbreak + \allowbreak \frac{1}{2!}T_{1}^{\mathrm{I}^2}T_{1}^{\mathrm{II}}T_{11}^{\mathrm{I,II}} \allowbreak + \allowbreak \frac{1}{2!}T_{1}^{\mathrm{I}^2}T_{1}^{\mathrm{II}}T_{12}^{\mathrm{I,II}} \allowbreak + \allowbreak \frac{1}{2!}T_{1}^{\mathrm{I}^2}T_{1}^{\mathrm{II}}T_{2}^{\mathrm{II}} \allowbreak + \allowbreak \frac{1}{2!}T_{1}^{\mathrm{I}^2}\frac{1}{2!}T_{1}^{\mathrm{II}^2} \allowbreak + \allowbreak \frac{1}{2!}T_{1}^{\mathrm{I}^2}\frac{1}{2!}T_{1}^{\mathrm{II}^2}T_{11}^{\mathrm{I,II}} \allowbreak + \allowbreak \frac{1}{2!}T_{1}^{\mathrm{I}^2}\frac{1}{3!}T_{1}^{\mathrm{II}^3} \allowbreak + \allowbreak \frac{1}{2!}T_{1}^{\mathrm{I}^2}T_{2}^{\mathrm{II}} \allowbreak + \allowbreak \frac{1}{2!}T_{1}^{\mathrm{I}^2}T_{2}^{\mathrm{II}}T_{11}^{\mathrm{I,II}} \allowbreak + \allowbreak \frac{1}{3!}T_{1}^{\mathrm{I}^3}T_{1}^{\mathrm{II}} \allowbreak + \allowbreak \frac{1}{3!}T_{1}^{\mathrm{I}^3}T_{1}^{\mathrm{II}}T_{2}^{\mathrm{II}} \allowbreak + \allowbreak \frac{1}{3!}T_{1}^{\mathrm{I}^3}\frac{1}{2!}T_{1}^{\mathrm{II}^2} \allowbreak + \allowbreak \frac{1}{3!}T_{1}^{\mathrm{I}^3}\frac{1}{3!}T_{1}^{\mathrm{II}^3} \allowbreak + \allowbreak \frac{1}{3!}T_{1}^{\mathrm{I}^3}T_{2}^{\mathrm{II}} \allowbreak + \allowbreak T_{1}^{\mathrm{II}}T_{11}^{\mathrm{I,II}} \allowbreak + \allowbreak T_{1}^{\mathrm{II}}T_{11}^{\mathrm{I,II}}T_{21}^{\mathrm{I,II}} \allowbreak + \allowbreak T_{1}^{\mathrm{II}}T_{12}^{\mathrm{I,II}} \allowbreak + \allowbreak T_{1}^{\mathrm{II}}T_{21}^{\mathrm{I,II}} \allowbreak + \allowbreak T_{1}^{\mathrm{II}}T_{22}^{\mathrm{I,II}} \allowbreak + \allowbreak \frac{1}{2!}T_{1}^{\mathrm{II}^2}T_{11}^{\mathrm{I,II}} \allowbreak + \allowbreak \frac{1}{2!}T_{1}^{\mathrm{II}^2}T_{21}^{\mathrm{I,II}} \allowbreak + \allowbreak T_{21}^{\mathrm{I,II}} \allowbreak + \allowbreak T_{22}^{\mathrm{I,II}} \allowbreak + \allowbreak T_{2}^{\mathrm{I}}T_{11}^{\mathrm{I,II}} \allowbreak + \allowbreak T_{2}^{\mathrm{I}}T_{12}^{\mathrm{I,II}} \allowbreak + \allowbreak T_{2}^{\mathrm{I}}T_{1}^{\mathrm{II}} \allowbreak + \allowbreak T_{2}^{\mathrm{I}}T_{1}^{\mathrm{II}}T_{11}^{\mathrm{I,II}} \allowbreak + \allowbreak T_{2}^{\mathrm{I}}T_{1}^{\mathrm{II}}T_{12}^{\mathrm{I,II}} \allowbreak + \allowbreak T_{2}^{\mathrm{I}}T_{1}^{\mathrm{II}}T_{2}^{\mathrm{II}} \allowbreak + \allowbreak T_{2}^{\mathrm{I}}\frac{1}{2!}T_{1}^{\mathrm{II}^2} \allowbreak + \allowbreak T_{2}^{\mathrm{I}}\frac{1}{2!}T_{1}^{\mathrm{II}^2}T_{11}^{\mathrm{I,II}} \allowbreak + \allowbreak T_{2}^{\mathrm{I}}\frac{1}{3!}T_{1}^{\mathrm{II}^3} \allowbreak + \allowbreak T_{2}^{\mathrm{I}}T_{2}^{\mathrm{II}} \allowbreak + \allowbreak T_{2}^{\mathrm{I}}T_{2}^{\mathrm{II}}T_{11}^{\mathrm{I,II}} \allowbreak + \allowbreak T_{2}^{\mathrm{II}}T_{11}^{\mathrm{I,II}} \allowbreak + \allowbreak T_{2}^{\mathrm{II}}T_{21}^{\mathrm{I,II}}  ]  \allowbreak + \allowbreak  [  (-T_{22}^{\mathrm{I,II}})  ]  \allowbreak V_{\mathrm{N}}^{\mathrm{I,II}} \allowbreak   [  1 \allowbreak + \allowbreak T_{11}^{\mathrm{I,II}} \allowbreak + \allowbreak T_{1}^{\mathrm{I}} \allowbreak + \allowbreak T_{1}^{\mathrm{I}}T_{1}^{\mathrm{II}} \allowbreak + \allowbreak T_{1}^{\mathrm{II}}  ]  \allowbreak + \allowbreak  [  (-T_{21}^{\mathrm{I,II}})  ]  \allowbreak V_{\mathrm{N}}^{\mathrm{I,II}} \allowbreak   [  1 \allowbreak + \allowbreak T_{11}^{\mathrm{I,II}} \allowbreak + \allowbreak T_{12}^{\mathrm{I,II}} \allowbreak + \allowbreak T_{1}^{\mathrm{I}} \allowbreak + \allowbreak T_{1}^{\mathrm{I}}T_{1}^{\mathrm{II}} \allowbreak + \allowbreak T_{1}^{\mathrm{I}}\frac{1}{2!}T_{1}^{\mathrm{II}^2} \allowbreak + \allowbreak T_{1}^{\mathrm{I}}T_{2}^{\mathrm{II}} \allowbreak + \allowbreak T_{1}^{\mathrm{II}} \allowbreak + \allowbreak T_{1}^{\mathrm{II}}T_{11}^{\mathrm{I,II}} \allowbreak + \allowbreak \frac{1}{2!}T_{1}^{\mathrm{II}^2} \allowbreak + \allowbreak T_{2}^{\mathrm{II}}  ]  \allowbreak + \allowbreak  [  (-T_{12}^{\mathrm{I,II}})  ]  \allowbreak V_{\mathrm{N}}^{\mathrm{I,II}} \allowbreak   [  1 \allowbreak + \allowbreak T_{11}^{\mathrm{I,II}} \allowbreak + \allowbreak T_{1}^{\mathrm{I}} \allowbreak + \allowbreak T_{1}^{\mathrm{I}}T_{11}^{\mathrm{I,II}} \allowbreak + \allowbreak T_{1}^{\mathrm{I}}T_{1}^{\mathrm{II}} \allowbreak + \allowbreak \frac{1}{2!}T_{1}^{\mathrm{I}^2} \allowbreak + \allowbreak \frac{1}{2!}T_{1}^{\mathrm{I}^2}T_{1}^{\mathrm{II}} \allowbreak + \allowbreak T_{1}^{\mathrm{II}} \allowbreak + \allowbreak T_{21}^{\mathrm{I,II}} \allowbreak + \allowbreak T_{2}^{\mathrm{I}} \allowbreak + \allowbreak T_{2}^{\mathrm{I}}T_{1}^{\mathrm{II}}  ]  \allowbreak + \allowbreak  [  \frac{1}{2!}T_{11}^{\mathrm{I,II}^2}  ]  \allowbreak V_{\mathrm{N}}^{\mathrm{I,II}} \allowbreak   [  1 \allowbreak + \allowbreak T_{11}^{\mathrm{I,II}} \allowbreak + \allowbreak T_{1}^{\mathrm{I}} \allowbreak + \allowbreak T_{1}^{\mathrm{I}}T_{1}^{\mathrm{II}} \allowbreak + \allowbreak T_{1}^{\mathrm{II}}  ]  \allowbreak + \allowbreak  [  (-T_{11}^{\mathrm{I,II}})  ]  \allowbreak V_{\mathrm{N}}^{\mathrm{I,II}} \allowbreak   [  1 \allowbreak + \allowbreak T_{11}^{\mathrm{I,II}} \allowbreak + \allowbreak T_{12}^{\mathrm{I,II}} \allowbreak + \allowbreak T_{1}^{\mathrm{I}} \allowbreak + \allowbreak T_{1}^{\mathrm{I}}T_{11}^{\mathrm{I,II}} \allowbreak + \allowbreak T_{1}^{\mathrm{I}}T_{12}^{\mathrm{I,II}} \allowbreak + \allowbreak T_{1}^{\mathrm{I}}T_{1}^{\mathrm{II}} \allowbreak + \allowbreak T_{1}^{\mathrm{I}}T_{1}^{\mathrm{II}}T_{11}^{\mathrm{I,II}} \allowbreak + \allowbreak T_{1}^{\mathrm{I}}\frac{1}{2!}T_{1}^{\mathrm{II}^2} \allowbreak + \allowbreak T_{1}^{\mathrm{I}}T_{2}^{\mathrm{II}} \allowbreak + \allowbreak \frac{1}{2!}T_{1}^{\mathrm{I}^2} \allowbreak + \allowbreak \frac{1}{2!}T_{1}^{\mathrm{I}^2}T_{1}^{\mathrm{II}} \allowbreak + \allowbreak \frac{1}{2!}T_{1}^{\mathrm{I}^2}\frac{1}{2!}T_{1}^{\mathrm{II}^2} \allowbreak + \allowbreak \frac{1}{2!}T_{1}^{\mathrm{I}^2}T_{2}^{\mathrm{II}} \allowbreak + \allowbreak T_{1}^{\mathrm{II}} \allowbreak + \allowbreak T_{1}^{\mathrm{II}}T_{11}^{\mathrm{I,II}} \allowbreak + \allowbreak T_{1}^{\mathrm{II}}T_{21}^{\mathrm{I,II}} \allowbreak + \allowbreak \frac{1}{2!}T_{1}^{\mathrm{II}^2} \allowbreak + \allowbreak T_{21}^{\mathrm{I,II}} \allowbreak + \allowbreak T_{22}^{\mathrm{I,II}} \allowbreak + \allowbreak T_{2}^{\mathrm{I}} \allowbreak + \allowbreak T_{2}^{\mathrm{I}}T_{1}^{\mathrm{II}} \allowbreak + \allowbreak T_{2}^{\mathrm{I}}\frac{1}{2!}T_{1}^{\mathrm{II}^2} \allowbreak + \allowbreak T_{2}^{\mathrm{I}}T_{2}^{\mathrm{II}} \allowbreak + \allowbreak T_{2}^{\mathrm{II}}  ]  \allowbreak + \allowbreak  [  (-T_{2}^{\mathrm{II}})  ]  \allowbreak V_{\mathrm{N}}^{\mathrm{I,II}} \allowbreak   [  T_{11}^{\mathrm{I,II}} \allowbreak + \allowbreak T_{1}^{\mathrm{I}} \allowbreak + \allowbreak T_{1}^{\mathrm{I}}T_{11}^{\mathrm{I,II}} \allowbreak + \allowbreak T_{1}^{\mathrm{I}}T_{1}^{\mathrm{II}} \allowbreak + \allowbreak T_{1}^{\mathrm{I}}T_{21}^{\mathrm{I,II}} \allowbreak + \allowbreak T_{1}^{\mathrm{I}}T_{2}^{\mathrm{I}} \allowbreak + \allowbreak T_{1}^{\mathrm{I}}T_{2}^{\mathrm{I}}T_{1}^{\mathrm{II}} \allowbreak + \allowbreak \frac{1}{2!}T_{1}^{\mathrm{I}^2} \allowbreak + \allowbreak \frac{1}{2!}T_{1}^{\mathrm{I}^2}T_{11}^{\mathrm{I,II}} \allowbreak + \allowbreak \frac{1}{2!}T_{1}^{\mathrm{I}^2}T_{1}^{\mathrm{II}} \allowbreak + \allowbreak \frac{1}{3!}T_{1}^{\mathrm{I}^3} \allowbreak + \allowbreak \frac{1}{3!}T_{1}^{\mathrm{I}^3}T_{1}^{\mathrm{II}} \allowbreak + \allowbreak T_{21}^{\mathrm{I,II}} \allowbreak + \allowbreak T_{2}^{\mathrm{I}} \allowbreak + \allowbreak T_{2}^{\mathrm{I}}T_{11}^{\mathrm{I,II}} \allowbreak + \allowbreak T_{2}^{\mathrm{I}}T_{1}^{\mathrm{II}}  ]  \allowbreak + \allowbreak  [  (-T_{1}^{\mathrm{II}})  ]  \allowbreak V_{\mathrm{N}}^{\mathrm{I,II}} \allowbreak   [  T_{11}^{\mathrm{I,II}} \allowbreak + \allowbreak T_{11}^{\mathrm{I,II}}T_{21}^{\mathrm{I,II}} \allowbreak + \allowbreak T_{12}^{\mathrm{I,II}} \allowbreak + \allowbreak T_{1}^{\mathrm{I}} \allowbreak + \allowbreak T_{1}^{\mathrm{I}}T_{11}^{\mathrm{I,II}} \allowbreak + \allowbreak T_{1}^{\mathrm{I}}T_{12}^{\mathrm{I,II}} \allowbreak + \allowbreak T_{1}^{\mathrm{I}}T_{1}^{\mathrm{II}} \allowbreak + \allowbreak T_{1}^{\mathrm{I}}T_{1}^{\mathrm{II}}T_{11}^{\mathrm{I,II}} \allowbreak + \allowbreak T_{1}^{\mathrm{I}}T_{1}^{\mathrm{II}}T_{21}^{\mathrm{I,II}} \allowbreak + \allowbreak T_{1}^{\mathrm{I}}\frac{1}{2!}T_{1}^{\mathrm{II}^2} \allowbreak + \allowbreak T_{1}^{\mathrm{I}}T_{21}^{\mathrm{I,II}} \allowbreak + \allowbreak T_{1}^{\mathrm{I}}T_{22}^{\mathrm{I,II}} \allowbreak + \allowbreak T_{1}^{\mathrm{I}}T_{2}^{\mathrm{I}} \allowbreak + \allowbreak T_{1}^{\mathrm{I}}T_{2}^{\mathrm{I}}T_{1}^{\mathrm{II}} \allowbreak + \allowbreak T_{1}^{\mathrm{I}}T_{2}^{\mathrm{I}}\frac{1}{2!}T_{1}^{\mathrm{II}^2} \allowbreak + \allowbreak T_{1}^{\mathrm{I}}T_{2}^{\mathrm{I}}T_{2}^{\mathrm{II}} \allowbreak + \allowbreak T_{1}^{\mathrm{I}}T_{2}^{\mathrm{II}} \allowbreak + \allowbreak \frac{1}{2!}T_{1}^{\mathrm{I}^2} \allowbreak + \allowbreak \frac{1}{2!}T_{1}^{\mathrm{I}^2}T_{11}^{\mathrm{I,II}} \allowbreak + \allowbreak \frac{1}{2!}T_{1}^{\mathrm{I}^2}T_{12}^{\mathrm{I,II}} \allowbreak + \allowbreak \frac{1}{2!}T_{1}^{\mathrm{I}^2}T_{1}^{\mathrm{II}} \allowbreak + \allowbreak \frac{1}{2!}T_{1}^{\mathrm{I}^2}T_{1}^{\mathrm{II}}T_{11}^{\mathrm{I,II}} \allowbreak + \allowbreak \frac{1}{2!}T_{1}^{\mathrm{I}^2}\frac{1}{2!}T_{1}^{\mathrm{II}^2} \allowbreak + \allowbreak \frac{1}{2!}T_{1}^{\mathrm{I}^2}T_{2}^{\mathrm{II}} \allowbreak + \allowbreak \frac{1}{3!}T_{1}^{\mathrm{I}^3} \allowbreak + \allowbreak \frac{1}{3!}T_{1}^{\mathrm{I}^3}T_{1}^{\mathrm{II}} \allowbreak + \allowbreak \frac{1}{3!}T_{1}^{\mathrm{I}^3}\frac{1}{2!}T_{1}^{\mathrm{II}^2} \allowbreak + \allowbreak \frac{1}{3!}T_{1}^{\mathrm{I}^3}T_{2}^{\mathrm{II}} \allowbreak + \allowbreak T_{1}^{\mathrm{II}}T_{11}^{\mathrm{I,II}} \allowbreak + \allowbreak T_{1}^{\mathrm{II}}T_{21}^{\mathrm{I,II}} \allowbreak + \allowbreak T_{21}^{\mathrm{I,II}} \allowbreak + \allowbreak T_{22}^{\mathrm{I,II}} \allowbreak + \allowbreak T_{2}^{\mathrm{I}} \allowbreak + \allowbreak T_{2}^{\mathrm{I}}T_{11}^{\mathrm{I,II}} \allowbreak + \allowbreak T_{2}^{\mathrm{I}}T_{12}^{\mathrm{I,II}} \allowbreak + \allowbreak T_{2}^{\mathrm{I}}T_{1}^{\mathrm{II}} \allowbreak + \allowbreak T_{2}^{\mathrm{I}}T_{1}^{\mathrm{II}}T_{11}^{\mathrm{I,II}} \allowbreak + \allowbreak T_{2}^{\mathrm{I}}\frac{1}{2!}T_{1}^{\mathrm{II}^2} \allowbreak + \allowbreak T_{2}^{\mathrm{I}}T_{2}^{\mathrm{II}}  ]  \allowbreak + \allowbreak  [  (-T_{1}^{\mathrm{II}})(-T_{21}^{\mathrm{I,II}})  ]  \allowbreak V_{\mathrm{N}}^{\mathrm{I,II}} \allowbreak   [  1 \allowbreak + \allowbreak T_{11}^{\mathrm{I,II}} \allowbreak + \allowbreak T_{1}^{\mathrm{I}} \allowbreak + \allowbreak T_{1}^{\mathrm{I}}T_{1}^{\mathrm{II}} \allowbreak + \allowbreak T_{1}^{\mathrm{II}}  ]  \allowbreak + \allowbreak  [  (-T_{11}^{\mathrm{I,II}})(-T_{1}^{\mathrm{II}})  ]  \allowbreak V_{\mathrm{N}}^{\mathrm{I,II}} \allowbreak   [  1 \allowbreak + \allowbreak T_{11}^{\mathrm{I,II}} \allowbreak + \allowbreak T_{1}^{\mathrm{I}} \allowbreak + \allowbreak T_{1}^{\mathrm{I}}T_{11}^{\mathrm{I,II}} \allowbreak + \allowbreak T_{1}^{\mathrm{I}}T_{1}^{\mathrm{II}} \allowbreak + \allowbreak \frac{1}{2!}T_{1}^{\mathrm{I}^2} \allowbreak + \allowbreak \frac{1}{2!}T_{1}^{\mathrm{I}^2}T_{1}^{\mathrm{II}} \allowbreak + \allowbreak T_{1}^{\mathrm{II}} \allowbreak + \allowbreak T_{21}^{\mathrm{I,II}} \allowbreak + \allowbreak T_{2}^{\mathrm{I}} \allowbreak + \allowbreak T_{2}^{\mathrm{I}}T_{1}^{\mathrm{II}}  ]  \allowbreak + \allowbreak  [  \frac{1}{2!}T_{1}^{\mathrm{II}^2}  ]  \allowbreak V_{\mathrm{N}}^{\mathrm{I,II}} \allowbreak   [  T_{11}^{\mathrm{I,II}} \allowbreak + \allowbreak T_{1}^{\mathrm{I}} \allowbreak + \allowbreak T_{1}^{\mathrm{I}}T_{11}^{\mathrm{I,II}} \allowbreak + \allowbreak T_{1}^{\mathrm{I}}T_{1}^{\mathrm{II}} \allowbreak + \allowbreak T_{1}^{\mathrm{I}}T_{21}^{\mathrm{I,II}} \allowbreak + \allowbreak T_{1}^{\mathrm{I}}T_{2}^{\mathrm{I}} \allowbreak + \allowbreak T_{1}^{\mathrm{I}}T_{2}^{\mathrm{I}}T_{1}^{\mathrm{II}} \allowbreak + \allowbreak \frac{1}{2!}T_{1}^{\mathrm{I}^2} \allowbreak + \allowbreak \frac{1}{2!}T_{1}^{\mathrm{I}^2}T_{11}^{\mathrm{I,II}} \allowbreak + \allowbreak \frac{1}{2!}T_{1}^{\mathrm{I}^2}T_{1}^{\mathrm{II}} \allowbreak + \allowbreak \frac{1}{3!}T_{1}^{\mathrm{I}^3} \allowbreak + \allowbreak \frac{1}{3!}T_{1}^{\mathrm{I}^3}T_{1}^{\mathrm{II}} \allowbreak + \allowbreak T_{21}^{\mathrm{I,II}} \allowbreak + \allowbreak T_{2}^{\mathrm{I}} \allowbreak + \allowbreak T_{2}^{\mathrm{I}}T_{11}^{\mathrm{I,II}} \allowbreak + \allowbreak T_{2}^{\mathrm{I}}T_{1}^{\mathrm{II}}  ]  \allowbreak + \allowbreak  [  (-T_{2}^{\mathrm{I}})  ]  \allowbreak V_{\mathrm{N}}^{\mathrm{I,II}} \allowbreak   [  T_{11}^{\mathrm{I,II}} \allowbreak + \allowbreak T_{12}^{\mathrm{I,II}} \allowbreak + \allowbreak T_{1}^{\mathrm{I}}T_{1}^{\mathrm{II}} \allowbreak + \allowbreak T_{1}^{\mathrm{I}}T_{1}^{\mathrm{II}}T_{2}^{\mathrm{II}} \allowbreak + \allowbreak T_{1}^{\mathrm{I}}\frac{1}{2!}T_{1}^{\mathrm{II}^2} \allowbreak + \allowbreak T_{1}^{\mathrm{I}}\frac{1}{3!}T_{1}^{\mathrm{II}^3} \allowbreak + \allowbreak T_{1}^{\mathrm{I}}T_{2}^{\mathrm{II}} \allowbreak + \allowbreak T_{1}^{\mathrm{II}} \allowbreak + \allowbreak T_{1}^{\mathrm{II}}T_{11}^{\mathrm{I,II}} \allowbreak + \allowbreak T_{1}^{\mathrm{II}}T_{12}^{\mathrm{I,II}} \allowbreak + \allowbreak T_{1}^{\mathrm{II}}T_{2}^{\mathrm{II}} \allowbreak + \allowbreak \frac{1}{2!}T_{1}^{\mathrm{II}^2} \allowbreak + \allowbreak \frac{1}{2!}T_{1}^{\mathrm{II}^2}T_{11}^{\mathrm{I,II}} \allowbreak + \allowbreak \frac{1}{3!}T_{1}^{\mathrm{II}^3} \allowbreak + \allowbreak T_{2}^{\mathrm{II}} \allowbreak + \allowbreak T_{2}^{\mathrm{II}}T_{11}^{\mathrm{I,II}}  ]  \allowbreak + \allowbreak  [  (-T_{2}^{\mathrm{I}})(-T_{2}^{\mathrm{II}})  ]  \allowbreak V_{\mathrm{N}}^{\mathrm{I,II}} \allowbreak   [  1 \allowbreak + \allowbreak T_{11}^{\mathrm{I,II}} \allowbreak + \allowbreak T_{1}^{\mathrm{I}} \allowbreak + \allowbreak T_{1}^{\mathrm{I}}T_{1}^{\mathrm{II}} \allowbreak + \allowbreak T_{1}^{\mathrm{II}}  ]  \allowbreak + \allowbreak  [  (-T_{1}^{\mathrm{II}})(-T_{2}^{\mathrm{I}})  ]  \allowbreak V_{\mathrm{N}}^{\mathrm{I,II}} \allowbreak   [  1 \allowbreak + \allowbreak T_{11}^{\mathrm{I,II}} \allowbreak + \allowbreak T_{12}^{\mathrm{I,II}} \allowbreak + \allowbreak T_{1}^{\mathrm{I}} \allowbreak + \allowbreak T_{1}^{\mathrm{I}}T_{1}^{\mathrm{II}} \allowbreak + \allowbreak T_{1}^{\mathrm{I}}\frac{1}{2!}T_{1}^{\mathrm{II}^2} \allowbreak + \allowbreak T_{1}^{\mathrm{I}}T_{2}^{\mathrm{II}} \allowbreak + \allowbreak T_{1}^{\mathrm{II}} \allowbreak + \allowbreak T_{1}^{\mathrm{II}}T_{11}^{\mathrm{I,II}} \allowbreak + \allowbreak \frac{1}{2!}T_{1}^{\mathrm{II}^2} \allowbreak + \allowbreak T_{2}^{\mathrm{II}}  ]  \allowbreak + \allowbreak  [  (-T_{2}^{\mathrm{I}})\frac{1}{2!}T_{1}^{\mathrm{II}^2}  ]  \allowbreak V_{\mathrm{N}}^{\mathrm{I,II}} \allowbreak   [  1 \allowbreak + \allowbreak T_{11}^{\mathrm{I,II}} \allowbreak + \allowbreak T_{1}^{\mathrm{I}} \allowbreak + \allowbreak T_{1}^{\mathrm{I}}T_{1}^{\mathrm{II}} \allowbreak + \allowbreak T_{1}^{\mathrm{II}}  ]  \allowbreak + \allowbreak  [  (-T_{1}^{\mathrm{I}})  ]  \allowbreak V_{\mathrm{N}}^{\mathrm{I,II}} \allowbreak   [  T_{11}^{\mathrm{I,II}} \allowbreak + \allowbreak T_{11}^{\mathrm{I,II}}T_{12}^{\mathrm{I,II}} \allowbreak + \allowbreak T_{12}^{\mathrm{I,II}} \allowbreak + \allowbreak T_{1}^{\mathrm{I}}T_{11}^{\mathrm{I,II}} \allowbreak + \allowbreak T_{1}^{\mathrm{I}}T_{12}^{\mathrm{I,II}} \allowbreak + \allowbreak T_{1}^{\mathrm{I}}T_{1}^{\mathrm{II}} \allowbreak + \allowbreak T_{1}^{\mathrm{I}}T_{1}^{\mathrm{II}}T_{11}^{\mathrm{I,II}} \allowbreak + \allowbreak T_{1}^{\mathrm{I}}T_{1}^{\mathrm{II}}T_{12}^{\mathrm{I,II}} \allowbreak + \allowbreak T_{1}^{\mathrm{I}}T_{1}^{\mathrm{II}}T_{2}^{\mathrm{II}} \allowbreak + \allowbreak T_{1}^{\mathrm{I}}\frac{1}{2!}T_{1}^{\mathrm{II}^2} \allowbreak + \allowbreak T_{1}^{\mathrm{I}}\frac{1}{2!}T_{1}^{\mathrm{II}^2}T_{11}^{\mathrm{I,II}} \allowbreak + \allowbreak T_{1}^{\mathrm{I}}\frac{1}{3!}T_{1}^{\mathrm{II}^3} \allowbreak + \allowbreak T_{1}^{\mathrm{I}}T_{2}^{\mathrm{II}} \allowbreak + \allowbreak T_{1}^{\mathrm{I}}T_{2}^{\mathrm{II}}T_{11}^{\mathrm{I,II}} \allowbreak + \allowbreak \frac{1}{2!}T_{1}^{\mathrm{I}^2}T_{1}^{\mathrm{II}} \allowbreak + \allowbreak \frac{1}{2!}T_{1}^{\mathrm{I}^2}T_{1}^{\mathrm{II}}T_{2}^{\mathrm{II}} \allowbreak + \allowbreak \frac{1}{2!}T_{1}^{\mathrm{I}^2}\frac{1}{2!}T_{1}^{\mathrm{II}^2} \allowbreak + \allowbreak \frac{1}{2!}T_{1}^{\mathrm{I}^2}\frac{1}{3!}T_{1}^{\mathrm{II}^3} \allowbreak + \allowbreak \frac{1}{2!}T_{1}^{\mathrm{I}^2}T_{2}^{\mathrm{II}} \allowbreak + \allowbreak T_{1}^{\mathrm{II}} \allowbreak + \allowbreak T_{1}^{\mathrm{II}}T_{11}^{\mathrm{I,II}} \allowbreak + \allowbreak T_{1}^{\mathrm{II}}T_{12}^{\mathrm{I,II}} \allowbreak + \allowbreak T_{1}^{\mathrm{II}}T_{21}^{\mathrm{I,II}} \allowbreak + \allowbreak T_{1}^{\mathrm{II}}T_{22}^{\mathrm{I,II}} \allowbreak + \allowbreak T_{1}^{\mathrm{II}}T_{2}^{\mathrm{II}} \allowbreak + \allowbreak \frac{1}{2!}T_{1}^{\mathrm{II}^2} \allowbreak + \allowbreak \frac{1}{2!}T_{1}^{\mathrm{II}^2}T_{11}^{\mathrm{I,II}} \allowbreak + \allowbreak \frac{1}{2!}T_{1}^{\mathrm{II}^2}T_{21}^{\mathrm{I,II}} \allowbreak + \allowbreak \frac{1}{3!}T_{1}^{\mathrm{II}^3} \allowbreak + \allowbreak T_{21}^{\mathrm{I,II}} \allowbreak + \allowbreak T_{22}^{\mathrm{I,II}} \allowbreak + \allowbreak T_{2}^{\mathrm{I}}T_{1}^{\mathrm{II}} \allowbreak + \allowbreak T_{2}^{\mathrm{I}}T_{1}^{\mathrm{II}}T_{2}^{\mathrm{II}} \allowbreak + \allowbreak T_{2}^{\mathrm{I}}\frac{1}{2!}T_{1}^{\mathrm{II}^2} \allowbreak + \allowbreak T_{2}^{\mathrm{I}}\frac{1}{3!}T_{1}^{\mathrm{II}^3} \allowbreak + \allowbreak T_{2}^{\mathrm{I}}T_{2}^{\mathrm{II}} \allowbreak + \allowbreak T_{2}^{\mathrm{II}} \allowbreak + \allowbreak T_{2}^{\mathrm{II}}T_{11}^{\mathrm{I,II}} \allowbreak + \allowbreak T_{2}^{\mathrm{II}}T_{21}^{\mathrm{I,II}}  ]  \allowbreak + \allowbreak  [  (-T_{12}^{\mathrm{I,II}})(-T_{1}^{\mathrm{I}})  ]  \allowbreak V_{\mathrm{N}}^{\mathrm{I,II}} \allowbreak   [  1 \allowbreak + \allowbreak T_{11}^{\mathrm{I,II}} \allowbreak + \allowbreak T_{1}^{\mathrm{I}} \allowbreak + \allowbreak T_{1}^{\mathrm{I}}T_{1}^{\mathrm{II}} \allowbreak + \allowbreak T_{1}^{\mathrm{II}}  ]  \allowbreak + \allowbreak  [  (-T_{11}^{\mathrm{I,II}})(-T_{1}^{\mathrm{I}})  ]  \allowbreak V_{\mathrm{N}}^{\mathrm{I,II}} \allowbreak   [  1 \allowbreak + \allowbreak T_{11}^{\mathrm{I,II}} \allowbreak + \allowbreak T_{12}^{\mathrm{I,II}} \allowbreak + \allowbreak T_{1}^{\mathrm{I}} \allowbreak + \allowbreak T_{1}^{\mathrm{I}}T_{1}^{\mathrm{II}} \allowbreak + \allowbreak T_{1}^{\mathrm{I}}\frac{1}{2!}T_{1}^{\mathrm{II}^2} \allowbreak + \allowbreak T_{1}^{\mathrm{I}}T_{2}^{\mathrm{II}} \allowbreak + \allowbreak T_{1}^{\mathrm{II}} \allowbreak + \allowbreak T_{1}^{\mathrm{II}}T_{11}^{\mathrm{I,II}} \allowbreak + \allowbreak \frac{1}{2!}T_{1}^{\mathrm{II}^2} \allowbreak + \allowbreak T_{2}^{\mathrm{II}}  ]  \allowbreak + \allowbreak  [  (-T_{1}^{\mathrm{I}})(-T_{2}^{\mathrm{II}})  ]  \allowbreak V_{\mathrm{N}}^{\mathrm{I,II}} \allowbreak   [  1 \allowbreak + \allowbreak T_{11}^{\mathrm{I,II}} \allowbreak + \allowbreak T_{1}^{\mathrm{I}} \allowbreak + \allowbreak T_{1}^{\mathrm{I}}T_{11}^{\mathrm{I,II}} \allowbreak + \allowbreak T_{1}^{\mathrm{I}}T_{1}^{\mathrm{II}} \allowbreak + \allowbreak \frac{1}{2!}T_{1}^{\mathrm{I}^2} \allowbreak + \allowbreak \frac{1}{2!}T_{1}^{\mathrm{I}^2}T_{1}^{\mathrm{II}} \allowbreak + \allowbreak T_{1}^{\mathrm{II}} \allowbreak + \allowbreak T_{21}^{\mathrm{I,II}} \allowbreak + \allowbreak T_{2}^{\mathrm{I}} \allowbreak + \allowbreak T_{2}^{\mathrm{I}}T_{1}^{\mathrm{II}}  ]  \allowbreak + \allowbreak  [  (-T_{1}^{\mathrm{I}})(-T_{1}^{\mathrm{II}})  ]  \allowbreak V_{\mathrm{N}}^{\mathrm{I,II}} \allowbreak   [  1 \allowbreak + \allowbreak T_{11}^{\mathrm{I,II}} \allowbreak + \allowbreak T_{12}^{\mathrm{I,II}} \allowbreak + \allowbreak T_{1}^{\mathrm{I}} \allowbreak + \allowbreak T_{1}^{\mathrm{I}}T_{11}^{\mathrm{I,II}} \allowbreak + \allowbreak T_{1}^{\mathrm{I}}T_{12}^{\mathrm{I,II}} \allowbreak + \allowbreak T_{1}^{\mathrm{I}}T_{1}^{\mathrm{II}} \allowbreak + \allowbreak T_{1}^{\mathrm{I}}T_{1}^{\mathrm{II}}T_{11}^{\mathrm{I,II}} \allowbreak + \allowbreak T_{1}^{\mathrm{I}}\frac{1}{2!}T_{1}^{\mathrm{II}^2} \allowbreak + \allowbreak T_{1}^{\mathrm{I}}T_{2}^{\mathrm{II}} \allowbreak + \allowbreak \frac{1}{2!}T_{1}^{\mathrm{I}^2} \allowbreak + \allowbreak \frac{1}{2!}T_{1}^{\mathrm{I}^2}T_{1}^{\mathrm{II}} \allowbreak + \allowbreak \frac{1}{2!}T_{1}^{\mathrm{I}^2}\frac{1}{2!}T_{1}^{\mathrm{II}^2} \allowbreak + \allowbreak \frac{1}{2!}T_{1}^{\mathrm{I}^2}T_{2}^{\mathrm{II}} \allowbreak + \allowbreak T_{1}^{\mathrm{II}} \allowbreak + \allowbreak T_{1}^{\mathrm{II}}T_{11}^{\mathrm{I,II}} \allowbreak + \allowbreak T_{1}^{\mathrm{II}}T_{21}^{\mathrm{I,II}} \allowbreak + \allowbreak \frac{1}{2!}T_{1}^{\mathrm{II}^2} \allowbreak + \allowbreak T_{21}^{\mathrm{I,II}} \allowbreak + \allowbreak T_{22}^{\mathrm{I,II}} \allowbreak + \allowbreak T_{2}^{\mathrm{I}} \allowbreak + \allowbreak T_{2}^{\mathrm{I}}T_{1}^{\mathrm{II}} \allowbreak + \allowbreak T_{2}^{\mathrm{I}}\frac{1}{2!}T_{1}^{\mathrm{II}^2} \allowbreak + \allowbreak T_{2}^{\mathrm{I}}T_{2}^{\mathrm{II}} \allowbreak + \allowbreak T_{2}^{\mathrm{II}}  ]  \allowbreak + \allowbreak  [  (-T_{11}^{\mathrm{I,II}})(-T_{1}^{\mathrm{I}})(-T_{1}^{\mathrm{II}})  ]  \allowbreak V_{\mathrm{N}}^{\mathrm{I,II}} \allowbreak   [  1 \allowbreak + \allowbreak T_{11}^{\mathrm{I,II}} \allowbreak + \allowbreak T_{1}^{\mathrm{I}} \allowbreak + \allowbreak T_{1}^{\mathrm{I}}T_{1}^{\mathrm{II}} \allowbreak + \allowbreak T_{1}^{\mathrm{II}}  ]  \allowbreak + \allowbreak  [  (-T_{1}^{\mathrm{I}})\frac{1}{2!}T_{1}^{\mathrm{II}^2}  ]  \allowbreak V_{\mathrm{N}}^{\mathrm{I,II}} \allowbreak   [  1 \allowbreak + \allowbreak T_{11}^{\mathrm{I,II}} \allowbreak + \allowbreak T_{1}^{\mathrm{I}} \allowbreak + \allowbreak T_{1}^{\mathrm{I}}T_{11}^{\mathrm{I,II}} \allowbreak + \allowbreak T_{1}^{\mathrm{I}}T_{1}^{\mathrm{II}} \allowbreak + \allowbreak \frac{1}{2!}T_{1}^{\mathrm{I}^2} \allowbreak + \allowbreak \frac{1}{2!}T_{1}^{\mathrm{I}^2}T_{1}^{\mathrm{II}} \allowbreak + \allowbreak T_{1}^{\mathrm{II}} \allowbreak + \allowbreak T_{21}^{\mathrm{I,II}} \allowbreak + \allowbreak T_{2}^{\mathrm{I}} \allowbreak + \allowbreak T_{2}^{\mathrm{I}}T_{1}^{\mathrm{II}}  ]  \allowbreak + \allowbreak  [  \frac{1}{2!}T_{1}^{\mathrm{I}^2}  ]  \allowbreak V_{\mathrm{N}}^{\mathrm{I,II}} \allowbreak   [  T_{11}^{\mathrm{I,II}} \allowbreak + \allowbreak T_{12}^{\mathrm{I,II}} \allowbreak + \allowbreak T_{1}^{\mathrm{I}}T_{1}^{\mathrm{II}} \allowbreak + \allowbreak T_{1}^{\mathrm{I}}T_{1}^{\mathrm{II}}T_{2}^{\mathrm{II}} \allowbreak + \allowbreak T_{1}^{\mathrm{I}}\frac{1}{2!}T_{1}^{\mathrm{II}^2} \allowbreak + \allowbreak T_{1}^{\mathrm{I}}\frac{1}{3!}T_{1}^{\mathrm{II}^3} \allowbreak + \allowbreak T_{1}^{\mathrm{I}}T_{2}^{\mathrm{II}} \allowbreak + \allowbreak T_{1}^{\mathrm{II}} \allowbreak + \allowbreak T_{1}^{\mathrm{II}}T_{11}^{\mathrm{I,II}} \allowbreak + \allowbreak T_{1}^{\mathrm{II}}T_{12}^{\mathrm{I,II}} \allowbreak + \allowbreak T_{1}^{\mathrm{II}}T_{2}^{\mathrm{II}} \allowbreak + \allowbreak \frac{1}{2!}T_{1}^{\mathrm{II}^2} \allowbreak + \allowbreak \frac{1}{2!}T_{1}^{\mathrm{II}^2}T_{11}^{\mathrm{I,II}} \allowbreak + \allowbreak \frac{1}{3!}T_{1}^{\mathrm{II}^3} \allowbreak + \allowbreak T_{2}^{\mathrm{II}} \allowbreak + \allowbreak T_{2}^{\mathrm{II}}T_{11}^{\mathrm{I,II}}  ]  \allowbreak + \allowbreak  [  (-T_{2}^{\mathrm{II}})\frac{1}{2!}T_{1}^{\mathrm{I}^2}  ]  \allowbreak V_{\mathrm{N}}^{\mathrm{I,II}} \allowbreak   [  1 \allowbreak + \allowbreak T_{11}^{\mathrm{I,II}} \allowbreak + \allowbreak T_{1}^{\mathrm{I}} \allowbreak + \allowbreak T_{1}^{\mathrm{I}}T_{1}^{\mathrm{II}} \allowbreak + \allowbreak T_{1}^{\mathrm{II}}  ]  \allowbreak + \allowbreak  [  (-T_{1}^{\mathrm{II}})\frac{1}{2!}T_{1}^{\mathrm{I}^2}  ]  \allowbreak V_{\mathrm{N}}^{\mathrm{I,II}} \allowbreak   [  1 \allowbreak + \allowbreak T_{11}^{\mathrm{I,II}} \allowbreak + \allowbreak T_{12}^{\mathrm{I,II}} \allowbreak + \allowbreak T_{1}^{\mathrm{I}} \allowbreak + \allowbreak T_{1}^{\mathrm{I}}T_{1}^{\mathrm{II}} \allowbreak + \allowbreak T_{1}^{\mathrm{I}}\frac{1}{2!}T_{1}^{\mathrm{II}^2} \allowbreak + \allowbreak T_{1}^{\mathrm{I}}T_{2}^{\mathrm{II}} \allowbreak + \allowbreak T_{1}^{\mathrm{II}} \allowbreak + \allowbreak T_{1}^{\mathrm{II}}T_{11}^{\mathrm{I,II}} \allowbreak + \allowbreak \frac{1}{2!}T_{1}^{\mathrm{II}^2} \allowbreak + \allowbreak T_{2}^{\mathrm{II}}  ]  \allowbreak + \allowbreak  [  \frac{1}{2!}T_{1}^{\mathrm{I}^2}\frac{1}{2!}T_{1}^{\mathrm{II}^2}  ]  \allowbreak V_{\mathrm{N}}^{\mathrm{I,II}} \allowbreak   [  1 \allowbreak + \allowbreak T_{11}^{\mathrm{I,II}} \allowbreak + \allowbreak T_{1}^{\mathrm{I}} \allowbreak + \allowbreak T_{1}^{\mathrm{I}}T_{1}^{\mathrm{II}} \allowbreak + \allowbreak T_{1}^{\mathrm{II}}  ]  \allowbreak + \allowbreak    \allowbreak \tilde{H}_{\mathrm{N}}^{\mathrm{II}} \allowbreak  [  T_{11}^{\mathrm{I,II}}T_{12}^{\mathrm{I,II}} \allowbreak + \allowbreak \frac{1}{2!}T_{12}^{\mathrm{I,II}^2} \allowbreak + \allowbreak T_{1}^{\mathrm{I}}T_{11}^{\mathrm{I,II}} \allowbreak + \allowbreak T_{1}^{\mathrm{I}}T_{12}^{\mathrm{I,II}} \allowbreak + \allowbreak T_{1}^{\mathrm{I}}T_{1}^{\mathrm{II}}T_{11}^{\mathrm{I,II}} \allowbreak + \allowbreak T_{1}^{\mathrm{I}}T_{1}^{\mathrm{II}}T_{12}^{\mathrm{I,II}} \allowbreak + \allowbreak T_{1}^{\mathrm{I}}T_{1}^{\mathrm{II}}T_{2}^{\mathrm{II}}T_{11}^{\mathrm{I,II}} \allowbreak + \allowbreak T_{1}^{\mathrm{I}}\frac{1}{2!}T_{1}^{\mathrm{II}^2}T_{11}^{\mathrm{I,II}} \allowbreak + \allowbreak T_{1}^{\mathrm{I}}\frac{1}{2!}T_{1}^{\mathrm{II}^2}T_{12}^{\mathrm{I,II}} \allowbreak + \allowbreak T_{1}^{\mathrm{I}}\frac{1}{3!}T_{1}^{\mathrm{II}^3}T_{11}^{\mathrm{I,II}} \allowbreak + \allowbreak T_{1}^{\mathrm{I}}T_{2}^{\mathrm{II}}T_{11}^{\mathrm{I,II}} \allowbreak + \allowbreak T_{1}^{\mathrm{I}}T_{2}^{\mathrm{II}}T_{12}^{\mathrm{I,II}} \allowbreak + \allowbreak \frac{1}{2!}T_{1}^{\mathrm{I}^2} \allowbreak + \allowbreak \frac{1}{2!}T_{1}^{\mathrm{I}^2}T_{1}^{\mathrm{II}} \allowbreak + \allowbreak \frac{1}{2!}T_{1}^{\mathrm{I}^2}T_{1}^{\mathrm{II}}T_{2}^{\mathrm{II}} \allowbreak + \allowbreak \frac{1}{2!}T_{1}^{\mathrm{I}^2}\frac{1}{2!}T_{1}^{\mathrm{II}^2} \allowbreak + \allowbreak \frac{1}{2!}T_{1}^{\mathrm{I}^2}\frac{1}{2!}T_{1}^{\mathrm{II}^2}T_{2}^{\mathrm{II}} \allowbreak + \allowbreak \frac{1}{2!}T_{1}^{\mathrm{I}^2}\frac{1}{3!}T_{1}^{\mathrm{II}^3} \allowbreak + \allowbreak \frac{1}{2!}T_{1}^{\mathrm{I}^2}\frac{1}{4!}T_{1}^{\mathrm{II}^4} \allowbreak + \allowbreak \frac{1}{2!}T_{1}^{\mathrm{I}^2}T_{2}^{\mathrm{II}} \allowbreak + \allowbreak \frac{1}{2!}T_{1}^{\mathrm{I}^2}\frac{1}{2!}T_{2}^{\mathrm{II}^2} \allowbreak + \allowbreak T_{1}^{\mathrm{II}}T_{11}^{\mathrm{I,II}}T_{12}^{\mathrm{I,II}} \allowbreak + \allowbreak T_{1}^{\mathrm{II}}T_{21}^{\mathrm{I,II}} \allowbreak + \allowbreak T_{1}^{\mathrm{II}}T_{22}^{\mathrm{I,II}} \allowbreak + \allowbreak T_{1}^{\mathrm{II}}T_{2}^{\mathrm{II}}T_{21}^{\mathrm{I,II}} \allowbreak + \allowbreak \frac{1}{2!}T_{1}^{\mathrm{II}^2}T_{21}^{\mathrm{I,II}} \allowbreak + \allowbreak \frac{1}{2!}T_{1}^{\mathrm{II}^2}T_{22}^{\mathrm{I,II}} \allowbreak + \allowbreak \frac{1}{3!}T_{1}^{\mathrm{II}^3}T_{21}^{\mathrm{I,II}} \allowbreak + \allowbreak T_{21}^{\mathrm{I,II}} \allowbreak + \allowbreak T_{22}^{\mathrm{I,II}} \allowbreak + \allowbreak T_{2}^{\mathrm{I}} \allowbreak + \allowbreak T_{2}^{\mathrm{I}}T_{1}^{\mathrm{II}} \allowbreak + \allowbreak T_{2}^{\mathrm{I}}T_{1}^{\mathrm{II}}T_{2}^{\mathrm{II}} \allowbreak + \allowbreak T_{2}^{\mathrm{I}}\frac{1}{2!}T_{1}^{\mathrm{II}^2} \allowbreak + \allowbreak T_{2}^{\mathrm{I}}\frac{1}{2!}T_{1}^{\mathrm{II}^2}T_{2}^{\mathrm{II}} \allowbreak + \allowbreak T_{2}^{\mathrm{I}}\frac{1}{3!}T_{1}^{\mathrm{II}^3} \allowbreak + \allowbreak T_{2}^{\mathrm{I}}\frac{1}{4!}T_{1}^{\mathrm{II}^4} \allowbreak + \allowbreak T_{2}^{\mathrm{I}}T_{2}^{\mathrm{II}} \allowbreak + \allowbreak T_{2}^{\mathrm{I}}\frac{1}{2!}T_{2}^{\mathrm{II}^2} \allowbreak + \allowbreak T_{2}^{\mathrm{II}}T_{21}^{\mathrm{I,II}} \allowbreak + \allowbreak T_{2}^{\mathrm{II}}T_{22}^{\mathrm{I,II}}  ]  \allowbreak + \allowbreak  [  (-T_{22}^{\mathrm{I,II}})  ]  \allowbreak \tilde{H}_{\mathrm{N}}^{\mathrm{II}} \allowbreak  [  1 \allowbreak + \allowbreak T_{1}^{\mathrm{II}} \allowbreak + \allowbreak \frac{1}{2!}T_{1}^{\mathrm{II}^2} \allowbreak + \allowbreak T_{2}^{\mathrm{II}}  ]  \allowbreak + \allowbreak  [  (-T_{21}^{\mathrm{I,II}})  ]  \allowbreak \tilde{H}_{\mathrm{N}}^{\mathrm{II}} \allowbreak  [  1 \allowbreak + \allowbreak T_{1}^{\mathrm{II}} \allowbreak + \allowbreak T_{1}^{\mathrm{II}}T_{2}^{\mathrm{II}} \allowbreak + \allowbreak \frac{1}{2!}T_{1}^{\mathrm{II}^2} \allowbreak + \allowbreak \frac{1}{3!}T_{1}^{\mathrm{II}^3} \allowbreak + \allowbreak T_{2}^{\mathrm{II}}  ]  \allowbreak + \allowbreak  [  (-T_{12}^{\mathrm{I,II}})  ]  \allowbreak \tilde{H}_{\mathrm{N}}^{\mathrm{II}} \allowbreak  [  T_{11}^{\mathrm{I,II}} \allowbreak + \allowbreak T_{12}^{\mathrm{I,II}} \allowbreak + \allowbreak T_{1}^{\mathrm{I}} \allowbreak + \allowbreak T_{1}^{\mathrm{I}}T_{1}^{\mathrm{II}} \allowbreak + \allowbreak T_{1}^{\mathrm{I}}\frac{1}{2!}T_{1}^{\mathrm{II}^2} \allowbreak + \allowbreak T_{1}^{\mathrm{I}}T_{2}^{\mathrm{II}} \allowbreak + \allowbreak T_{1}^{\mathrm{II}}T_{11}^{\mathrm{I,II}}  ]  \allowbreak + \allowbreak  [  \frac{1}{2!}T_{11}^{\mathrm{I,II}^2}  ]  \allowbreak \tilde{H}_{\mathrm{N}}^{\mathrm{II}} \allowbreak  [  1 \allowbreak + \allowbreak T_{1}^{\mathrm{II}} \allowbreak + \allowbreak \frac{1}{2!}T_{1}^{\mathrm{II}^2} \allowbreak + \allowbreak T_{2}^{\mathrm{II}}  ]  \allowbreak + \allowbreak  [  (-T_{11}^{\mathrm{I,II}})  ]  \allowbreak \tilde{H}_{\mathrm{N}}^{\mathrm{II}} \allowbreak  [  T_{11}^{\mathrm{I,II}} \allowbreak + \allowbreak T_{12}^{\mathrm{I,II}} \allowbreak + \allowbreak T_{1}^{\mathrm{I}} \allowbreak + \allowbreak T_{1}^{\mathrm{I}}T_{1}^{\mathrm{II}} \allowbreak + \allowbreak T_{1}^{\mathrm{I}}T_{1}^{\mathrm{II}}T_{2}^{\mathrm{II}} \allowbreak + \allowbreak T_{1}^{\mathrm{I}}\frac{1}{2!}T_{1}^{\mathrm{II}^2} \allowbreak + \allowbreak T_{1}^{\mathrm{I}}\frac{1}{3!}T_{1}^{\mathrm{II}^3} \allowbreak + \allowbreak T_{1}^{\mathrm{I}}T_{2}^{\mathrm{II}} \allowbreak + \allowbreak T_{1}^{\mathrm{II}}T_{11}^{\mathrm{I,II}} \allowbreak + \allowbreak T_{1}^{\mathrm{II}}T_{12}^{\mathrm{I,II}} \allowbreak + \allowbreak \frac{1}{2!}T_{1}^{\mathrm{II}^2}T_{11}^{\mathrm{I,II}} \allowbreak + \allowbreak T_{2}^{\mathrm{II}}T_{11}^{\mathrm{I,II}}  ]  \allowbreak + \allowbreak  [  (-T_{2}^{\mathrm{II}})  ]  \allowbreak \tilde{H}_{\mathrm{N}}^{\mathrm{II}} \allowbreak  [  T_{1}^{\mathrm{I}}T_{11}^{\mathrm{I,II}} \allowbreak + \allowbreak T_{1}^{\mathrm{I}}T_{12}^{\mathrm{I,II}} \allowbreak + \allowbreak T_{1}^{\mathrm{I}}T_{1}^{\mathrm{II}}T_{11}^{\mathrm{I,II}} \allowbreak + \allowbreak \frac{1}{2!}T_{1}^{\mathrm{I}^2} \allowbreak + \allowbreak \frac{1}{2!}T_{1}^{\mathrm{I}^2}T_{1}^{\mathrm{II}} \allowbreak + \allowbreak \frac{1}{2!}T_{1}^{\mathrm{I}^2}\frac{1}{2!}T_{1}^{\mathrm{II}^2} \allowbreak + \allowbreak \frac{1}{2!}T_{1}^{\mathrm{I}^2}T_{2}^{\mathrm{II}} \allowbreak + \allowbreak T_{1}^{\mathrm{II}}T_{21}^{\mathrm{I,II}} \allowbreak + \allowbreak T_{21}^{\mathrm{I,II}} \allowbreak + \allowbreak T_{22}^{\mathrm{I,II}} \allowbreak + \allowbreak T_{2}^{\mathrm{I}} \allowbreak + \allowbreak T_{2}^{\mathrm{I}}T_{1}^{\mathrm{II}} \allowbreak + \allowbreak T_{2}^{\mathrm{I}}\frac{1}{2!}T_{1}^{\mathrm{II}^2} \allowbreak + \allowbreak T_{2}^{\mathrm{I}}T_{2}^{\mathrm{II}}  ]  \allowbreak + \allowbreak  [  (-T_{1}^{\mathrm{II}})  ]  \allowbreak \tilde{H}_{\mathrm{N}}^{\mathrm{II}} \allowbreak  [  T_{11}^{\mathrm{I,II}}T_{12}^{\mathrm{I,II}} \allowbreak + \allowbreak T_{1}^{\mathrm{I}}T_{11}^{\mathrm{I,II}} \allowbreak + \allowbreak T_{1}^{\mathrm{I}}T_{12}^{\mathrm{I,II}} \allowbreak + \allowbreak T_{1}^{\mathrm{I}}T_{1}^{\mathrm{II}}T_{11}^{\mathrm{I,II}} \allowbreak + \allowbreak T_{1}^{\mathrm{I}}T_{1}^{\mathrm{II}}T_{12}^{\mathrm{I,II}} \allowbreak + \allowbreak T_{1}^{\mathrm{I}}\frac{1}{2!}T_{1}^{\mathrm{II}^2}T_{11}^{\mathrm{I,II}} \allowbreak + \allowbreak T_{1}^{\mathrm{I}}T_{2}^{\mathrm{II}}T_{11}^{\mathrm{I,II}} \allowbreak + \allowbreak \frac{1}{2!}T_{1}^{\mathrm{I}^2} \allowbreak + \allowbreak \frac{1}{2!}T_{1}^{\mathrm{I}^2}T_{1}^{\mathrm{II}} \allowbreak + \allowbreak \frac{1}{2!}T_{1}^{\mathrm{I}^2}T_{1}^{\mathrm{II}}T_{2}^{\mathrm{II}} \allowbreak + \allowbreak \frac{1}{2!}T_{1}^{\mathrm{I}^2}\frac{1}{2!}T_{1}^{\mathrm{II}^2} \allowbreak + \allowbreak \frac{1}{2!}T_{1}^{\mathrm{I}^2}\frac{1}{3!}T_{1}^{\mathrm{II}^3} \allowbreak + \allowbreak \frac{1}{2!}T_{1}^{\mathrm{I}^2}T_{2}^{\mathrm{II}} \allowbreak + \allowbreak T_{1}^{\mathrm{II}}T_{21}^{\mathrm{I,II}} \allowbreak + \allowbreak T_{1}^{\mathrm{II}}T_{22}^{\mathrm{I,II}} \allowbreak + \allowbreak \frac{1}{2!}T_{1}^{\mathrm{II}^2}T_{21}^{\mathrm{I,II}} \allowbreak + \allowbreak T_{21}^{\mathrm{I,II}} \allowbreak + \allowbreak T_{22}^{\mathrm{I,II}} \allowbreak + \allowbreak T_{2}^{\mathrm{I}} \allowbreak + \allowbreak T_{2}^{\mathrm{I}}T_{1}^{\mathrm{II}} \allowbreak + \allowbreak T_{2}^{\mathrm{I}}T_{1}^{\mathrm{II}}T_{2}^{\mathrm{II}} \allowbreak + \allowbreak T_{2}^{\mathrm{I}}\frac{1}{2!}T_{1}^{\mathrm{II}^2} \allowbreak + \allowbreak T_{2}^{\mathrm{I}}\frac{1}{3!}T_{1}^{\mathrm{II}^3} \allowbreak + \allowbreak T_{2}^{\mathrm{I}}T_{2}^{\mathrm{II}} \allowbreak + \allowbreak T_{2}^{\mathrm{II}}T_{21}^{\mathrm{I,II}}  ]  \allowbreak + \allowbreak  [  (-T_{1}^{\mathrm{II}})(-T_{21}^{\mathrm{I,II}})  ]  \allowbreak \tilde{H}_{\mathrm{N}}^{\mathrm{II}} \allowbreak  [  1 \allowbreak + \allowbreak T_{1}^{\mathrm{II}} \allowbreak + \allowbreak \frac{1}{2!}T_{1}^{\mathrm{II}^2} \allowbreak + \allowbreak T_{2}^{\mathrm{II}}  ]  \allowbreak + \allowbreak  [  (-T_{11}^{\mathrm{I,II}})(-T_{1}^{\mathrm{II}})  ]  \allowbreak \tilde{H}_{\mathrm{N}}^{\mathrm{II}} \allowbreak  [  T_{11}^{\mathrm{I,II}} \allowbreak + \allowbreak T_{12}^{\mathrm{I,II}} \allowbreak + \allowbreak T_{1}^{\mathrm{I}} \allowbreak + \allowbreak T_{1}^{\mathrm{I}}T_{1}^{\mathrm{II}} \allowbreak + \allowbreak T_{1}^{\mathrm{I}}\frac{1}{2!}T_{1}^{\mathrm{II}^2} \allowbreak + \allowbreak T_{1}^{\mathrm{I}}T_{2}^{\mathrm{II}} \allowbreak + \allowbreak T_{1}^{\mathrm{II}}T_{11}^{\mathrm{I,II}}  ]  \allowbreak + \allowbreak  [  \frac{1}{2!}T_{1}^{\mathrm{II}^2}  ]  \allowbreak \tilde{H}_{\mathrm{N}}^{\mathrm{II}} \allowbreak  [  T_{1}^{\mathrm{I}}T_{11}^{\mathrm{I,II}} \allowbreak + \allowbreak T_{1}^{\mathrm{I}}T_{12}^{\mathrm{I,II}} \allowbreak + \allowbreak T_{1}^{\mathrm{I}}T_{1}^{\mathrm{II}}T_{11}^{\mathrm{I,II}} \allowbreak + \allowbreak \frac{1}{2!}T_{1}^{\mathrm{I}^2} \allowbreak + \allowbreak \frac{1}{2!}T_{1}^{\mathrm{I}^2}T_{1}^{\mathrm{II}} \allowbreak + \allowbreak \frac{1}{2!}T_{1}^{\mathrm{I}^2}\frac{1}{2!}T_{1}^{\mathrm{II}^2} \allowbreak + \allowbreak \frac{1}{2!}T_{1}^{\mathrm{I}^2}T_{2}^{\mathrm{II}} \allowbreak + \allowbreak T_{1}^{\mathrm{II}}T_{21}^{\mathrm{I,II}} \allowbreak + \allowbreak T_{21}^{\mathrm{I,II}} \allowbreak + \allowbreak T_{22}^{\mathrm{I,II}} \allowbreak + \allowbreak T_{2}^{\mathrm{I}} \allowbreak + \allowbreak T_{2}^{\mathrm{I}}T_{1}^{\mathrm{II}} \allowbreak + \allowbreak T_{2}^{\mathrm{I}}\frac{1}{2!}T_{1}^{\mathrm{II}^2} \allowbreak + \allowbreak T_{2}^{\mathrm{I}}T_{2}^{\mathrm{II}}  ]  \allowbreak + \allowbreak  [  (-T_{2}^{\mathrm{I}})  ]  \allowbreak \tilde{H}_{\mathrm{N}}^{\mathrm{II}} \allowbreak  [  1 \allowbreak + \allowbreak T_{1}^{\mathrm{II}} \allowbreak + \allowbreak T_{1}^{\mathrm{II}}T_{2}^{\mathrm{II}} \allowbreak + \allowbreak \frac{1}{2!}T_{1}^{\mathrm{II}^2} \allowbreak + \allowbreak \frac{1}{2!}T_{1}^{\mathrm{II}^2}T_{2}^{\mathrm{II}} \allowbreak + \allowbreak \frac{1}{3!}T_{1}^{\mathrm{II}^3} \allowbreak + \allowbreak \frac{1}{4!}T_{1}^{\mathrm{II}^4} \allowbreak + \allowbreak T_{2}^{\mathrm{II}} \allowbreak + \allowbreak \frac{1}{2!}T_{2}^{\mathrm{II}^2}  ]  \allowbreak + \allowbreak  [  (-T_{2}^{\mathrm{I}})(-T_{2}^{\mathrm{II}})  ]  \allowbreak \tilde{H}_{\mathrm{N}}^{\mathrm{II}} \allowbreak  [  1 \allowbreak + \allowbreak T_{1}^{\mathrm{II}} \allowbreak + \allowbreak \frac{1}{2!}T_{1}^{\mathrm{II}^2} \allowbreak + \allowbreak T_{2}^{\mathrm{II}}  ]  \allowbreak + \allowbreak  [  (-T_{1}^{\mathrm{II}})(-T_{2}^{\mathrm{I}})  ]  \allowbreak \tilde{H}_{\mathrm{N}}^{\mathrm{II}} \allowbreak  [  1 \allowbreak + \allowbreak T_{1}^{\mathrm{II}} \allowbreak + \allowbreak T_{1}^{\mathrm{II}}T_{2}^{\mathrm{II}} \allowbreak + \allowbreak \frac{1}{2!}T_{1}^{\mathrm{II}^2} \allowbreak + \allowbreak \frac{1}{3!}T_{1}^{\mathrm{II}^3} \allowbreak + \allowbreak T_{2}^{\mathrm{II}}  ]  \allowbreak + \allowbreak  [  (-T_{2}^{\mathrm{I}})\frac{1}{2!}T_{1}^{\mathrm{II}^2}  ]  \allowbreak \tilde{H}_{\mathrm{N}}^{\mathrm{II}} \allowbreak  [  1 \allowbreak + \allowbreak T_{1}^{\mathrm{II}} \allowbreak + \allowbreak \frac{1}{2!}T_{1}^{\mathrm{II}^2} \allowbreak + \allowbreak T_{2}^{\mathrm{II}}  ]  \allowbreak + \allowbreak  [  (-T_{1}^{\mathrm{I}})  ]  \allowbreak \tilde{H}_{\mathrm{N}}^{\mathrm{II}} \allowbreak  [  T_{11}^{\mathrm{I,II}} \allowbreak + \allowbreak T_{12}^{\mathrm{I,II}} \allowbreak + \allowbreak T_{1}^{\mathrm{I}} \allowbreak + \allowbreak T_{1}^{\mathrm{I}}T_{1}^{\mathrm{II}} \allowbreak + \allowbreak T_{1}^{\mathrm{I}}T_{1}^{\mathrm{II}}T_{2}^{\mathrm{II}} \allowbreak + \allowbreak T_{1}^{\mathrm{I}}\frac{1}{2!}T_{1}^{\mathrm{II}^2} \allowbreak + \allowbreak T_{1}^{\mathrm{I}}\frac{1}{2!}T_{1}^{\mathrm{II}^2}T_{2}^{\mathrm{II}} \allowbreak + \allowbreak T_{1}^{\mathrm{I}}\frac{1}{3!}T_{1}^{\mathrm{II}^3} \allowbreak + \allowbreak T_{1}^{\mathrm{I}}\frac{1}{4!}T_{1}^{\mathrm{II}^4} \allowbreak + \allowbreak T_{1}^{\mathrm{I}}T_{2}^{\mathrm{II}} \allowbreak + \allowbreak T_{1}^{\mathrm{I}}\frac{1}{2!}T_{2}^{\mathrm{II}^2} \allowbreak + \allowbreak T_{1}^{\mathrm{II}}T_{11}^{\mathrm{I,II}} \allowbreak + \allowbreak T_{1}^{\mathrm{II}}T_{12}^{\mathrm{I,II}} \allowbreak + \allowbreak T_{1}^{\mathrm{II}}T_{2}^{\mathrm{II}}T_{11}^{\mathrm{I,II}} \allowbreak + \allowbreak \frac{1}{2!}T_{1}^{\mathrm{II}^2}T_{11}^{\mathrm{I,II}} \allowbreak + \allowbreak \frac{1}{2!}T_{1}^{\mathrm{II}^2}T_{12}^{\mathrm{I,II}} \allowbreak + \allowbreak \frac{1}{3!}T_{1}^{\mathrm{II}^3}T_{11}^{\mathrm{I,II}} \allowbreak + \allowbreak T_{2}^{\mathrm{II}}T_{11}^{\mathrm{I,II}} \allowbreak + \allowbreak T_{2}^{\mathrm{II}}T_{12}^{\mathrm{I,II}}  ]  \allowbreak + \allowbreak  [  (-T_{12}^{\mathrm{I,II}})(-T_{1}^{\mathrm{I}})  ]  \allowbreak \tilde{H}_{\mathrm{N}}^{\mathrm{II}} \allowbreak  [  1 \allowbreak + \allowbreak T_{1}^{\mathrm{II}} \allowbreak + \allowbreak \frac{1}{2!}T_{1}^{\mathrm{II}^2} \allowbreak + \allowbreak T_{2}^{\mathrm{II}}  ]  \allowbreak + \allowbreak  [  (-T_{11}^{\mathrm{I,II}})(-T_{1}^{\mathrm{I}})  ]  \allowbreak \tilde{H}_{\mathrm{N}}^{\mathrm{II}} \allowbreak  [  1 \allowbreak + \allowbreak T_{1}^{\mathrm{II}} \allowbreak + \allowbreak T_{1}^{\mathrm{II}}T_{2}^{\mathrm{II}} \allowbreak + \allowbreak \frac{1}{2!}T_{1}^{\mathrm{II}^2} \allowbreak + \allowbreak \frac{1}{3!}T_{1}^{\mathrm{II}^3} \allowbreak + \allowbreak T_{2}^{\mathrm{II}}  ]  \allowbreak + \allowbreak  [  (-T_{1}^{\mathrm{I}})(-T_{2}^{\mathrm{II}})  ]  \allowbreak \tilde{H}_{\mathrm{N}}^{\mathrm{II}} \allowbreak  [  T_{11}^{\mathrm{I,II}} \allowbreak + \allowbreak T_{12}^{\mathrm{I,II}} \allowbreak + \allowbreak T_{1}^{\mathrm{I}} \allowbreak + \allowbreak T_{1}^{\mathrm{I}}T_{1}^{\mathrm{II}} \allowbreak + \allowbreak T_{1}^{\mathrm{I}}\frac{1}{2!}T_{1}^{\mathrm{II}^2} \allowbreak + \allowbreak T_{1}^{\mathrm{I}}T_{2}^{\mathrm{II}} \allowbreak + \allowbreak T_{1}^{\mathrm{II}}T_{11}^{\mathrm{I,II}}  ]  \allowbreak + \allowbreak  [  (-T_{1}^{\mathrm{I}})(-T_{1}^{\mathrm{II}})  ]  \allowbreak \tilde{H}_{\mathrm{N}}^{\mathrm{II}} \allowbreak  [  T_{11}^{\mathrm{I,II}} \allowbreak + \allowbreak T_{12}^{\mathrm{I,II}} \allowbreak + \allowbreak T_{1}^{\mathrm{I}} \allowbreak + \allowbreak T_{1}^{\mathrm{I}}T_{1}^{\mathrm{II}} \allowbreak + \allowbreak T_{1}^{\mathrm{I}}T_{1}^{\mathrm{II}}T_{2}^{\mathrm{II}} \allowbreak + \allowbreak T_{1}^{\mathrm{I}}\frac{1}{2!}T_{1}^{\mathrm{II}^2} \allowbreak + \allowbreak T_{1}^{\mathrm{I}}\frac{1}{3!}T_{1}^{\mathrm{II}^3} \allowbreak + \allowbreak T_{1}^{\mathrm{I}}T_{2}^{\mathrm{II}} \allowbreak + \allowbreak T_{1}^{\mathrm{II}}T_{11}^{\mathrm{I,II}} \allowbreak + \allowbreak T_{1}^{\mathrm{II}}T_{12}^{\mathrm{I,II}} \allowbreak + \allowbreak \frac{1}{2!}T_{1}^{\mathrm{II}^2}T_{11}^{\mathrm{I,II}} \allowbreak + \allowbreak T_{2}^{\mathrm{II}}T_{11}^{\mathrm{I,II}}  ]  \allowbreak + \allowbreak  [  (-T_{11}^{\mathrm{I,II}})(-T_{1}^{\mathrm{I}})(-T_{1}^{\mathrm{II}})  ]  \allowbreak \tilde{H}_{\mathrm{N}}^{\mathrm{II}} \allowbreak  [  1 \allowbreak + \allowbreak T_{1}^{\mathrm{II}} \allowbreak + \allowbreak \frac{1}{2!}T_{1}^{\mathrm{II}^2} \allowbreak + \allowbreak T_{2}^{\mathrm{II}}  ]  \allowbreak + \allowbreak  [  (-T_{1}^{\mathrm{I}})\frac{1}{2!}T_{1}^{\mathrm{II}^2}  ]  \allowbreak \tilde{H}_{\mathrm{N}}^{\mathrm{II}} \allowbreak  [  T_{11}^{\mathrm{I,II}} \allowbreak + \allowbreak T_{12}^{\mathrm{I,II}} \allowbreak + \allowbreak T_{1}^{\mathrm{I}} \allowbreak + \allowbreak T_{1}^{\mathrm{I}}T_{1}^{\mathrm{II}} \allowbreak + \allowbreak T_{1}^{\mathrm{I}}\frac{1}{2!}T_{1}^{\mathrm{II}^2} \allowbreak + \allowbreak T_{1}^{\mathrm{I}}T_{2}^{\mathrm{II}} \allowbreak + \allowbreak T_{1}^{\mathrm{II}}T_{11}^{\mathrm{I,II}}  ]  \allowbreak + \allowbreak  [  \frac{1}{2!}T_{1}^{\mathrm{I}^2}  ]  \allowbreak \tilde{H}_{\mathrm{N}}^{\mathrm{II}} \allowbreak  [  1 \allowbreak + \allowbreak T_{1}^{\mathrm{II}} \allowbreak + \allowbreak T_{1}^{\mathrm{II}}T_{2}^{\mathrm{II}} \allowbreak + \allowbreak \frac{1}{2!}T_{1}^{\mathrm{II}^2} \allowbreak + \allowbreak \frac{1}{2!}T_{1}^{\mathrm{II}^2}T_{2}^{\mathrm{II}} \allowbreak + \allowbreak \frac{1}{3!}T_{1}^{\mathrm{II}^3} \allowbreak + \allowbreak \frac{1}{4!}T_{1}^{\mathrm{II}^4} \allowbreak + \allowbreak T_{2}^{\mathrm{II}} \allowbreak + \allowbreak \frac{1}{2!}T_{2}^{\mathrm{II}^2}  ]  \allowbreak + \allowbreak  [  (-T_{2}^{\mathrm{II}})\frac{1}{2!}T_{1}^{\mathrm{I}^2}  ]  \allowbreak \tilde{H}_{\mathrm{N}}^{\mathrm{II}} \allowbreak  [  1 \allowbreak + \allowbreak T_{1}^{\mathrm{II}} \allowbreak + \allowbreak \frac{1}{2!}T_{1}^{\mathrm{II}^2} \allowbreak + \allowbreak T_{2}^{\mathrm{II}}  ]  \allowbreak + \allowbreak  [  (-T_{1}^{\mathrm{II}})\frac{1}{2!}T_{1}^{\mathrm{I}^2}  ]  \allowbreak \tilde{H}_{\mathrm{N}}^{\mathrm{II}} \allowbreak  [  1 \allowbreak + \allowbreak T_{1}^{\mathrm{II}} \allowbreak + \allowbreak T_{1}^{\mathrm{II}}T_{2}^{\mathrm{II}} \allowbreak + \allowbreak \frac{1}{2!}T_{1}^{\mathrm{II}^2} \allowbreak + \allowbreak \frac{1}{3!}T_{1}^{\mathrm{II}^3} \allowbreak + \allowbreak T_{2}^{\mathrm{II}}  ]  \allowbreak + \allowbreak  [  \frac{1}{2!}T_{1}^{\mathrm{I}^2}\frac{1}{2!}T_{1}^{\mathrm{II}^2}  ]  \allowbreak \tilde{H}_{\mathrm{N}}^{\mathrm{II}} \allowbreak  [  1 \allowbreak + \allowbreak T_{1}^{\mathrm{II}} \allowbreak + \allowbreak \frac{1}{2!}T_{1}^{\mathrm{II}^2} \allowbreak + \allowbreak T_{2}^{\mathrm{II}}  ]  \allowbreak + \allowbreak    \allowbreak \tilde{H}_{\mathrm{N}}^{\mathrm{I}} \allowbreak  [  T_{11}^{\mathrm{I,II}}T_{21}^{\mathrm{I,II}} \allowbreak + \allowbreak T_{12}^{\mathrm{I,II}} \allowbreak + \allowbreak T_{1}^{\mathrm{I}}T_{11}^{\mathrm{I,II}}T_{21}^{\mathrm{I,II}} \allowbreak + \allowbreak T_{1}^{\mathrm{I}}T_{12}^{\mathrm{I,II}} \allowbreak + \allowbreak T_{1}^{\mathrm{I}}T_{1}^{\mathrm{II}}T_{11}^{\mathrm{I,II}} \allowbreak + \allowbreak T_{1}^{\mathrm{I}}T_{1}^{\mathrm{II}}T_{21}^{\mathrm{I,II}} \allowbreak + \allowbreak T_{1}^{\mathrm{I}}\frac{1}{2!}T_{1}^{\mathrm{II}^2} \allowbreak + \allowbreak T_{1}^{\mathrm{I}}T_{22}^{\mathrm{I,II}} \allowbreak + \allowbreak T_{1}^{\mathrm{I}}T_{2}^{\mathrm{I}}T_{12}^{\mathrm{I,II}} \allowbreak + \allowbreak T_{1}^{\mathrm{I}}T_{2}^{\mathrm{I}}T_{1}^{\mathrm{II}}T_{11}^{\mathrm{I,II}} \allowbreak + \allowbreak T_{1}^{\mathrm{I}}T_{2}^{\mathrm{I}}\frac{1}{2!}T_{1}^{\mathrm{II}^2} \allowbreak + \allowbreak T_{1}^{\mathrm{I}}T_{2}^{\mathrm{I}}T_{2}^{\mathrm{II}} \allowbreak + \allowbreak T_{1}^{\mathrm{I}}T_{2}^{\mathrm{II}} \allowbreak + \allowbreak \frac{1}{2!}T_{1}^{\mathrm{I}^2}T_{12}^{\mathrm{I,II}} \allowbreak + \allowbreak \frac{1}{2!}T_{1}^{\mathrm{I}^2}T_{1}^{\mathrm{II}}T_{11}^{\mathrm{I,II}} \allowbreak + \allowbreak \frac{1}{2!}T_{1}^{\mathrm{I}^2}T_{1}^{\mathrm{II}}T_{21}^{\mathrm{I,II}} \allowbreak + \allowbreak \frac{1}{2!}T_{1}^{\mathrm{I}^2}\frac{1}{2!}T_{1}^{\mathrm{II}^2} \allowbreak + \allowbreak \frac{1}{2!}T_{1}^{\mathrm{I}^2}T_{22}^{\mathrm{I,II}} \allowbreak + \allowbreak \frac{1}{2!}T_{1}^{\mathrm{I}^2}T_{2}^{\mathrm{I}}\frac{1}{2!}T_{1}^{\mathrm{II}^2} \allowbreak + \allowbreak \frac{1}{2!}T_{1}^{\mathrm{I}^2}T_{2}^{\mathrm{I}}T_{2}^{\mathrm{II}} \allowbreak + \allowbreak \frac{1}{2!}T_{1}^{\mathrm{I}^2}T_{2}^{\mathrm{II}} \allowbreak + \allowbreak \frac{1}{3!}T_{1}^{\mathrm{I}^3}T_{12}^{\mathrm{I,II}} \allowbreak + \allowbreak \frac{1}{3!}T_{1}^{\mathrm{I}^3}T_{1}^{\mathrm{II}}T_{11}^{\mathrm{I,II}} \allowbreak + \allowbreak \frac{1}{3!}T_{1}^{\mathrm{I}^3}\frac{1}{2!}T_{1}^{\mathrm{II}^2} \allowbreak + \allowbreak \frac{1}{3!}T_{1}^{\mathrm{I}^3}T_{2}^{\mathrm{II}} \allowbreak + \allowbreak \frac{1}{4!}T_{1}^{\mathrm{I}^4}\frac{1}{2!}T_{1}^{\mathrm{II}^2} \allowbreak + \allowbreak \frac{1}{4!}T_{1}^{\mathrm{I}^4}T_{2}^{\mathrm{II}} \allowbreak + \allowbreak T_{1}^{\mathrm{II}}T_{11}^{\mathrm{I,II}} \allowbreak + \allowbreak T_{1}^{\mathrm{II}}T_{21}^{\mathrm{I,II}} \allowbreak + \allowbreak \frac{1}{2!}T_{1}^{\mathrm{II}^2} \allowbreak + \allowbreak \frac{1}{2!}T_{21}^{\mathrm{I,II}^2} \allowbreak + \allowbreak T_{22}^{\mathrm{I,II}} \allowbreak + \allowbreak T_{2}^{\mathrm{I}}T_{12}^{\mathrm{I,II}} \allowbreak + \allowbreak T_{2}^{\mathrm{I}}T_{1}^{\mathrm{II}}T_{11}^{\mathrm{I,II}} \allowbreak + \allowbreak T_{2}^{\mathrm{I}}T_{1}^{\mathrm{II}}T_{21}^{\mathrm{I,II}} \allowbreak + \allowbreak T_{2}^{\mathrm{I}}\frac{1}{2!}T_{1}^{\mathrm{II}^2} \allowbreak + \allowbreak T_{2}^{\mathrm{I}}T_{22}^{\mathrm{I,II}} \allowbreak + \allowbreak T_{2}^{\mathrm{I}}T_{2}^{\mathrm{II}} \allowbreak + \allowbreak \frac{1}{2!}T_{2}^{\mathrm{I}^2}\frac{1}{2!}T_{1}^{\mathrm{II}^2} \allowbreak + \allowbreak \frac{1}{2!}T_{2}^{\mathrm{I}^2}T_{2}^{\mathrm{II}} \allowbreak + \allowbreak T_{2}^{\mathrm{II}}  ]  \allowbreak + \allowbreak  [  (-T_{22}^{\mathrm{I,II}})  ]  \allowbreak \tilde{H}_{\mathrm{N}}^{\mathrm{I}} \allowbreak  [  1 \allowbreak + \allowbreak T_{1}^{\mathrm{I}} \allowbreak + \allowbreak \frac{1}{2!}T_{1}^{\mathrm{I}^2} \allowbreak + \allowbreak T_{2}^{\mathrm{I}}  ]  \allowbreak + \allowbreak  [  (-T_{21}^{\mathrm{I,II}})  ]  \allowbreak \tilde{H}_{\mathrm{N}}^{\mathrm{I}} \allowbreak  [  T_{11}^{\mathrm{I,II}} \allowbreak + \allowbreak T_{1}^{\mathrm{I}}T_{11}^{\mathrm{I,II}} \allowbreak + \allowbreak T_{1}^{\mathrm{I}}T_{1}^{\mathrm{II}} \allowbreak + \allowbreak \frac{1}{2!}T_{1}^{\mathrm{I}^2}T_{1}^{\mathrm{II}} \allowbreak + \allowbreak T_{1}^{\mathrm{II}} \allowbreak + \allowbreak T_{21}^{\mathrm{I,II}} \allowbreak + \allowbreak T_{2}^{\mathrm{I}}T_{1}^{\mathrm{II}}  ]  \allowbreak + \allowbreak  [  (-T_{12}^{\mathrm{I,II}})  ]  \allowbreak \tilde{H}_{\mathrm{N}}^{\mathrm{I}} \allowbreak  [  1 \allowbreak + \allowbreak T_{1}^{\mathrm{I}} \allowbreak + \allowbreak T_{1}^{\mathrm{I}}T_{2}^{\mathrm{I}} \allowbreak + \allowbreak \frac{1}{2!}T_{1}^{\mathrm{I}^2} \allowbreak + \allowbreak \frac{1}{3!}T_{1}^{\mathrm{I}^3} \allowbreak + \allowbreak T_{2}^{\mathrm{I}}  ]  \allowbreak + \allowbreak  [  \frac{1}{2!}T_{11}^{\mathrm{I,II}^2}  ]  \allowbreak \tilde{H}_{\mathrm{N}}^{\mathrm{I}} \allowbreak  [  1 \allowbreak + \allowbreak T_{1}^{\mathrm{I}} \allowbreak + \allowbreak \frac{1}{2!}T_{1}^{\mathrm{I}^2} \allowbreak + \allowbreak T_{2}^{\mathrm{I}}  ]  \allowbreak + \allowbreak  [  (-T_{11}^{\mathrm{I,II}})  ]  \allowbreak \tilde{H}_{\mathrm{N}}^{\mathrm{I}} \allowbreak  [  T_{11}^{\mathrm{I,II}} \allowbreak + \allowbreak T_{1}^{\mathrm{I}}T_{11}^{\mathrm{I,II}} \allowbreak + \allowbreak T_{1}^{\mathrm{I}}T_{1}^{\mathrm{II}} \allowbreak + \allowbreak T_{1}^{\mathrm{I}}T_{21}^{\mathrm{I,II}} \allowbreak + \allowbreak T_{1}^{\mathrm{I}}T_{2}^{\mathrm{I}}T_{1}^{\mathrm{II}} \allowbreak + \allowbreak \frac{1}{2!}T_{1}^{\mathrm{I}^2}T_{11}^{\mathrm{I,II}} \allowbreak + \allowbreak \frac{1}{2!}T_{1}^{\mathrm{I}^2}T_{1}^{\mathrm{II}} \allowbreak + \allowbreak \frac{1}{3!}T_{1}^{\mathrm{I}^3}T_{1}^{\mathrm{II}} \allowbreak + \allowbreak T_{1}^{\mathrm{II}} \allowbreak + \allowbreak T_{21}^{\mathrm{I,II}} \allowbreak + \allowbreak T_{2}^{\mathrm{I}}T_{11}^{\mathrm{I,II}} \allowbreak + \allowbreak T_{2}^{\mathrm{I}}T_{1}^{\mathrm{II}}  ]  \allowbreak + \allowbreak  [  (-T_{2}^{\mathrm{II}})  ]  \allowbreak \tilde{H}_{\mathrm{N}}^{\mathrm{I}} \allowbreak  [  1 \allowbreak + \allowbreak T_{1}^{\mathrm{I}} \allowbreak + \allowbreak T_{1}^{\mathrm{I}}T_{2}^{\mathrm{I}} \allowbreak + \allowbreak \frac{1}{2!}T_{1}^{\mathrm{I}^2} \allowbreak + \allowbreak \frac{1}{2!}T_{1}^{\mathrm{I}^2}T_{2}^{\mathrm{I}} \allowbreak + \allowbreak \frac{1}{3!}T_{1}^{\mathrm{I}^3} \allowbreak + \allowbreak \frac{1}{4!}T_{1}^{\mathrm{I}^4} \allowbreak + \allowbreak T_{2}^{\mathrm{I}} \allowbreak + \allowbreak \frac{1}{2!}T_{2}^{\mathrm{I}^2}  ]  \allowbreak + \allowbreak  [  (-T_{1}^{\mathrm{II}})  ]  \allowbreak \tilde{H}_{\mathrm{N}}^{\mathrm{I}} \allowbreak  [  T_{11}^{\mathrm{I,II}} \allowbreak + \allowbreak T_{1}^{\mathrm{I}}T_{11}^{\mathrm{I,II}} \allowbreak + \allowbreak T_{1}^{\mathrm{I}}T_{1}^{\mathrm{II}} \allowbreak + \allowbreak T_{1}^{\mathrm{I}}T_{21}^{\mathrm{I,II}} \allowbreak + \allowbreak T_{1}^{\mathrm{I}}T_{2}^{\mathrm{I}}T_{11}^{\mathrm{I,II}} \allowbreak + \allowbreak T_{1}^{\mathrm{I}}T_{2}^{\mathrm{I}}T_{1}^{\mathrm{II}} \allowbreak + \allowbreak \frac{1}{2!}T_{1}^{\mathrm{I}^2}T_{11}^{\mathrm{I,II}} \allowbreak + \allowbreak \frac{1}{2!}T_{1}^{\mathrm{I}^2}T_{1}^{\mathrm{II}} \allowbreak + \allowbreak \frac{1}{2!}T_{1}^{\mathrm{I}^2}T_{21}^{\mathrm{I,II}} \allowbreak + \allowbreak \frac{1}{2!}T_{1}^{\mathrm{I}^2}T_{2}^{\mathrm{I}}T_{1}^{\mathrm{II}} \allowbreak + \allowbreak \frac{1}{3!}T_{1}^{\mathrm{I}^3}T_{11}^{\mathrm{I,II}} \allowbreak + \allowbreak \frac{1}{3!}T_{1}^{\mathrm{I}^3}T_{1}^{\mathrm{II}} \allowbreak + \allowbreak \frac{1}{4!}T_{1}^{\mathrm{I}^4}T_{1}^{\mathrm{II}} \allowbreak + \allowbreak T_{1}^{\mathrm{II}} \allowbreak + \allowbreak T_{21}^{\mathrm{I,II}} \allowbreak + \allowbreak T_{2}^{\mathrm{I}}T_{11}^{\mathrm{I,II}} \allowbreak + \allowbreak T_{2}^{\mathrm{I}}T_{1}^{\mathrm{II}} \allowbreak + \allowbreak T_{2}^{\mathrm{I}}T_{21}^{\mathrm{I,II}} \allowbreak + \allowbreak \frac{1}{2!}T_{2}^{\mathrm{I}^2}T_{1}^{\mathrm{II}}  ]  \allowbreak + \allowbreak  [  (-T_{1}^{\mathrm{II}})(-T_{21}^{\mathrm{I,II}})  ]  \allowbreak \tilde{H}_{\mathrm{N}}^{\mathrm{I}} \allowbreak  [  1 \allowbreak + \allowbreak T_{1}^{\mathrm{I}} \allowbreak + \allowbreak \frac{1}{2!}T_{1}^{\mathrm{I}^2} \allowbreak + \allowbreak T_{2}^{\mathrm{I}}  ]  \allowbreak + \allowbreak  [  (-T_{11}^{\mathrm{I,II}})(-T_{1}^{\mathrm{II}})  ]  \allowbreak \tilde{H}_{\mathrm{N}}^{\mathrm{I}} \allowbreak  [  1 \allowbreak + \allowbreak T_{1}^{\mathrm{I}} \allowbreak + \allowbreak T_{1}^{\mathrm{I}}T_{2}^{\mathrm{I}} \allowbreak + \allowbreak \frac{1}{2!}T_{1}^{\mathrm{I}^2} \allowbreak + \allowbreak \frac{1}{3!}T_{1}^{\mathrm{I}^3} \allowbreak + \allowbreak T_{2}^{\mathrm{I}}  ]  \allowbreak + \allowbreak  [  \frac{1}{2!}T_{1}^{\mathrm{II}^2}  ]  \allowbreak \tilde{H}_{\mathrm{N}}^{\mathrm{I}} \allowbreak  [  1 \allowbreak + \allowbreak T_{1}^{\mathrm{I}} \allowbreak + \allowbreak T_{1}^{\mathrm{I}}T_{2}^{\mathrm{I}} \allowbreak + \allowbreak \frac{1}{2!}T_{1}^{\mathrm{I}^2} \allowbreak + \allowbreak \frac{1}{2!}T_{1}^{\mathrm{I}^2}T_{2}^{\mathrm{I}} \allowbreak + \allowbreak \frac{1}{3!}T_{1}^{\mathrm{I}^3} \allowbreak + \allowbreak \frac{1}{4!}T_{1}^{\mathrm{I}^4} \allowbreak + \allowbreak T_{2}^{\mathrm{I}} \allowbreak + \allowbreak \frac{1}{2!}T_{2}^{\mathrm{I}^2}  ]  \allowbreak + \allowbreak  [  (-T_{2}^{\mathrm{I}})  ]  \allowbreak \tilde{H}_{\mathrm{N}}^{\mathrm{I}} \allowbreak  [  T_{12}^{\mathrm{I,II}} \allowbreak + \allowbreak T_{1}^{\mathrm{I}}T_{12}^{\mathrm{I,II}} \allowbreak + \allowbreak T_{1}^{\mathrm{I}}T_{1}^{\mathrm{II}}T_{11}^{\mathrm{I,II}} \allowbreak + \allowbreak T_{1}^{\mathrm{I}}\frac{1}{2!}T_{1}^{\mathrm{II}^2} \allowbreak + \allowbreak T_{1}^{\mathrm{I}}T_{2}^{\mathrm{II}} \allowbreak + \allowbreak \frac{1}{2!}T_{1}^{\mathrm{I}^2}\frac{1}{2!}T_{1}^{\mathrm{II}^2} \allowbreak + \allowbreak \frac{1}{2!}T_{1}^{\mathrm{I}^2}T_{2}^{\mathrm{II}} \allowbreak + \allowbreak T_{1}^{\mathrm{II}}T_{11}^{\mathrm{I,II}} \allowbreak + \allowbreak T_{1}^{\mathrm{II}}T_{21}^{\mathrm{I,II}} \allowbreak + \allowbreak \frac{1}{2!}T_{1}^{\mathrm{II}^2} \allowbreak + \allowbreak T_{22}^{\mathrm{I,II}} \allowbreak + \allowbreak T_{2}^{\mathrm{I}}\frac{1}{2!}T_{1}^{\mathrm{II}^2} \allowbreak + \allowbreak T_{2}^{\mathrm{I}}T_{2}^{\mathrm{II}} \allowbreak + \allowbreak T_{2}^{\mathrm{II}}  ]  \allowbreak + \allowbreak  [  (-T_{2}^{\mathrm{I}})(-T_{2}^{\mathrm{II}})  ]  \allowbreak \tilde{H}_{\mathrm{N}}^{\mathrm{I}} \allowbreak  [  1 \allowbreak + \allowbreak T_{1}^{\mathrm{I}} \allowbreak + \allowbreak \frac{1}{2!}T_{1}^{\mathrm{I}^2} \allowbreak + \allowbreak T_{2}^{\mathrm{I}}  ]  \allowbreak + \allowbreak  [  (-T_{1}^{\mathrm{II}})(-T_{2}^{\mathrm{I}})  ]  \allowbreak \tilde{H}_{\mathrm{N}}^{\mathrm{I}} \allowbreak  [  T_{11}^{\mathrm{I,II}} \allowbreak + \allowbreak T_{1}^{\mathrm{I}}T_{11}^{\mathrm{I,II}} \allowbreak + \allowbreak T_{1}^{\mathrm{I}}T_{1}^{\mathrm{II}} \allowbreak + \allowbreak \frac{1}{2!}T_{1}^{\mathrm{I}^2}T_{1}^{\mathrm{II}} \allowbreak + \allowbreak T_{1}^{\mathrm{II}} \allowbreak + \allowbreak T_{21}^{\mathrm{I,II}} \allowbreak + \allowbreak T_{2}^{\mathrm{I}}T_{1}^{\mathrm{II}}  ]  \allowbreak + \allowbreak  [  (-T_{2}^{\mathrm{I}})\frac{1}{2!}T_{1}^{\mathrm{II}^2}  ]  \allowbreak \tilde{H}_{\mathrm{N}}^{\mathrm{I}} \allowbreak  [  1 \allowbreak + \allowbreak T_{1}^{\mathrm{I}} \allowbreak + \allowbreak \frac{1}{2!}T_{1}^{\mathrm{I}^2} \allowbreak + \allowbreak T_{2}^{\mathrm{I}}  ]  \allowbreak + \allowbreak  [  (-T_{1}^{\mathrm{I}})  ]  \allowbreak \tilde{H}_{\mathrm{N}}^{\mathrm{I}} \allowbreak  [  T_{11}^{\mathrm{I,II}}T_{21}^{\mathrm{I,II}} \allowbreak + \allowbreak T_{12}^{\mathrm{I,II}} \allowbreak + \allowbreak T_{1}^{\mathrm{I}}T_{12}^{\mathrm{I,II}} \allowbreak + \allowbreak T_{1}^{\mathrm{I}}T_{1}^{\mathrm{II}}T_{11}^{\mathrm{I,II}} \allowbreak + \allowbreak T_{1}^{\mathrm{I}}T_{1}^{\mathrm{II}}T_{21}^{\mathrm{I,II}} \allowbreak + \allowbreak T_{1}^{\mathrm{I}}\frac{1}{2!}T_{1}^{\mathrm{II}^2} \allowbreak + \allowbreak T_{1}^{\mathrm{I}}T_{22}^{\mathrm{I,II}} \allowbreak + \allowbreak T_{1}^{\mathrm{I}}T_{2}^{\mathrm{I}}\frac{1}{2!}T_{1}^{\mathrm{II}^2} \allowbreak + \allowbreak T_{1}^{\mathrm{I}}T_{2}^{\mathrm{I}}T_{2}^{\mathrm{II}} \allowbreak + \allowbreak T_{1}^{\mathrm{I}}T_{2}^{\mathrm{II}} \allowbreak + \allowbreak \frac{1}{2!}T_{1}^{\mathrm{I}^2}T_{12}^{\mathrm{I,II}} \allowbreak + \allowbreak \frac{1}{2!}T_{1}^{\mathrm{I}^2}T_{1}^{\mathrm{II}}T_{11}^{\mathrm{I,II}} \allowbreak + \allowbreak \frac{1}{2!}T_{1}^{\mathrm{I}^2}\frac{1}{2!}T_{1}^{\mathrm{II}^2} \allowbreak + \allowbreak \frac{1}{2!}T_{1}^{\mathrm{I}^2}T_{2}^{\mathrm{II}} \allowbreak + \allowbreak \frac{1}{3!}T_{1}^{\mathrm{I}^3}\frac{1}{2!}T_{1}^{\mathrm{II}^2} \allowbreak + \allowbreak \frac{1}{3!}T_{1}^{\mathrm{I}^3}T_{2}^{\mathrm{II}} \allowbreak + \allowbreak T_{1}^{\mathrm{II}}T_{11}^{\mathrm{I,II}} \allowbreak + \allowbreak T_{1}^{\mathrm{II}}T_{21}^{\mathrm{I,II}} \allowbreak + \allowbreak \frac{1}{2!}T_{1}^{\mathrm{II}^2} \allowbreak + \allowbreak T_{22}^{\mathrm{I,II}} \allowbreak + \allowbreak T_{2}^{\mathrm{I}}T_{12}^{\mathrm{I,II}} \allowbreak + \allowbreak T_{2}^{\mathrm{I}}T_{1}^{\mathrm{II}}T_{11}^{\mathrm{I,II}} \allowbreak + \allowbreak T_{2}^{\mathrm{I}}\frac{1}{2!}T_{1}^{\mathrm{II}^2} \allowbreak + \allowbreak T_{2}^{\mathrm{I}}T_{2}^{\mathrm{II}} \allowbreak + \allowbreak T_{2}^{\mathrm{II}}  ]  \allowbreak + \allowbreak  [  (-T_{12}^{\mathrm{I,II}})(-T_{1}^{\mathrm{I}})  ]  \allowbreak \tilde{H}_{\mathrm{N}}^{\mathrm{I}} \allowbreak  [  1 \allowbreak + \allowbreak T_{1}^{\mathrm{I}} \allowbreak + \allowbreak \frac{1}{2!}T_{1}^{\mathrm{I}^2} \allowbreak + \allowbreak T_{2}^{\mathrm{I}}  ]  \allowbreak + \allowbreak  [  (-T_{11}^{\mathrm{I,II}})(-T_{1}^{\mathrm{I}})  ]  \allowbreak \tilde{H}_{\mathrm{N}}^{\mathrm{I}} \allowbreak  [  T_{11}^{\mathrm{I,II}} \allowbreak + \allowbreak T_{1}^{\mathrm{I}}T_{11}^{\mathrm{I,II}} \allowbreak + \allowbreak T_{1}^{\mathrm{I}}T_{1}^{\mathrm{II}} \allowbreak + \allowbreak \frac{1}{2!}T_{1}^{\mathrm{I}^2}T_{1}^{\mathrm{II}} \allowbreak + \allowbreak T_{1}^{\mathrm{II}} \allowbreak + \allowbreak T_{21}^{\mathrm{I,II}} \allowbreak + \allowbreak T_{2}^{\mathrm{I}}T_{1}^{\mathrm{II}}  ]  \allowbreak + \allowbreak  [  (-T_{1}^{\mathrm{I}})(-T_{2}^{\mathrm{II}})  ]  \allowbreak \tilde{H}_{\mathrm{N}}^{\mathrm{I}} \allowbreak  [  1 \allowbreak + \allowbreak T_{1}^{\mathrm{I}} \allowbreak + \allowbreak T_{1}^{\mathrm{I}}T_{2}^{\mathrm{I}} \allowbreak + \allowbreak \frac{1}{2!}T_{1}^{\mathrm{I}^2} \allowbreak + \allowbreak \frac{1}{3!}T_{1}^{\mathrm{I}^3} \allowbreak + \allowbreak T_{2}^{\mathrm{I}}  ]  \allowbreak + \allowbreak  [  (-T_{1}^{\mathrm{I}})(-T_{1}^{\mathrm{II}})  ]  \allowbreak \tilde{H}_{\mathrm{N}}^{\mathrm{I}} \allowbreak  [  T_{11}^{\mathrm{I,II}} \allowbreak + \allowbreak T_{1}^{\mathrm{I}}T_{11}^{\mathrm{I,II}} \allowbreak + \allowbreak T_{1}^{\mathrm{I}}T_{1}^{\mathrm{II}} \allowbreak + \allowbreak T_{1}^{\mathrm{I}}T_{21}^{\mathrm{I,II}} \allowbreak + \allowbreak T_{1}^{\mathrm{I}}T_{2}^{\mathrm{I}}T_{1}^{\mathrm{II}} \allowbreak + \allowbreak \frac{1}{2!}T_{1}^{\mathrm{I}^2}T_{11}^{\mathrm{I,II}} \allowbreak + \allowbreak \frac{1}{2!}T_{1}^{\mathrm{I}^2}T_{1}^{\mathrm{II}} \allowbreak + \allowbreak \frac{1}{3!}T_{1}^{\mathrm{I}^3}T_{1}^{\mathrm{II}} \allowbreak + \allowbreak T_{1}^{\mathrm{II}} \allowbreak + \allowbreak T_{21}^{\mathrm{I,II}} \allowbreak + \allowbreak T_{2}^{\mathrm{I}}T_{11}^{\mathrm{I,II}} \allowbreak + \allowbreak T_{2}^{\mathrm{I}}T_{1}^{\mathrm{II}}  ]  \allowbreak + \allowbreak  [  (-T_{11}^{\mathrm{I,II}})(-T_{1}^{\mathrm{I}})(-T_{1}^{\mathrm{II}})  ]  \allowbreak \tilde{H}_{\mathrm{N}}^{\mathrm{I}} \allowbreak  [  1 \allowbreak + \allowbreak T_{1}^{\mathrm{I}} \allowbreak + \allowbreak \frac{1}{2!}T_{1}^{\mathrm{I}^2} \allowbreak + \allowbreak T_{2}^{\mathrm{I}}  ]  \allowbreak + \allowbreak  [  (-T_{1}^{\mathrm{I}})\frac{1}{2!}T_{1}^{\mathrm{II}^2}  ]  \allowbreak \tilde{H}_{\mathrm{N}}^{\mathrm{I}} \allowbreak  [  1 \allowbreak + \allowbreak T_{1}^{\mathrm{I}} \allowbreak + \allowbreak T_{1}^{\mathrm{I}}T_{2}^{\mathrm{I}} \allowbreak + \allowbreak \frac{1}{2!}T_{1}^{\mathrm{I}^2} \allowbreak + \allowbreak \frac{1}{3!}T_{1}^{\mathrm{I}^3} \allowbreak + \allowbreak T_{2}^{\mathrm{I}}  ]  \allowbreak + \allowbreak  [  \frac{1}{2!}T_{1}^{\mathrm{I}^2}  ]  \allowbreak \tilde{H}_{\mathrm{N}}^{\mathrm{I}} \allowbreak  [  T_{12}^{\mathrm{I,II}} \allowbreak + \allowbreak T_{1}^{\mathrm{I}}T_{12}^{\mathrm{I,II}} \allowbreak + \allowbreak T_{1}^{\mathrm{I}}T_{1}^{\mathrm{II}}T_{11}^{\mathrm{I,II}} \allowbreak + \allowbreak T_{1}^{\mathrm{I}}\frac{1}{2!}T_{1}^{\mathrm{II}^2} \allowbreak + \allowbreak T_{1}^{\mathrm{I}}T_{2}^{\mathrm{II}} \allowbreak + \allowbreak \frac{1}{2!}T_{1}^{\mathrm{I}^2}\frac{1}{2!}T_{1}^{\mathrm{II}^2} \allowbreak + \allowbreak \frac{1}{2!}T_{1}^{\mathrm{I}^2}T_{2}^{\mathrm{II}} \allowbreak + \allowbreak T_{1}^{\mathrm{II}}T_{11}^{\mathrm{I,II}} \allowbreak + \allowbreak T_{1}^{\mathrm{II}}T_{21}^{\mathrm{I,II}} \allowbreak + \allowbreak \frac{1}{2!}T_{1}^{\mathrm{II}^2} \allowbreak + \allowbreak T_{22}^{\mathrm{I,II}} \allowbreak + \allowbreak T_{2}^{\mathrm{I}}\frac{1}{2!}T_{1}^{\mathrm{II}^2} \allowbreak + \allowbreak T_{2}^{\mathrm{I}}T_{2}^{\mathrm{II}} \allowbreak + \allowbreak T_{2}^{\mathrm{II}}  ]  \allowbreak + \allowbreak  [  (-T_{2}^{\mathrm{II}})\frac{1}{2!}T_{1}^{\mathrm{I}^2}  ]  \allowbreak \tilde{H}_{\mathrm{N}}^{\mathrm{I}} \allowbreak  [  1 \allowbreak + \allowbreak T_{1}^{\mathrm{I}} \allowbreak + \allowbreak \frac{1}{2!}T_{1}^{\mathrm{I}^2} \allowbreak + \allowbreak T_{2}^{\mathrm{I}}  ]  \allowbreak + \allowbreak  [  (-T_{1}^{\mathrm{II}})\frac{1}{2!}T_{1}^{\mathrm{I}^2}  ]  \allowbreak \tilde{H}_{\mathrm{N}}^{\mathrm{I}} \allowbreak  [  T_{11}^{\mathrm{I,II}} \allowbreak + \allowbreak T_{1}^{\mathrm{I}}T_{11}^{\mathrm{I,II}} \allowbreak + \allowbreak T_{1}^{\mathrm{I}}T_{1}^{\mathrm{II}} \allowbreak + \allowbreak \frac{1}{2!}T_{1}^{\mathrm{I}^2}T_{1}^{\mathrm{II}} \allowbreak + \allowbreak T_{1}^{\mathrm{II}} \allowbreak + \allowbreak T_{21}^{\mathrm{I,II}} \allowbreak + \allowbreak T_{2}^{\mathrm{I}}T_{1}^{\mathrm{II}}  ]  \allowbreak + \allowbreak  [  \frac{1}{2!}T_{1}^{\mathrm{I}^2}\frac{1}{2!}T_{1}^{\mathrm{II}^2}  ]  \allowbreak \tilde{H}_{\mathrm{N}}^{\mathrm{I}} \allowbreak  [  1 \allowbreak + \allowbreak T_{1}^{\mathrm{I}} \allowbreak + \allowbreak \frac{1}{2!}T_{1}^{\mathrm{I}^2} \allowbreak + \allowbreak T_{2}^{\mathrm{I}}  ] \vert 0^{\mathrm{I}}0^{\mathrm{II}} \rangle = 0$

%% file: appendix_b.tex
\begin{table}[ht]
  \begin{center}
   \caption{\textbf{Exponents of the Gaussian basis functions used the calculations.}}
   \label{tab:basis}
    \begin{tabular}{c c c c c c c c}
    \hline
     $k=0.0001$ & $k=0.0010$ & $k=0.0100$ & $k=0.1000$ & $k=0.2500$ & $k=0.5000$ & $k=1.0000$ & $k=5.0000$ \\
      \hline
     5.50$\times 10^{-2}$   &   9.32$\times 10^{-2}$   &   1.38$\times 10^{-1}$   &   4.36$\times 10^{-1}$   &   1.00$\times 10^{-1}$   &   1.96$\times 10^{-1}$   &   1.16                   &   6.21$\times 10^{-1}$   \\
     1.66$\times 10^{-2}$   &   3.84$\times 10^{-2}$   &   8.31$\times 10^{-2}$   &   2.63$\times 10^{-1}$   &   1.58$\times 10^{-1}$   &   2.63$\times 10^{-1}$   &   7.62$\times 10^{-1}$   &   8.34$\times 10^{-1}$   \\
     5.00$\times 10^{-3}$   &   1.58$\times 10^{-2}$   &   5.00$\times 10^{-2}$   &   1.58$\times 10^{-1}$   &   2.50$\times 10^{-1}$   &   3.54$\times 10^{-1}$   &   5.00$\times 10^{-1}$   &   1.12   \\
     1.51$\times 10^{-3}$   &   6.51$\times 10^{-3}$   &   3.01$\times 10^{-2}$   &   9.51$\times 10^{-2}$   &   3.95$\times 10^{-1}$   &   4.75$\times 10^{-1}$   &   3.28$\times 10^{-1}$   &   1.50   \\
     4.55$\times 10^{-4}$   &   2.68$\times 10^{-3}$   &   1.81$\times 10^{-2}$   &   5.73$\times 10^{-2}$   &   6.25$\times 10^{-1}$   &   6.38$\times 10^{-1}$   &   2.16$\times 10^{-1}$   &   2.02   \\
     1.37$\times 10^{-4}$   &   1.10$\times 10^{-3}$   &   1.09$\times 10^{-2}$   &   3.45$\times 10^{-2}$   &   9.88$\times 10^{-1}$   &   8.56$\times 10^{-1}$   &   1.42$\times 10^{-1}$   &   2.71   \\
     4.13$\times 10^{-5}$   &   4.54$\times 10^{-4}$   &   6.57$\times 10^{-3}$   &   2.08$\times 10^{-2}$   &   1.56                   &   1.15                   &   9.29$\times 10^{-2}$   &   3.64   \\
     1.25$\times 10^{-5}$   &   1.87$\times 10^{-4}$   &   3.95$\times 10^{-3}$   &   1.25$\times 10^{-2}$   &   2.47                   &   1.54                   &   6.10$\times 10^{-2}$   &   4.89   \\
     3.76$\times 10^{-6}$   &   7.70$\times 10^{-5}$   &   2.38$\times 10^{-3}$   &   7.52$\times 10^{-3}$   &   3.91                   &   2.07                   &   4.00$\times 10^{-2}$   &   6.57   \\
     1.13$\times 10^{-6}$   &   3.17$\times 10^{-5}$   &   1.43$\times 10^{-3}$   &   4.53$\times 10^{-3}$   &   6.18                   &   2.79                   &   2.63$\times 10^{-2}$   &   8.82   \\ 
     \hline 
    \end{tabular}
  \end{center}  
\end{table} 